\documentclass{aa}  

\usepackage{natbib}
\usepackage{subfigure}

\usepackage{graphicx}
\usepackage{txfonts}
\let\ACMmaketitle=\maketitle
\renewcommand{\maketitle}{\begingroup\let\footnote=\thanks \ACMmaketitle\endgroup}

\begin{document}

   \title{Characterisation of stellar activity of M dwarfs. I. Long-timescale variability in a large sample and detection of new cycles\footnote{Full Table A.1 is only available at the CDS via anonymous ftp to cdsarc.u-strasbg.fr (130.79.128.5) or via http://cdsarc.u-strasbg.fr/viz-bin/cat/J/A+A/XXX/AX}}

   \titlerunning{Characterisation of stellar activity of M dwarfs. I. Long-timescale variability}

\author{L. Mignon \inst{1,2}, N. Meunier \inst{1}, X. Delfosse \inst{1}, X. Bonfils \inst{1}, N. C. Santos \inst{3,4}, T. Forveille \inst{1}, G. Gaisné \inst{1}, N. Astudillo-Defru \inst{5}, C. Lovis \inst{2}, and S. Udry \inst{2}
}
\authorrunning{L. Mignon et al.}
 
\institute{
Univ. Grenoble Alpes, CNRS, IPAG, F-38000 Grenoble, France 
\and
Observatoire astronomique de l'Université de Genève, 51 chemin Pegasis, 1290 Versoix, Switzerland
\and Instituto de Astrofísica e Ciências do Espaço, Universidade do Porto, CAUP, Rua das Estrelas, 4150-762 Porto, Portugal
\and Departamento de Física e Astronomia, Faculdade de Ciências, Universidade do Porto, Rua Campo Alegre, 4169-007 Porto, Portugal
\and Departamento de Matemática y Física Aplicadas, Universidad Católica de la Santísima Concepción, Alonso de Rivera 2850, Concepción, Chile
}

   \date{Received June 13, 2022; accepted February 8, 2023}

\abstract{M dwarfs are active stars that exhibit variability in chromospheric emission and photometry at short and long timescales, including long cycles that are related to dynamo processes. This activity also impacts the search for exoplanets because it affects the radial velocities.}  
{We analysed a large sample of 177 M dwarfs observed with HARPS during the period 2003-2020 in order to characterise the long-term variability of these stars. We compared the variability obtained in three chromospheric activity indices (Ca II H \& K, the Na D doublet, and H$\alpha$) and with ASAS photometry. 
}
{We focused on the detailed analysis of the chromospheric emission based on linear, quadratic, and sinusoidal models. We used various tools to estimate the significance of the variability and to quantify the improvement brought by the models. In addition, we analysed complementary photometric time series for the most variable stars to be able to provide a broader view of the long-term variability in M dwarfs. }
{We find that most stars are significantly variable, even the quietest stars. 
Most stars in our sample (75\%) exhibit a long-term variability, which manifests itself mostly through linear or quadratic variability, although the true behaviour may be more complex. We found significant variability with estimated timescales for 24 stars, and estimated the lower limit for a possible cycle period for an additional  9 stars that were not previously published. We found evidence of complex variability because more than one long-term timescale may be present for at least 12 stars, together with significant differences between the behaviour of the three activity indices. This complexity may also be the source of the discrepancies observed between previous publications.}
{We conclude that long-term variability is present for all spectral types and activity level in M dwarfs, without a significant trend with spectral type or mean activity level.}
    \keywords{Stars: activity -- Stars: chromospheres - techniques: spectroscopy -- planetary systems} 
 
   \maketitle
%

\section{Introduction}

Stellar activity is an important research topic in itself, allowing us to probe the stellar structure and evolution. It is also closely linked to studies of exoplanets. In this context, it is particularly important to better characterise the activity of M dwarfs because interest  in a search for exoplanets around these stars is currently considerable, particularly, to study planets in their habitable zones (HZ) and to characterise their core density and atmosphere \cite[]{doyon14,beichman14,snellen15,lovis17}. Accordingly, numerous exoplanet surveys are conducted out around M dwarfs, whether by radial velocity \cite[hereafter 
RV;
e.g.][]{zechmeister09,bonfils13,mignon22a} or by transit \cite[e.g.][]{kopparapu13,dressing15,hsu2020}.
However, RVs are strongly impacted by stellar activity by spot and plage contrasts, but also through inhibition of the convection   in plages \cite[e.g.][]{saar99,meunier10,dumusque14}, with an impact on the rotational timescale. The rotational signal frequently even dominates the RV time series, for example GJ~205 \cite[][]{hebrard2016modelling,cortes23}, for GJ~388  \cite[][]{morin08,carmona22},  and Proxima Centauri \cite[][]{faria22}.
The inhibition of convection also impacts longer timescales (activity cycles). Other processes can also impact long-term variability, such as the meridional circulation \cite[][]{makarov10,meunier20c}, and possibly granulation or supergranulation   \cite[][]{dumusque11b,meunier15,meunier19e,meunier20b}, but the small-scale flows are expected to be smaller in M stars than in FGK stars \cite[][]{allendeprieto13,beeck13,beeck13a,tremblay13,meunier17,meunier17b}.  These processes are reviewed in \cite{meunier21b}. Stellar activity can lead to the publication of exoplanets that are later invalidated \cite[e.g.][]{robertson14,robertson14b,santos14}, showing that it is necessary to take stellar activity into account \cite[e.g.][]{bonfils07}.
Furthermore, stellar activity through the effect of winds and coronal mass ejections might impact the atmosphere and then the habitability conditions
\cite[e.g.][]{ip00,buccino06,buccino07,vonbloh07,jardine13,vidotto13,vidotto15,strugarek16}, or it might provoke the erosion of planetary atmospheres \cite[e.g.][]{attia21}.

The fraction of active stars is much higher among M dwarfs than among solar-type stars due to two effects: M dwarfs have much longer rotational braking times (\cite{delfosse98,barnes03,delorme11}), and 
they show stronger activity emission for a given rotation period \cite[e.g.][]{kiraga07}.
Their activity is characterised by strong chromospheric and corona emissions with a dependence on spectral type \cite[][]{reiners10,reiners12,astudillo17}, and their level of activity exhibits a strong correlation with the stellar rotational period, whether observed in X-band luminosity \cite[e.g.][]{pizzolato03,kiraga07,wright11}  or
through a chromospheric activity estimator such as the H$\alpha$ or Ca II lines \cite[e.g.][]{astudillo17,wright18}.
This suggests that a dynamo process is at work inside these stars. 
The usual solar-type $\alpha-\omega$ dynamo \cite[][]{parker55,steenbeck69,robinson82,saar99,charbonneau10} is  based on an interaction between differential rotation and turbulence, with a strong role of the velocity gradients 
in the tachocline that is present only in partially convective stars  \cite[][]{spiegel92,charbonneau97,dikpati05}. 
Due to the transition between partially convective stars and fully convective stars around spectral type M3.5, or a mass around 0.35M$_\odot$ \cite[][]{chabrier97},
this role of the tachocline  might be questioned in fully convective stars \cite[][]{barnes03}, for which $\alpha^2$ dynamos, which are more widely distributed and lack a long-term cycle, were expected. However, \cite{chabrier06}  showed that it was possible to generate large-scale fields for these stars as well, and \cite{browning08} modelled 3D dynamos able to generate strong fields without any tachocline.  \cite{yadav15}, \cite{yadav16}, and \cite{wright16} later found that a tachocline was not as critical, with the possibility of generating solar-type cycles even for fully convective stars.

Long- and mid-term activity of FGK stars is relatively well characterised, in particular because the Sun is a good model. In addition, long chromospheric emission surveys have been performed \cite[][]{wilson78,baliunas85,baliunas95}, and more recently with HARPS \cite[][]{lovis11b}, and have been combined with photometric analysis \cite[][]{radick98,lockwood07,radick18}. For example, this allowed relating cycle  and rotation periods, which can then be interpreted in terms of dynamo actions \cite[e.g.][]{bohm07,saar11}. However, it is not as well characterised  for M dwarfs, and there is very likely a large diversity. Properties very different from solar-type stars have also been observed, such as very long-lived stable spots, for example for GJ~674 \cite[][]{bonfils07,cortes23}, or GJ~388 \cite[][]{morin08,carmona22},
and a few other stars \cite[][]{robertson20}. Long-term chromospheric variability has been analysed in limited samples  \cite[][]{gomes11,gomes12,robertson13}, leading to 10-20 stars with a trend or quadratic variability. Cycles have also been published for M dwarfs, some of them based on a large sample of stars, mostly in photometry 
 \cite[][]{savanov12,vida13,vida14,suarez16,kuker19}, and sometimes in combination with a spectroscopic analysis \cite[][]{suarez18}. 
 To summarise, cycles have been observed for both fast and slow rotators,
and for both partially and fully convective stars \cite[as shown e.g. by the sample of stars in ][see below for a more complete list]{buccino11,ibanez19,ibanez19b,ibanez20}. 
 However, we note some incompatibilities  in the publications for a given star. A complex correlation between activity indicators is also observed because cycles or long-term variability is not always detected in all time series, including some anti-correlation between activity indicators \cite[][]{cincunegui07} depending on the bandpass used to compute index (at least for FGK stars), as shown in \citet[][]{gomes22}, which shows that the analysis and interpretation of this long-term variability is not straightforward, as shown for FGK stars \cite[][]{gomes14,meunier22}.

Our objective is first to characterise the long-term chromospheric variability of M dwarfs based on the largest sample so far, regardless of whether it is cyclic. We then wish to compare these properties with photometric variability whenever possible to build a global view of this long-term variability. We also address the dependence on stellar parameters. Finally, we produce a list of stars with interesting variability, which is useful for exoplanet studies and as targets for future stellar variability observations. 
The outline of this paper is the following. 
In Sect.~2 we present the stellar sample, the chromospheric emission indices (Ca, Na, and H$\alpha$) used in the analysis, and the selection process for the data. 
We present the results in Sect.~3 and conclude in Sect.~4.

\section{Data and analysis}

In this section, we first describe the stellar sample and the HARPS data used in the analysis. 
The activity indices are defined first, and the selection process of the spectra are defined second. 

\subsection{M dwarfs observed with HARPS}


We first built a list of known M dwarfs by crossing two main catalogues, \cite{gaidos2014} and \cite{winters2014solar}. The latter was updated with \cite{winters2021}, which did not add any target, and was then completed with the release of the Gliese and Jahreiss catalogues \cite[][]{stauffer2010accurate} and the 100 closest systems gathered by the Research Consortium On Nearby Stars (RECONS\footnote{\url{http://www.recons.org/TOP100.posted.htm}}) \cite[][]{henry2018}. 
This list of stars will also be used in a companion paper \cite[][]{mignon22a} in which we study the RVs of M dwarfs in the solar neighbourhood.

We then crossed this list  of M dwarfs of the solar neighbourhood with the database of public spectra taken with the high-accuracy RV planet searcher (HARPS) installed on the ESO 3.6 m telescope at La Silla Observatory in Chile \cite{mayor03}.  Each spectrum covers the 380-690 nm spectral range, spread over 71 orders, with a median resolving power of 115000, and is wavelength calibrated with a high accuracy. 
All observations are downloaded on the ESO archive (spectral data products query form\footnote{\url{http://archive.eso.org/wdb/wdb/adp/phase3_spectral/form}}) and come from several observation programs of M dwarfs (listed in the acknowledgements): this represents about 15 000 spectra from January 2003 to June 2019 for 490 stars.

\begin{figure}
\includegraphics{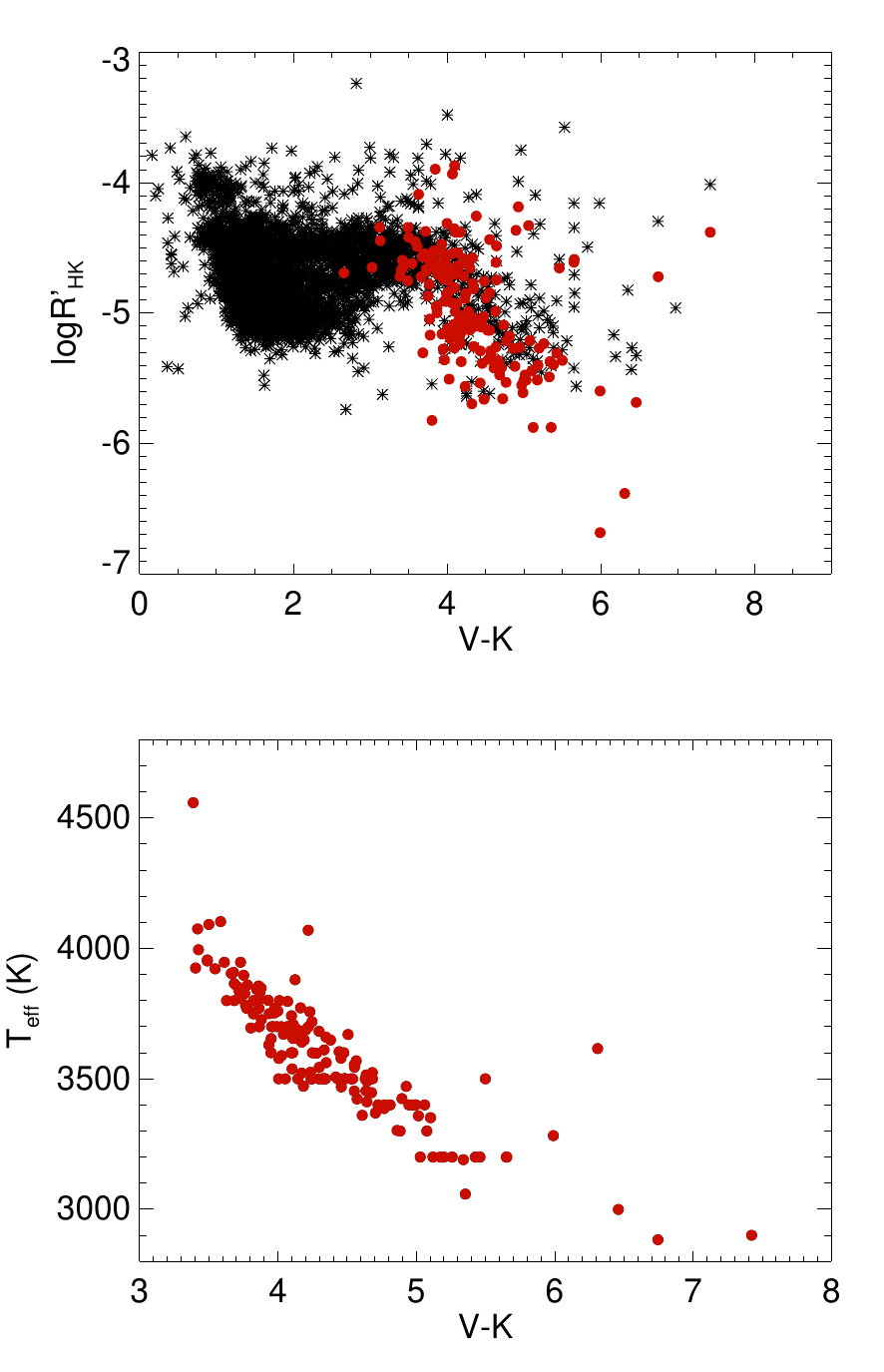}
\caption{$\log R'_{HK}$ vs V-K for our sample of 177 M dwarfs (red) and for the sample of FGKM stars of \cite{borosaikia18} (black). The lower panel shows T$_{\rm eff}$ vs V-K for stars in our M sample. 
}
\label{fsample}
\end{figure}

\begin{figure}
\includegraphics{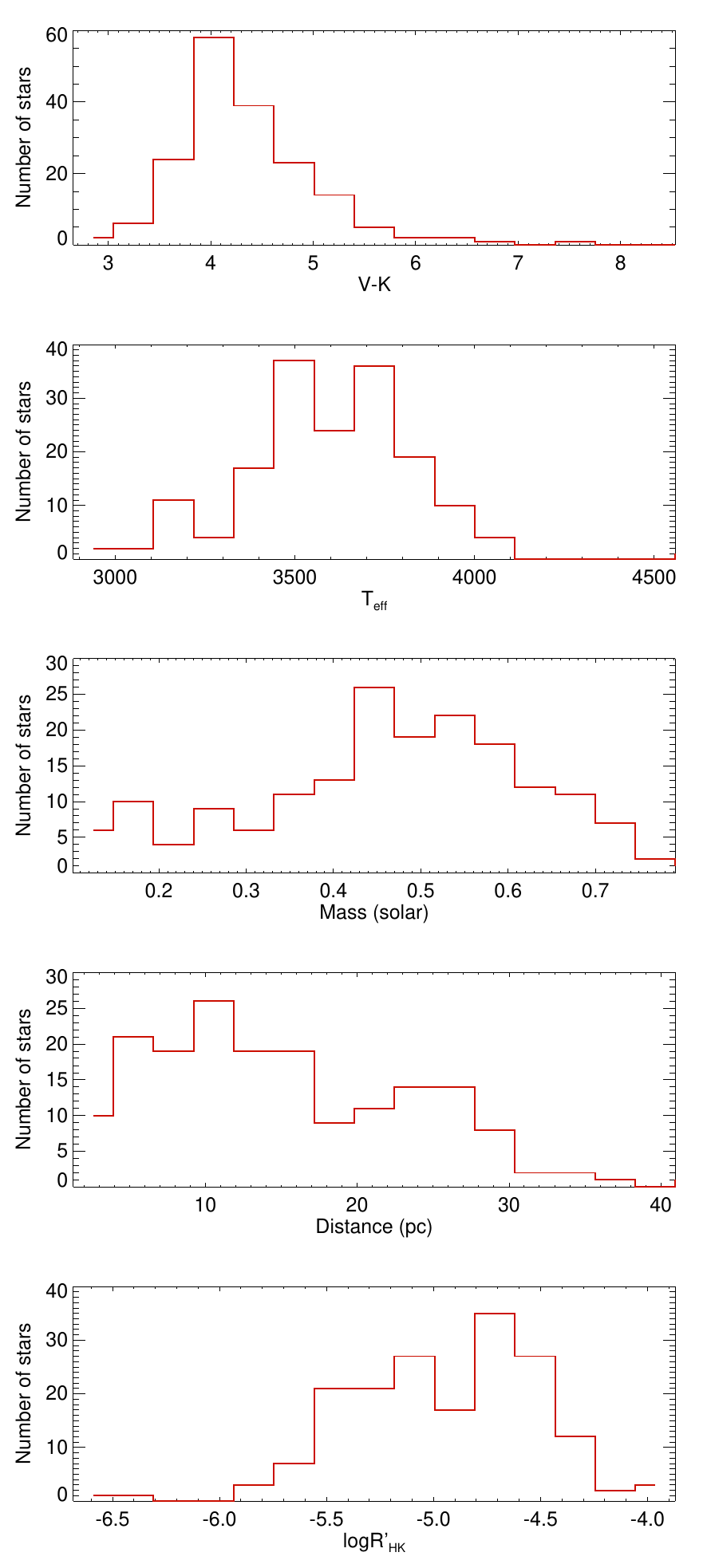}
\caption{
Distribution of V-K, T$_{\rm eff}$, stellar mass, distance, and $\log R'_{HK}$ (from top to bottom) for our sample of 177 M dwarfs.
}
\label{sample2}
\end{figure}

\begin{figure}
\includegraphics{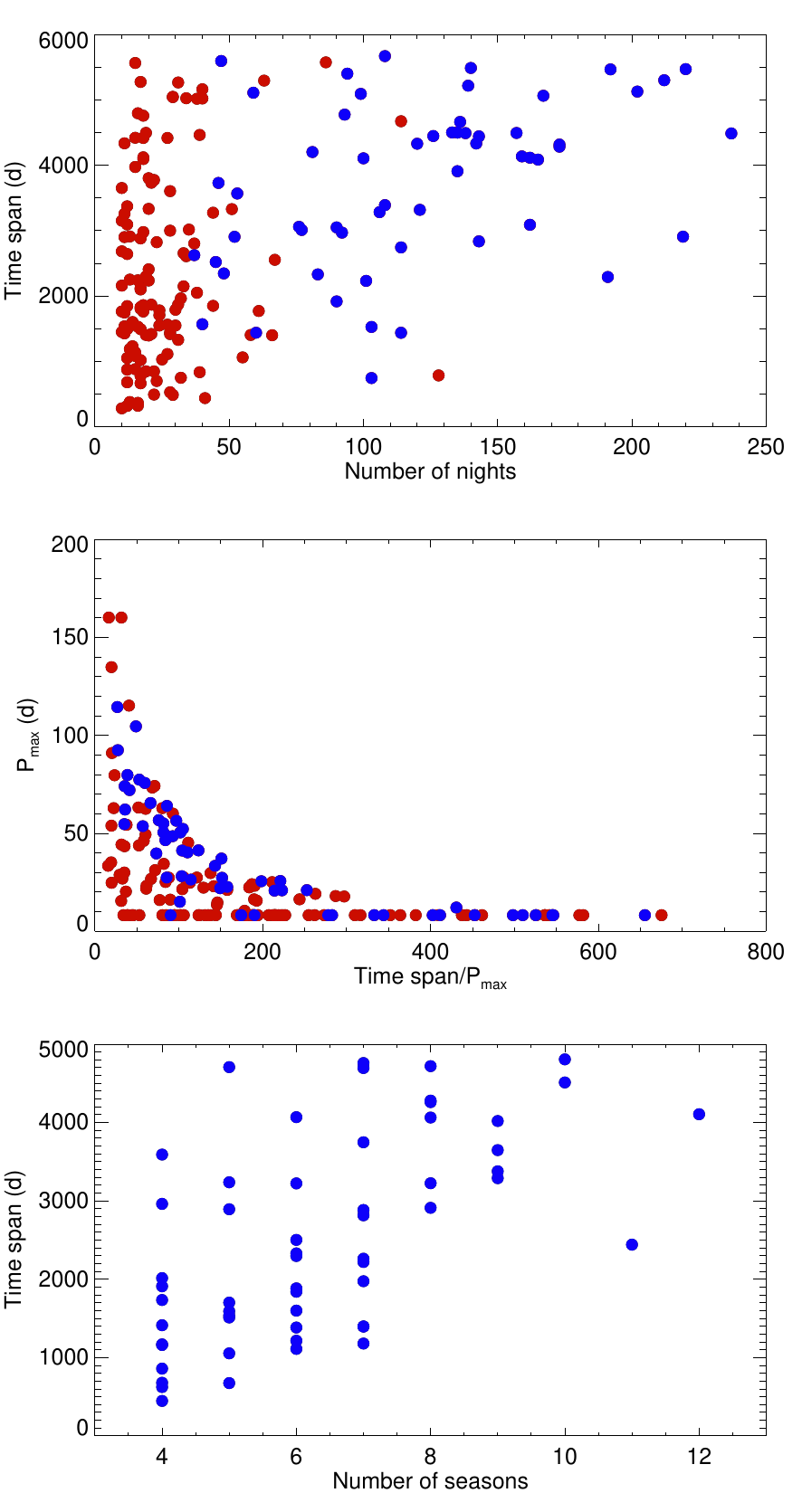}
\caption{Temporal coverage vs number of independent nights with observations (upper panel), 
maximum rotational period (P$_{\rm max}$) vs ratio of the time span and P$_{\rm max}$ (middle panel), and temporal coverage vs number of independent seasons 
with observations for the season sub-sample (lower panel).  
In the two first panels, the season sub-sample is shown in blue, and all other stars in our sample are shown in red.
}
\label{sample3}
\end{figure}

\subsection{Activity indicators}

To estimate the chromospheric contribution of the activity, we computed the S-index in three emission lines: the H$\alpha$ line, the H and K line of ionised calcium, and the sodium doublet. The Na lines form in the lower chromosphere \cite[][]{mauas94,andretta97,fontenla16}, the Ca II H and K lines in the middle chromosphere \cite[][]{mauas94} to the upper chromosphere \cite[][]{mauas00,houdebine09,fontenla16} and the H$\alpha$ line in the upper chromosphere \cite[][]{mauas94,fuhrmeister05,houdebine09,fontenla16,leenaarts12} 

We integrated  the emission in the three bands in the windows given in Table~\ref{tab_band}, which was then divided by the flux into  two continuum bands for each index. Wavelengths and bandwidths were taken from \cite{kurster03}, \cite{gomes11}, and  \cite{astudillo17}.  The Ca index was corrected to ensure consistency with the Mount Wilson program with the law derived in \cite{astudillo17} from the comparison with \cite{wright04},

\begin{equation}
S_{M.W.} = 1.053 · S_{HARPS} + 0.026,
\end{equation}

where $S_{HARPS}$ was obtained from the HARPS spectra, and $S_{M.W.}$ is the corrected value. 
Uncertainties on each index were computed with the quadratic sum of the read-out noise and the flux photon noise in the bands and lines through propagation from the uncertainties in each pixel to the computation of the ratio of the flux defining the indices described above, as in \cite{astudillo17}. The spectra with the lowest signal to noise ratio (hereafter S/N) may bias the computation of the indices, and their impact is evaluated in Sect.~2.3 and  Appendix~\ref{selection_app}. 

We also  computed the classical $\log R'_{HK}$ estimator by using the empiric relations established by \cite{astudillo17} based on V-K, which provides reliable photospheric and bolometric corrections for M dwarfs. We can therefore compare the average activity levels of the M dwarfs in our sample using the average $\log R'_{HK}$ for each star.

\begin{table*}
\caption{Wavelength bands}
\label{tab_band}
\begin{center}
\renewcommand{\footnoterule}{}  
\begin{tabular}{lllll}
\hline
Index  & Band 1 & Band 2 & Blue continuum & Red Continuum \\
\hline
Ca & 3933.663$\pm$1.09 (Ca II K) & 3968.469$\pm$1.09 (Ca II H) &  3891.07--3911.07 & 3991.07--4011.07 \\
Na &   5889.95$\pm$0.25 (NaD2) & 5895.92$\pm$0.25 (NaD1) & 5860--5870 & 5904--5908 \\
H$\alpha$ & 6562.808$\pm$0.8 & - & 6545.49--6556.25 & 6575.93--6584.68 \\
\hline
\end{tabular}
\end{center}
\tablefoot{Ca H and K: triangular function. All wavelengths are in $\AA.$}
\end{table*}

\subsection{Selection process}\label{select}

Starting with the complete sample of spectra, a selection process must be implemented to guarantee that the series are sufficiently complete and of good quality to allow a long-term temporal analysis in three steps.
We first removed stars with a time coverage too short compared to the rotational modulation for a long-term analysis: we used a threshold of six times the rotation period. Second, we eliminated poor spectra based on the signal-to-noise ratio (S/N) on the activity indicators, with a threshold of one and an a posteriori verification for the lowest S/N values. Finally, we identified and eliminated outliers (e.g. flares). 
The details can be found in Appendix~\ref{selection_app}.

We then binned the activity indicator time series to one point per night. In most cases, there were one to two spectra per night, and only five stars in our final sample were observed more than ten times per night during at least one night, with a maximum of four such nights: these observations could have a lower S/N than the remaining time series, and their sampling is discussed in Appendix~\ref{selection_app}.

\subsection{Final sample}

After these selections, our main final sample is composed of 177 stars with a time coverage ranging from 280 to 5675~days (median value of 2685~days). Of these stars, the temporal coverage for 156 (88 \%) of them is longer than 1000~days, and 117 stars (66 \%) have more than 20 nights of observations. The full time series (binned over each night) of the observation of this sample is analysed in Sect.~3.
The stellar properties (V-K and T$_{\rm eff}$) of the full sample were extracted from the CDS\footnote{SIMBAD Astronomical Database (\url{simbad.u-strasbg.fr})} and are shown in Table~\ref{tab_targets} together with the corresponding references. Distances were mostly taken from GAIA \cite[][]{gaia18}, and masses were computed from the mass-luminosity relation \cite[][]{delfosse00}.
This final sample of 177 stars represents 9942 nights of observations and an average of 56 nights of observation per star, ranging from 10 (the minimum number of nights of 10 was also applied after the different steps described in the previous section) to 237 nights. 
They are listed in Table~\ref{tab_targets} in App.~\ref{sample}. 
Their properties are shown in Fig.~\ref{fsample} and Fig.~\ref{sample2}.

\subsection{Season definition}
\label{SectSeason}

To focus on the long-term variability, we also defined seasons to apply statistical tests on binned time series:  We therefore defined a sub-sample of stars with observations during enough seasons.  
We defined these seasons as bins
of 150~days (to average the rotational modulation as best possible) with at least five observations (150 days is the typical maximum limit for the rotation period of M dwarfs \cite[][]{newton18}), 
and gaps between observations shorter than 40~days inside a 150-day bin. We kept stars with at least four seasons of observations, which led to 57 stars in this sub-sample. We had between 4 and 12 seasons, with a median of 6. We found that 36 stars have at least 6 seasons and 14 stars have more than 8 seasons.  The span corresponding to these season time series was computed, and it is frequently smaller than the total span 
because the beginning or the end of the time series does not necessarily fall into a season, that is, some observation nights lay outside the seasons.
This seasonal  span ranges from 447 to 4806~days, with a median of 2296~days. 
There can be more than one season per year on average over the whole time series, hence a time span longer than the number of seasons times 365 d for a few stars. The properties of the final sample and of this sub-sample in terms of time span and number of nights are shown in  Fig.~\ref{sample3}. For this sub-sample of 57 stars, the season time series is also analysed in Sects.~3.1 and 3.2 to compare it with the analysis of the full time series of the sample of 177 stars. 


\section{Results}\label{sectRes}

We first present the results derived from the analysis of the chromospheric indices, first considering the global variability, and then studying the long-term variability based on different models in more detail. Our results are then compared to the literature and to the analysis of photometric time series for the most interesting stars in our sample.

\subsection{Global variability: Constant model}

The objective of this section is to determine whether the stars of our sample are globally variable at short and/or long timescales. For this purpose, we computed the $\chi^2$ probability of the constant model, hereafter p$\chi^2_0$, that is, we considered that the time series exhibits some scatter without taking  the possible timescale of evolution of the signal into account. This probability  allowed us to estimate the probability for the observed scatter to be due to the identified sources of noise, and has the advantage over the F-test probability \cite[which is frequently used in the literature for these stars; e.g.][]{gomes11,gomes12} to take the individual uncertainties into account,  which are highly variable within a given time series. 

 In our sample of 177 full time series, 
the probability of having
stable activity indicators in at least one index  is lower than 1\% for more than 99\% of the stars,
and 94\% of the stars have a probability below 1\% for all indices. This means that almost all of them show a significant variability (i.e. when we consider data with photon and readout noise, a constant model is not sufficient to explain them).  The variability for all stars in the three activity indicators is indicated in Table~\ref{tab_nuits} (Appendix~\ref{globvar}).

 For the season sub-sample, the analysis of the season time series leads to slightly lower percentages when a given activity indicator is considered. This indicates that some stars may be dominated only by short-term variability. The percentages remain high, however, and all stars in the season sub-sample correspond to a probability lower than 1\% to be constant.
We conclude that almost all stars are variable, that is, even when they are usually considered as quiet stars due to their low average activity level, for example with $\log R'_{HK}$ below -5. 


\subsection{Linear and quadratic models}

We first describe the method we implemented to analyse the time series with linear and quadratic models. We then present the results for the sample of 177 stars and for the season sub-sample. 

\subsubsection{Method}

We now  compare the time series with linear and quadratic models:  We assumed that the activity indicators vary linearly or quadratically with time. The quadratic function was chosen to represent long-term non-linear variability in a very simple way. It may model variability over a fraction of a potential cycle. Models like this were  explored by \cite{gomes11} and \cite{gomes12} for about 30 stars. \cite{gomes11} found that for 20\%  of these stars, a 
quadratic model better explained the behaviour of the data than a linear model (suggesting a cycle) based on the F-test probability. Trends in activity indices  were also  observed, for example for GJ~433 \cite[][]{delfosse13}, and for a few stars in \cite{robertson13}, suggesting very long periods. 

We first estimated the $\chi^2$ probability  
when fitting our data with these models for the  177-star sample and for the sub-sample of season time series.  As previously set out, a low $\chi^2_1$ ( $\chi^2_2$) probability means that the linear (quadratic)  model is not sufficient  to explain the dispersion of the time series. A large majority of probabilities are very low: up to 96\% of the time series cannot be described by a linear model for at least one of the three indices used.

Even though the linear and quadratic models are usually not sufficient (as shown by the very low $\chi_2$  probabilities), they may still be dominant in some cases when the  time series is described. 
We quantitatively checked the model improvement from constant to linear model and that from the linear to quadratic model by using a combination of two criteria.
First, the F-test probability, to quantify the significance of the improvement, allows us to check whether a model is statistically better than another \cite[][]{cumming99}. As described in \cite{cumming99} and \cite{zechmeister09b}, a low probability indicates a statistically significant improvement from model k to j, while a probability close to one means that the dispersion between the residuals of the two models is the same and they have a similar $\chi^2$. 
A second criterion, F$_{\rm red}$, allows us to quantify the amplitude of the improvement: F$_{\rm red}$ is equal to the ratio of the difference between the $\chi_2$ of the two models and the $\chi_2$ of the first model. An F$_{\rm red}$ value close to one implies a huge improvement between the new tested model and the model used for comparison, while an F$_{\rm red}$ value close zero implies a very slight improvement, and a negative value implies that there is no improvement at all.

Our purpose was therefore not to try to validate the model as the only model that can explain the observed variability, but to identify the dominant model among the three models we tested.
We therefore  combined the information on the 
improvement brought by the linear model compared to the constant model (notation 1/0) and by the quadratic model compared to the linear model (notation 2/1).
In the following, we consider two typical thresholds for the F$_{\rm red}$ indicator, 0.5 (for a strong improvement, hereafter called the strict threshold), and 0.2 (for a weak improvement, hereafter the soft threshold).
We used these thresholds to place each star into one of the three following categories, estimated separately for each activity index:

\begin{itemize}
\item{Linear category: Stars with F$_{\rm red\:1/0}$  higher than the threshold (0.2 or 0.5, i.e. soft or strict), p(F)$_{\rm 1/0}$ lower than 5\%, and either F$_{\rm red\:2/1}$ lower than 0.2, or p(F)$_{\rm 2/1}$ higher than 5\% (i.e. the quadratic fit does not improve compared to the linear fit). 
}
\item{Quadratic category: Stars that are not in the linear category, and with either F$_{\rm red\:2/0}$  higher than the threshold and p(F)$_{\rm 2/0}$ lower than 5\%, or F$_{\rm red\:2/1}$  higher than the threshold and p(F)$_{\rm 2/1}$ lower than 5\%. As before, the threshold on the F$_{\rm red}$ values was either 0.2 or 0.5. 
}
\item{No category: all other stars are in this category.}
\end{itemize}

We applied these criteria to all activity indicator time series of the sample of 177 stars, which allowed us to classify each star for each activity indicator. We also applied these criteria to  the season time series for the sub-sample of 57 stars, which also provides another classification for the star. The stars in the linear or quadratic category from the sample or the sub-sample analyses are shown in Table~\ref{tab_recap}.

\subsubsection{Results for linear and quadratic variability}

Of the season sub-sample, 72\% of the stars are in either the linear or quadratic category for one index according to the strict threshold, and 88\% are in the category of the soft threshold. 
When we consider the sample of 177 stars, the percentages are lower because of the short-term variability. This may prevent them from reaching higher values of F$_{\rm red}$:  20\% of the stars are in either the linear or quadratic category for at least one index (69\% with the soft threshold), however. We did not observe any significant dependence of these rates on the activity level (defined by the average $\log R'_{HK}$ of the star) or on  T$_{\rm eff}$.

\begin{figure}
\includegraphics{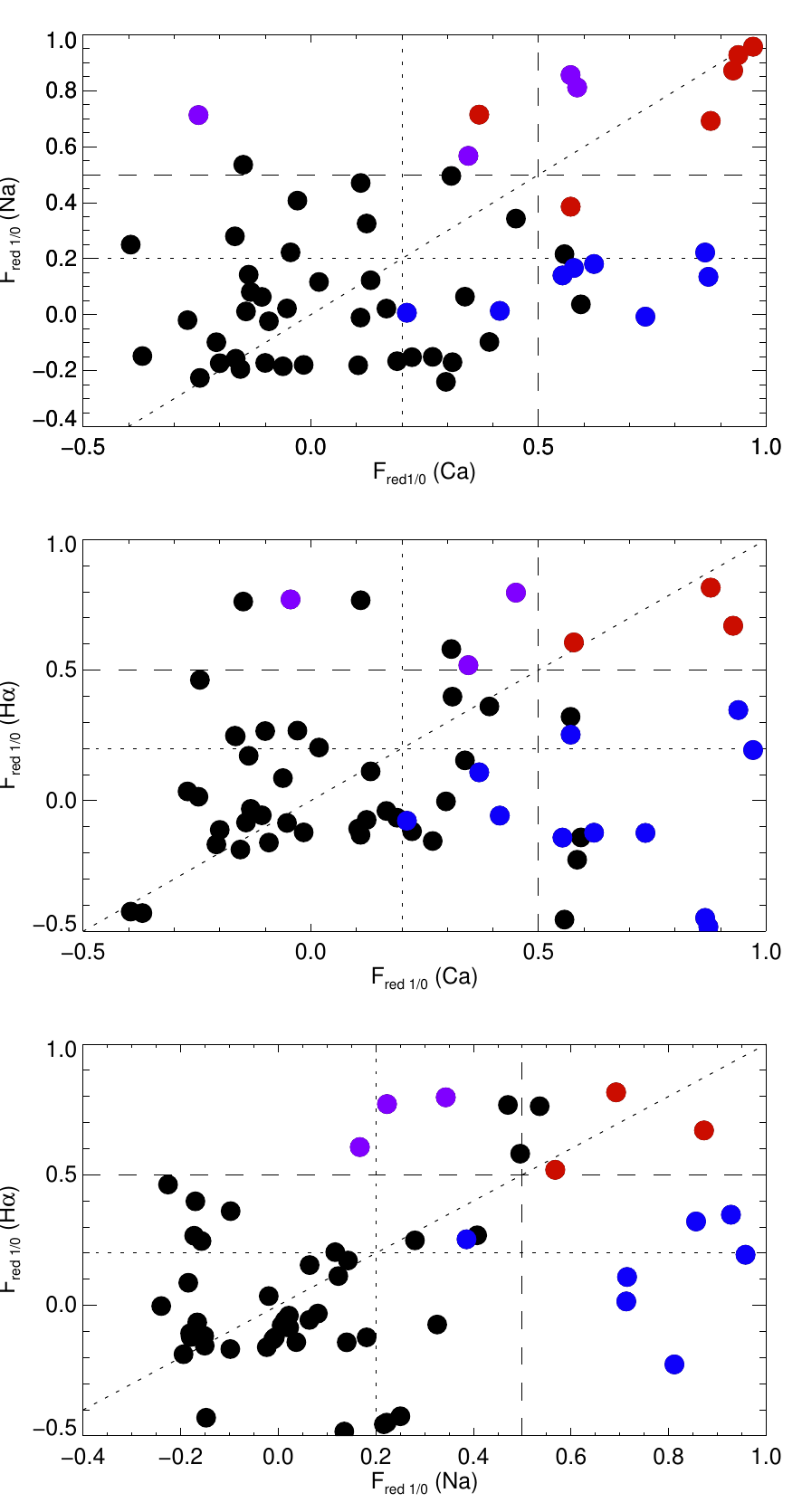}
\caption{F$_{\rm red\:1/0}$  for Na vs Ca (upper panel) for H$\alpha$ vs Ca (middle panel) and for H$\alpha$ vs Na (lower panel) for the season sub-sample. 
The colour-code indicates the degree of significance: p(F) of the x-axis and of the y-axis below 5\% (red),  pF of the x-axis  below 5\% (blue), and p(F) of the y-axis below 5\% (purple). The dotted and dashed lines correspond to the  soft and strict  thresholds (0.2 and 0.5) used in the analysis.
}
\label{fred_ind}

\end{figure}

We  point out that the indices are frequently different. 
This is sometimes due to the chosen threshold when the star is at the limit, but it often corresponds to fundamentally different configurations. This is illustrated in Fig.~\ref{fred_ind},  which compares the F$_{\rm red}$ values for the three indices. They are weakly correlated or not correlated at all. This means that a different classification for a star in different indices is not dominated by a threshold effect.  As a consequence, almost no star is in a given category (e.g. linear) for all activity indicators, even regarding the season sub-sample. We conclude that the different activity indicators often exhibit a different long-term variability.

The two categories we defined are sensitive to the observational history of the star. For example, a star may show a linear or quadratic trend, but may have an underlying periodic variability that cannot be detected because of gaps in the temporal sampling. A star may also appear here in the linear category, but could have exhibited a quadratic variability if it had been observed for longer. These categories remain extremely useful in several ways, however. First, the high percentages are indicative of a general long-term variability. Second, stars in these two categories are then flagged as very interesting targets for long-term variability studies. Last but not least, the knowledge of these flagged stars is also crucial for a search for exoplanets around these stars because it is important to be very cautious when stellar variability like this is present.

The unambiguous identification of a cycle  requires measurements showing its repetition over several periods.  This requires data taken over a long period of time. Even if the time coverage is not sufficient for some stars, however, our data can be used to estimate a  minimum cycle period if present. Our linear or quadratic fit analysis is not sufficient to guarantee that the signal is periodic or even quasi-periodic.
However, we can conclude that if there is a linear trend over a long period, it still gives an idea of the typical timescale of the evolution, which is an interesting characterisation and may also provide clues for future observations of these stars. Out of the 128 stars   in the linear or quadratic category for at least one activity indicator, the previous criteria allowed us to estimate that some variability above a typical timescale $\tau_{\rm min}$ is very likely to be present for 25 of these stars, as indicated in Table~\ref{tab_recap}. The $\tau_{\rm min}$ values were determined as in \citet[][]{ibanez19}, with simulated time series of indices based on the Sun series, scaled to each star, and covering a wide range of periods. $\tau_{\rm min}$ corresponds to the minimum period for which the observed F$_{\rm red}$ values are reached for at least 1\% of the realisations at a given period. The full process is described in Appendix~\ref{pmin}.

\begin{table*}
\caption{Stars with the best constraints from our chromospheric emission analysis}
\label{tab_recap}
\begin{center}
\renewcommand{\footnoterule}{}  
\begin{tabular}{ll|l|ll|l}
\hline
  \multicolumn{2}{c|}{strict LIN or QUAD } & P$_{\rm min}$ & \multicolumn{2}{c|}{ (strict LIN or QUAD) and P$_{\rm min}$} & None\\
  \hline
   \multicolumn{6}{c}{Stars with no chromospheric long-term peak above the 1\% fap} \\
     \hline
   GJ~91       & GJ~2066 (2P$_{\rm phot}$)  & GJ~213 & GJ~436 &  & all other  \\
    GJ~93      & GJ~3018 & GJ~341 (Tr$_{\rm phot}$) & GJ~849 (L)$^{(a)}$  &  &  stars \\
GJ~163         & GJ~3082  & GJ~551 (L) (P$_{\rm phot}$) & &  &   \\
GJ~173         & GJ~3148 (P$_{\rm phot}$)   & GJ~9482$^{(a)}$  &  &  &   \\
GJ~179         & GJ~3307 &  &  &  & (91 stars) \\
GJ~180 (P$_{\rm phot}$) & GJ~3404 &  &  &  &  \\
GJ~208$^{(a)}$ & GJ~3470  &  &  &  &  \\ 
GJ~390        & GJ~3502  & &  &  &  \\
GJ~406 (L)$^{(a)}$        & GJ~3528  &  &  &  &  \\
GJ~645           & GJ~3804 &  &  &  &  \\
GJ~682     &  GJ~3874 &  &  &  &  \\
GJ~699 (L)   &  GJ~4001$^{(a)}$  &  &  &  &  \\     
GJ~740 (L)         &  GJ~4092 &  &  &  &  \\
GJ~864        &  GJ~4100 &  &  &  &  \\
GJ~1001          & GJ~9133 &  &  &  &  \\     
 GJ~1132            & GJ~9360 &  &  &  &  \\ 
 GJ~2056 (P$_{\rm phot}$)   &  &  &  &  &  \\
   \hline
   \multicolumn{6}{c}{Stars with a chromospheric long-term peak above the 1\% fap, validated after steps 1-3} \\
   \hline
GJ~191 (P$_{\rm phot}$) & &  GJ~887  & GJ~87 (P$_{\rm phot}$)$^{(a)}$ & GJ~701 & GJ~514 (L) (P$_{\rm phot}$)$^{(a)}$ \\
GJ~433 &  &  & GJ~221 (P$_{\rm phot}$)$^{(a,b)}$ & GJ~752A (L) (P$_{\rm phot}$)$^{(a)}$  & GJ~693  \\
GJ~447 (L) (P$_{\rm phot}$)     &  &  & GJ~229 (L) (2P$_{\rm phot}$)  & GJ~880  (2P$_{\rm phot}$)$^{(b)}$  & GJ~754 \\
GJ~581 (L) (P$_{\rm phot}$)   &  &  & GJ~273 (L) (P$_{\rm phot}$)  & GJ~9425 (2P$_{\rm phot}$) & GJ~2126$^{(a)}$ \\
GJ~588 (L)    &  &  & GJ~317 (L) & GJ~9592& LP816-60 (L) \\
GJ~676A (L) (P$_{\rm phot}$)    &  &  & GJ~536 &  &  \\
GJ~832 (L) (P$_{\rm phot}$)$^{(a)}$       &  &  & GJ~628 (L) (2P$_{\rm phot}$) &  &  \\     
GJ~876 (P$_{\rm phot}$)    &  &  &  &  & \\
\hline
    \multicolumn{6}{c}{Stars with a chromospheric long-term peak above the 1\% fap, not validated after steps 1-3} \\
  \hline
 GJ~696 (P$_{\rm phot}$)$^{(a)}$  &  &  GJ~654 & GJ~149B &  & GJ~54.1 \\
GJ~3256    &  &  GJ~3822 & GJ~176 (L)  &  & GJ~361 \\
GJ~3543     &  &  & GJ~393 (P$_{\rm phot}$) &  & GJ~369 \\
  &  &  & GJ~908 (P$_{\rm phot}$)$^{(b)}$  &  & GJ~667C (L) \\
   &  &  & GJ~1075 (P$_{\rm phot}$)$^{(b)}$ &  &  GJ~680 (L) (2P$_{\rm phot}$) \\
 & & & GJ~3135 &  & GJ~3053 \\
  & & & GJ~3341 &  & GJ~3293 \\
  & & & &  & GJ~3634$^{(a)}$  \\
    \hline
\end{tabular}
\end{center}
\tablefoot{The three first columns identify how we classified the star, using the strict criterion (F$_{\rm red}>$ 0.5) with the linear or quadratic analysis in at least one chromospheric index, and/or if this criteria has allowed us to determine a minimum period (see {Sect.~3.2}). Validation of peaks when using the sinusoidal model is performed in Sect.~5.2 and corresponds here to steps 1-3, in order to keep interesting stars for the photometric analysis.  (L) indicates that a long-term cycle has already been published for that star (see Table~\ref{tab_Pcyc} for the references). P$_{\rm phot}$, 2P$_{\rm phot}$ , or Tr$_{\rm phot}$ indicate that we  fitted one or two sinusoidal functions on the photometric time series, with a peak above the 1\% fap. (a) indicates peaks in the photometry with a period longer  than the temporal coverage of the observations, and (b) indicates that all peaks in the chromospheric analysis have a  period longer than the temporal coverage (i.e. did not pass step 4). All  stars are from the nightly sample. 
We refer to section 3.2 for the definition of the LIN and QUAD categories.
}
\end{table*}

\subsection{Sinusoidal model}

\subsubsection{Method}

 Finally, in this section, we tested periodic models  for the sample of 177 full time series for each activity indicator. We used the GLS periodograms presented in \cite{zechmeister09b}, and we considered that a signal is convincing if its false-alarm probability (fap), determined with a classical bootstrap method on 1000 permutations, was below 1\%. 
As emphasised by \cite{kuker19}, the notion of a cycle or period is only meaningful if at least two periods are observed, and it also needs to be well sampled, otherwise, the repeatability  is not guaranteed. Our data rarely follow these conditions, and the identified periods may therefore not correspond to a unique description of the stellar variability or may be the dominant timescale, but they allow us to study the complex variability that is observed in many cases, in particular, the significant difference in behaviour between the indices for some stars. 

For a maximum peak above the fap level, we fit and subtracted a sinusoidal signal 
on the time series and repeated the process on the residual. We stopped the process after the third iteration. Fits were obtained by using a least-squared optimization function that also gives the covariance matrix of the fitted parameters. We used diagonal terms of the matrix as uncertainties on the fitted parameters because we considered uncorrelated parameters. This periodic analysis of the time series may return the rotation period and their harmonics, as well as the long-term variation.

\subsubsection{Results of the periodic search}

\begin{figure}
\includegraphics{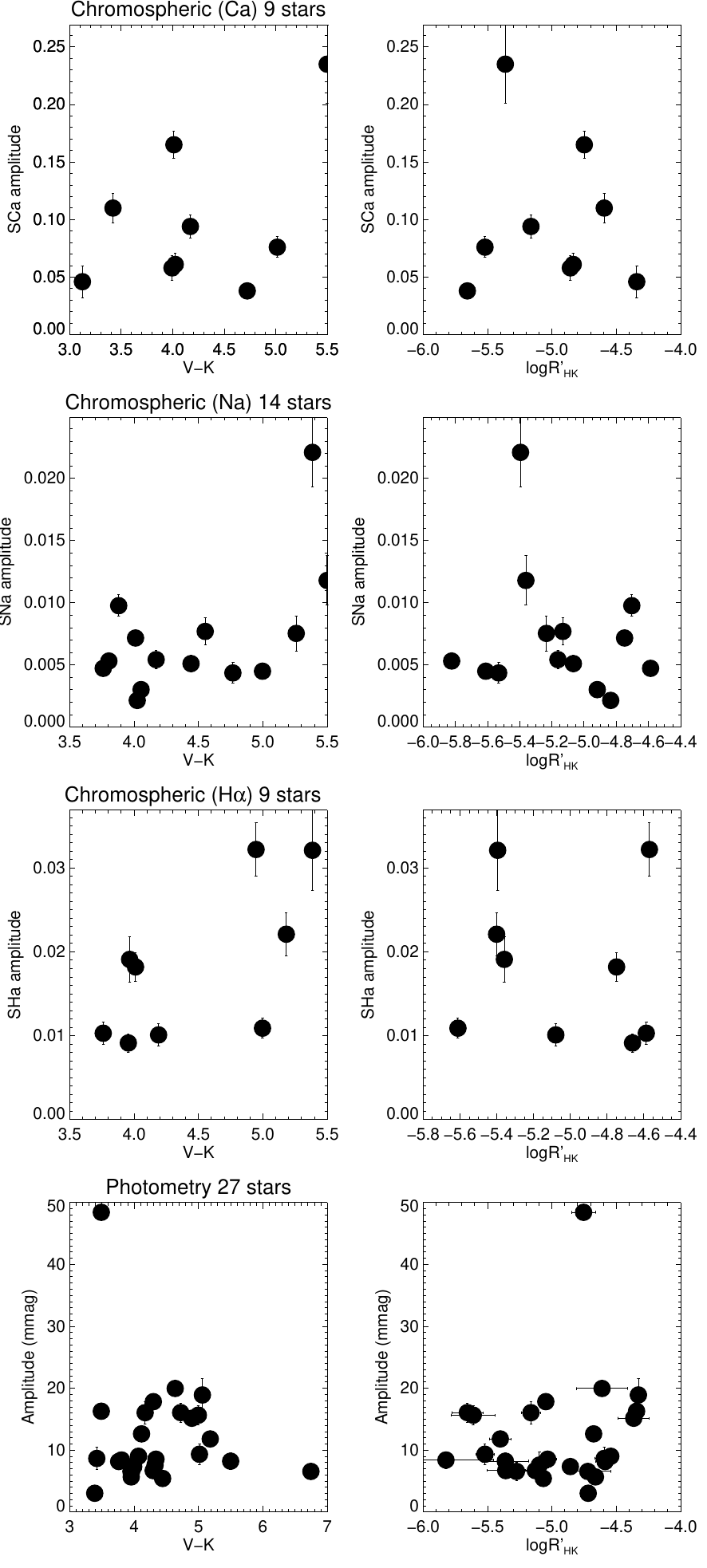}
\caption{Cycle amplitudes for the three chromospheric analysis (three first lines of the panels) and for our photometric analysis (bottom panels) vs V-K (left) and $\log R'_{HK}$ (right).}
\label{fig_ampl}
\end{figure}

We identified 74 stars for which one peak or several peak lie above the 1\% fap for at least one activity index, either short term, long term, or both. The highest or only peak  in the periodogram corresponds to a short-term (<200 days) period for 29 stars, and is a long-term period for 23 stars (>200 days). For all other stars, the main peak may be at short or long periods depending on the analysed index, with at least one peak dominated by the short-term peak and another dominated by the long-term peak.
We focus here on stars with long-term periodicities, corresponding to 44 objects, including the 23 stars for which the main peaks in the periodogram are systematically long period, and 19 stars exhibit a mix of short and long variability periods with significant peaks.

We then adopted  the following approach to validate or refute these peaks, which we consider to be indicative of the typical timescales of the long-term variability:

\begin{itemize}
    \item Step 1: Three stars (GJ~754, GJ~3053, and GJ~3135) were removed from this analysis because the periodograms were significantly impacted when nights including some observations with a very low S/N were removed (see Appendix~\ref{selection_app}). 
    \item Step 2: We eliminated any detection for which the fitted peak was at the period of a strong peak in the periodogram of the window function. This led us to eliminate the peak detected in Ca for two stars, GJ~908 and GJ~191. 
    \item Step 3: The time series contains gaps, and for some of these, we detected a periodicity longer than their time span \cite[although cycles corresponding to this configuration were published before;][]{suarez16}. The detected periods could therefore be due to aliases, with a completely different true periodicity. Based on synthetic time series with variability injected at different timescales between 200 and 20000 days and a cycle shape based on the average solar cycle (average of all solar cycles from the SIDC spot number)\footnote{This takes into account the possible complex variability corresponding to a non-sinusoidal shape, although tests have been also made with a sinusoidal, which led to less conservative results, however. It does not take into account the possible variability in amplitude of consecutive cycles.}, an amplitude and short-term dispersion similar to the observed one, we identified periods corresponding to aliases. They were identified by the peaks above the fap level at periods different from the injected period: We considered that an alias is present when more than 1\% of the realisations for a given period leads to an alias. We note that the validation may be different from one activity indicator to the next for a given star, despite the same temporal sampling. This is likely to be due to two effects: first, the detected period and the variability in general between different indices, and second, the different short-term dispersion relative to the amplitude of the detected long-term signal from one index to the next. This emphasizes the likely difference in behaviour between activity indices. As a consequence, several periods were also eliminated because too many aliases were present in those simulations, that is, GJ~54.1, GJ~176, and GJ~273 in H$\alpha$, GJ~361 and GJ~880 in Ca and H$\alpha$, GJ~3634 and LP816-60 in Ca and H$\alpha$, GJ~514 in H$\alpha$, and GJ~581 in Na. In addition, a few peaks were eliminated because the uncertainty on the period was so large that the fit did not provide any information. We also eliminated the fitted periods for GJ~149B, GJ~369, GJ~696, GJ~1075, GJ~3822, GJ~3256, and GJ~3543 because the synthetic time series did not provide enough peaks above the fap for a reliable validation, that is, the simulated time series did not lead to enough significant peaks, even when the injected period corresponded to the fitted period. 
    \item{Step 4: Finally, we removed all periods longer than the coverage of the observations from the final list. The list of eliminated periods, indicted as timescales $\tau$, is provided in Table~\ref{tab_elim}: Their period is not robust because the coverage is poor, but it is of interest for exoplanet searches and future observations of these stars.  }
\end{itemize}

Fig.~\ref{fig_periodo} illustrates  the periodograms for three representative stars. The Ca periodogram of GJ~54.1 exhibits long-term peaks above the 1~\% fap threshold, but the highest peak was invalidated at step 3 due to the strong possibility of aliases from other periods. The highest Ca peak for GJ~273, on the other hand, was validated, but not in H$\alpha$ (invalidated at step 3). Finally, the highest long-term peak was validated for the three indices for GJ~9592.

We  validated 32 long-term periods for 24 stars, which represent 14\% of the stars in our nightly sample of 177 stars. 
They are listed in Table~\ref{tab_Pcyc}.  Table~\ref{tab_cyc2} summarises the percentages depending on the activity index. Since our sample was heterogeneous and some stars were not as well sampled as others (although they all covered a long time span), this is a lower bound of the long-term periodicities. 
As pointed out above, these periods, although dominant for our temporal sampling, may not have been the dominant period if the star had been observed with a longer time span or a better phase coverage, and therefore, they may not correspond to a cycle. We therefore considered them typical timescales $\tau$ of the variability of these stars in the following. We eliminated possible aliases, and 
all validated peaks correspond to a very low fap level (most below 0.1\%), and are therefore very significant. The improvement brought by the model is also good in many cases. Compared to a constant model, the F$_{\rm red}$ values are  between 0.12 and 0.69 (0.5 representing an improvement of the reduced $\chi^2$ of typically a factor of two), with an median  of 0.37. 
We note that the $\chi^2$ probability is always extremely low, meaning that the long-term sinusoidal model is never sufficient, even for high values of F$_{\rm red}$, showing that there is a complex underlying behaviour of the time series (in particular due to the short-term variability in many cases). 
This is also not surprising, since we do not expect the long-term variability to be purely sinusoidal, as shown by the best-known example, the Sun, which shows a steeper increase in the cycle compared to a slower decrease after the cycle maximum.

We find stars with a significant long-term variability over the whole range in V-K and in $\log R'_{HK}$. The amplitude of the variability  and period derived from the fits versus V-K and $\log R'_{HK}$ do not show any trend either, as shown in Fig.~\ref{fig_ampl}. There may be a small trend showing  smaller amplitudes of the signal for the quietest stars, but the sample is too small to conclude. 
This behaviour differs from FGK stars, for which there is a large dispersion for active stars and an upper envelope increasing with average $\log R'_{HK}$ \cite[][]{lovis11b,meunier19}.

As pointed out above and shown in Table~\ref{tab_Pcyc}, in several cases, periods were validated in more than one activity index 
(one star with the three indices, and six stars with two indices), but often with different values. Only for three out of these seven stars (GJ~536, GJ~754, and GJ~832) are the periods similar, otherwise, there is always one index with a period that is completely different from the other(s). \cite{gomes11}, based on linear and quadratic models, also found that the variability was usually not significant in all activity indices. A visual examination (Fig.~\ref{plot_serie} in Appendix~\ref{plot_serie}) shows that in some cases, the behaviour is indeed very  different, in particular, the very long-term variability (trend) not seen for all indices: for example, for GJ~87, the long-term trend clearly seen in Ca is not seen in H$\alpha$. The difference can also be due to the short-term variation seen in one indicator and not in the other, impacting the dominant periodicity (e.g. GJ~676 A, GJ~752 A, and GJ~9592): it is possible that both periodicities are present. 
In some cases, the behaviours of  the two indices are very similar, and it is likely that this is the limit between two periods that are not aliases of each other, but probably arise becuase both periodicities are present (although our sampling does not allow us to characterise this aspect in more detail here).
Finally, the slightly lower number of validated periods in Ca may be due to  the lower S/N of the spectra, but intrinsically, a different behaviour impacts the results as well.

\begin{table}
\caption{Percentage of stars with a detected period from the periodogram analysis or a $\tau_{\rm min}$ estimate.} 
\label{tab_cyc2}
\begin{center}
\renewcommand{\footnoterule}{}  
\begin{tabular}{lll}
\hline
Index  & Validated peak  &  $\tau_{\rm min}$ from linear  \\
   & $>$fap 1\%  & and quadratic trends\\
\hline
Ca &   9 (5.1\%) &   15 (8.5\%) \\
Na  &  14 (7.9\%)  &   13 (7.3\%) \\
H$\alpha$   &  11 (6.2\%) & 10 (5.5\%)\\
At least one    &  24 (14.1\%) & 25 (14.1\%) \\
All     &  1 (0.6\%) &  4 (2.2\%)\\
\hline
\end{tabular}
\end{center}
\tablefoot{Number of stars (out of the 177 stars in the nightly sample) with a validated long-term  chromospheric emission peak (see  Sect.~4.3) or an estimated minimum period from the linear and quadratic analysis (see Sect.~4.2).
 We refer to Sect.~3.2 for the definition of the linear and quadratic categories.
}
\end{table}

\subsection{Comparison with published cycles based on chromospheric activity}

\begin{table*}
\caption{Compatibility of our periods and published cycles  based on chromospheric indicators.}
\label{tab_compat}
\begin{center}
\renewcommand{\footnoterule}{}  
\begin{tabular}{llllllll}
\hline
Star & Activity & Fitted & Reference & Fitted & n $\sigma$ & Percentage & Status \\
     & Index & Period (d) &           & period (d) &           & in our fitted  &    \\
      & (this paper) &  &  & (published) & range & \\
\hline
GJ~273      & Ca & 1975$\pm$49  & (4, Ca)  & 1013 &  $>$3$\sigma$ & 0.2\% & not compatible \\
GJ~581    & Ca & 1419$\pm$72  & (1,  H$\alpha$) & 1633$\pm$93  & 2$\sigma$ & 7.9\% & compatible \\
    & Ca &   & (6,  Na) & 1407 & 1$\sigma$ & 36.7\% & compatible\\
GJ~832     & Ca & 2357$\pm$69  & (5, Na) & 1726 & 3$\sigma$ & 0.1\% & not compatible\\
    & Na & 2346$\pm$76  & (5,  Na) &  & 3$\sigma$  & 0.2\% & not compatible\\
GJ~447     &   Ca & 2064$\pm$120  & (2, Na+Ca) &  1956$\pm$98  & 1$\sigma$ & 41\% & compatible \\
    & Na  &  806$\pm$16 &  (2, Na+Ca) &  1956$\pm$98  &  $>$ 3$\sigma$ & 0\% & not compatible\\
GJ~676A    & Na & 740$\pm$15 & (5, Na) & 1224 & $>$3$\sigma$ & 0\% & not compatible\\
    & H$\alpha$ & 2982$\pm$269  & (5, Na) &   & $>$3$\sigma$ & 0\% & not compatible\\
GJ~752A  & Na & 1255$\pm$36  & (3, Ca)  & 2704$\pm$289   & 2$\sigma$ & 0\% & not compatible\\
GJ~229      & H$\alpha$ & 713$\pm$8 & (3, Ca) & 1627$\pm$23   & $>$3$\sigma$ & 0\% & not compatible \\
\hline
\end{tabular}
\end{center}
\tablefoot{The activity index in the second column corresponds to our fitted period, indicated in the third column. For clarity, the period is not repeated when it is similar to the period indicated above. The published periods (fifth column) may have been obtained with a different indicator, indicated next to the reference number: (1) \cite{robertson13}, (2) \cite{ibanez19b}, (3) \cite{buccino11}, (4) \cite{perdelwitz21}, and (5) \cite{gomes12}. \cite{buccino11} used photometry combined to the Ca index.  
For \cite{gomes12}, we used an arbitrary uncertainty of 10\% of the published periods. n$\sigma$ indicates to which level the fitted period agrees with the published period. 
The percentage is derived from the analysis of synthetic time series made for the published period (based on their estimated uncertainties) and producing a highest peak within 1$\sigma$ of our fitted period (see text for more detail). 
}
\end{table*}

We now compare the timescales we found in our sample with cycle periods previously published for these stars that are also based on chromospheric indicators. They are listed in Table~\ref{tab_compat}, which lists seven stars. We directly compared these fitted periods and their uncertainties. In addition, we performed simulations on synthetic series similar to those described in Sect.~3.3, but using the published period as input, to verify whether the percentage of the highest peaks (out of the realisations with a highest peak above the fap) is compatible with our analysis. This percentage is indicated in Table~\ref{tab_compat}, and it indicates a compatibility when it is high. 
The periods are compatible for only two stars: GJ~581 (different activity indicators) and GJ~447 (only for our Ca measurement, and comparison with Ca+Na measurements). For six stars and one of our GJ~447 measurements, we find an incompatibility between our measurements and published cycles or only marginal compatibility. In some cases (e.g. GJ~273 in Ca and GJ~832 in Na), both measurements correspond to the same activity indicator. This shows that there may be an underlying complexity with more than one timescalet, although the temporal sampling makes it difficult to characterise the variability more precisely. In other cases, the two measurements were made with different activity indicators (e.g. GJ~229 in Ca and H$\alpha$, or GJ~752A in Ca and Na). This suggests a possible different behaviour of the activity indicators, which is also a result of our analysis of Ca, Na and H$\alpha$ indicators observed with similar temporal samplings.

\subsection{Comparison with photometry}
\label{SectPhot}

Our results show a complex relation between the activity indicators. Furthermore, a comparison of our chromospheric timescales with previously published periods derived from photometric light curves (\cite{suarez16}), which can be made for all stars from Table~\ref{tab_compat} except for GJ~880 and LP816-60, and for four additional stars (GJ~317, GJ~514, GJ~588, and  GJ~628), is also highly incompatible between timescales. A few previously published periods based on different approaches also show these incompatibilities, for example GJ~581, for which \cite{robertson13} found a period of 1633$\pm$93 days in H$\alpha,$ while \cite{suarez16} found 2265$\pm$329 days based on photometry.

Because the compatibility rate between publications is very low, a second step for investigating this issue is therefore a detailed comparison of the results we obtained based on the chromospheric time series and in photometry. We retrieved photometric data from the public archive of the All Sky Automatic Survey \cite[ASAS;][]{pojmanski97,pojmanski02} for the following stars: First, those for which we identified a long-term peak above the fap in the periodogram of the chomospheric time series, whether validated in Sect.~5.2 or not. Second, those with a constraint on $\tau_{\rm min}$ estimated from the linear and quadratic analysis in Sect.~4.2. Third, those within a linear or quadratic category when the strict threshold (F$_{\rm red}>$0.5) in at least one of the samples (nightly or seasons) was considered and with at least one activity index. The objective here was to perform an extraction from the archive based on the argument that these stars very likely have long-term variability. They are summarised in Table~\ref{tab_recap}. 
Stars with unreliable photometry were then removed (magnitude depending on the aperture, suggesting the impact of another star close to the line of sight), leading to the analysis of 79 stars. As in \cite{buccino11}, we selected only quality A and B data, from which we produced a weighted (by the uncertainties) average of the different apertures. To eliminate outliers, the dispersion was first estimated from a Gaussian fit on the distribution of values. Data outside the 3 $\sigma$ were then eliminated. The GLS periodograms of the time series were computed, as was the fap at the 1\% level. The  highest peak was identified, and if it was above the 1\% fap, was fitted using a sinusoidal function. When the periodogram is dominated by the rotation modulation while a long-term peak above the fap is also present, a two-sinusoidal fit was performed to be able to provide the long-term period. The fits were primarily made on all points, but were also checked on 100-daybinned data for robustness. We note that conversely to the chromospheric indicator data set, the temporal sampling has very few gaps so that the confusion with aliases is much less likely. 

We identified 32 stars with at least a long-term photometric period, several of which were not published so far. 
For 20 stars, the fits were performed with a single sinusoidal function.  Those with a period shorter than the coverage of the observations (12 stars) are listed in Table~\ref{tab_Pcyc}, and the others were eliminated from the analysis and are listed in Table~\ref{tab_elim}.
In addition, we identified 12 stars with a more complex behaviour, for which there are two long-term peaks above the fap at 1\%, and where the two-sinusoidal function was also statistically better than a single-sinusoidal fit. A few stars like this were studied by \cite{suarez16}, \cite{ibanez19}, and \cite{kuker19}, showing that two periodicities might be present simultaneously. A period was published for 6 of these 12 stars, and two different periods in two different publications for GJ~229 \cite[][]{buccino11,suarez16}. In 6 cases, the longest period is longer than the temporal coverage, so that only the shortest period is shown in Table~\ref{tab_Pcyc}: The longest period is only listed in Table~\ref{tab_elim}. Overall, we validated periods in photometry for 24 stars, including at least one for 14 stars out of the 24 with at least a validated chromospheric period. 

The unambiguous multiple significant periodicities in the photometric time series suggest that this complex signal mightbe related to the difference in behaviour between different chromospheric indicators if they are sensitive to different processes at different timescales. This should be explored further with better-sampled chromospheric data and a better correspondence with the temporal coverage of the photometric time series. Finally, we also note a few strong disagreements between chromospheric and photospheric time series: a notable case is GJ~9592, for which a clear new long-term variability is derived from our chromospheric analysis, but which is very flat in photometry. Conversely, we observe a clear long-term variability in photometry for certain stars, but this variability is not observed in the chromospheric time series despite a good temporal coverage, either because it is absent, or because it is completely dominated by short-term variability that is absent in photometry, for example GJ~551.

All fits are shown in Fig.~\ref{plot_serie} (Appendix~\ref{plot_serie}), were the chromospheric and photometric variability are compared. 
We compared stars for stars for which we  validated at least a chromospheric timescale with the photometric timescale, following a procedure similar to what was presented in the previous section. As before,  a few stars agree and many others disagree. In addition, for some stars, it is compatible for one of the activity indices, but not for the others.


\section{Conclusions}

We analysed a very large sample of chromospheric emission time series of M dwarfs in three activity indices (Ca, Na, and H$\alpha$) and characterised their long-term variability using various models (linear, quadratic, and sinusoidal) and tools. We  systematically quantified the improvement brought by the models, and estimated the typical timescales of the long-term variability or lower boundary whenever possible. The results were compared with the properties derived from photometric time series and with published results.  The derived periods are not necessarily true periods. Given the temporal sampling of the observations, we found a well-sampled periodicity over two cycles for GJ1075, GJ3148, and GJ2066, which lack a previous publication. We obtained the following results.

All stars are variable in chromospheric emission at one scale or another, and almost all stars are variable on long timescales. Despite their usual classification as quiet or active depending on their average activity level, even the quiet stars exhibit some measurable variability, including long-term cycles. The long-term variability dominates for many stars 
(31\% of the stars have a peak above the 1\% fap in at least one chromospheric index),
while others  (39\%) are dominated by the short-term variability. This is the subject of a future paper.

We find long-term timescales ranging from several years to more than 20 years.
For two-thirds of the sample, a linear or quadratic model provides an important improvement and is often dominant. It is never sufficient to model the variability, however, which is always more complex than these simple models. These percentages are impacted by the heterogeneous temporal sampling in our data-sets because better-sampled stars might exhibit a more complex (e.g. periodic) behaviour if they had been better sampled: it is indicative of the strong presence of long-term variability, however, and indicates that these stars have a strong long-term variability, which is important when searching for exoplanets. 

The relation between chromospheric indices is complex,
suggesting dominant periods that frequently different for the different chromospheric indices:  this is particularly important when these activity indicators are used to eliminate faps in RV exoplanet detections because they do not provide equivalent information. 
This is reinforced by the fact that this may be due to a complex underlying variability at different timescales simultaneously, as suggested by the complementary analysis of the photometry for these stars. Some of the differences between H$\alpha$ time series and the other indicators (Ca and Na) might also be due to the way in which the H$\alpha$ indicator is computed \cite[][]{gomes22}, at least for FGK stars.  A more detailed analysis of the correlations between activity indices will be provided in a future
paper. 
Finally, we do not observe any trend between a long-term variability on spectral type and the average activity level (determined by the average $\log R'_{HK}$).

\begin{acknowledgements}

This work has been supported by a grant from LabEx OSUG@2020 (Investissements d'avenir - ANR10LABX56). 
This work was supported by the "Programme National de Physique Stellaire" (PNPS) of CNRS/INSU co-funded by CEA and CNES.
This work was supported by the Programme National de Plan\'etologie (PNP) of CNRS/INSU, co-funded by CNES. We thank the Solar Influences Data analysis Center (SILSO data/image, Royal Observatory of Belgium, Brussels). 
The HARPS data have been retrieved from the ESO archive at http://archive.eso.org/wdb/wdb/adp/phase3\_spectral/form.
This research has made use of the SIMBAD database, operated at CDS, Strasbourg, France.
NCS was supported by the European Research Council through the grant agreement 101052347 (FIERCE) and by FCT - Fundação para a Ciência e a Tecnologia through national funds and by FEDER through COMPETE2020 - Programa Operacional Competitividade e Internacionalização by these grants: UIDB/04434/2020; UIDP/04434/2020.
ESO program IDs (the name of the PI is indicated) corresponding to data used in this paper are: 
 095.C-0718 (Albrecht)
 096.C-0082 191.C-0505 (Anglada-Escude) 
 100.C-0884 (Astudillo Defru) 
 0101.D-0494 (Berdinas) 
 082.C-0718 180.C-0886  183.C-0437 191.C-0873 198.C-0873 1102.C-0339 (Bonfils) 
 078.D-0245 (Dall) 
 075.D-0614 (Debernardi)
 096.C-0499  098.C-0518 0100.C-0487 (Diaz) 
 076.C-0279 (Galland) 
 074.C-0037     075.C-0202 076.C-0010 (Guenther) 
 096.C-0876 097.C-0390 (Haswell) 
 078.C-0044     (Hébrard)
 097.C-0090 (Kuerster)   
 0104.C-0863 (Jeffers)
 089.C-0006 (Lachaume) 
 089.C-0739 098.C-0739 099.C-0205 0104.C-0418 192.C-0224 (Lagrange)  
 097.C-0864 (Lannier) 
 085.C-0019 087.C-0831 089.C-0732 090.C-0421 091.C-0034 093.C-0409 095.C-0551 096.C-0460 098.C-0366 099.C-0458 0100.C-0097 0101.C-0379  0102.C-0558 0103.C-0432  196.C-1006 (LoCurto)
 072.C-0488  077.C-0364 (Mayor) 
 076.C-0155 (Melo)  
 089.C-0050 (Pepe) 
 185.D-0056 (Poretti)  
 074.C-0364 (Robichon) 
 084.C-0228 090.C-0395 (Ruiz)    
 086.C-0284 (Santos) 
 079.C-0463 (Sterzik) 
 0100.C-0414 (Trifonov) 
 183.C-0972 192.C-0852 (Udry) 
 060.A-9036 60.A-9709.
\end{acknowledgements}

%
%

\bibliographystyle{aa}
\bibliography{biblio}

\begin{thebibliography}{129}
\expandafter\ifx\csname natexlab\endcsname\relax\def\natexlab#1{#1}\fi

\bibitem[{{Allende Prieto} {et~al.}(2013){Allende Prieto}, {Koesterke},
  {Ludwig}, {Freytag}, \& {Caffau}}]{allendeprieto13}
{Allende Prieto}, C., {Koesterke}, L., {Ludwig}, H.-G., {Freytag}, B., \&
  {Caffau}, E. 2013, \aap, 550, A103

\bibitem[{{Andretta} {et~al.}(1997){Andretta}, {Doyle}, \&
  {Byrne}}]{andretta97}
{Andretta}, V., {Doyle}, J.~G., \& {Byrne}, P.~B. 1997, \aap, 322, 266

\bibitem[{{Astudillo-Defru} {et~al.}(2017){Astudillo-Defru}, {Delfosse},
  {Bonfils}, {Forveille}, {Lovis}, \& {Rameau}}]{astudillo17}
{Astudillo-Defru}, N., {Delfosse}, X., {Bonfils}, X., {et~al.} 2017, \aap, 600,
  A13

\bibitem[{{Attia} {et~al.}(2021){Attia}, {Bourrier}, {Eggenberger},
  {Mordasini}, {Beust}, \& {Ehrenreich}}]{attia21}
{Attia}, O., {Bourrier}, V., {Eggenberger}, P., {et~al.} 2021, \aap, 647, A40

\bibitem[{{Baliunas} {et~al.}(1995){Baliunas}, {Donahue}, {Soon}, {Horne},
  {Frazer}, {Woodard-Eklund}, {Bradford}, {Rao}, {Wilson}, {Zhang}, {Bennett},
  {Briggs}, {Carroll}, {Duncan}, {Figueroa}, {Lanning}, {Misch}, {Mueller},
  {Noyes}, {Poppe}, {Porter}, {Robinson}, {Russell}, {Shelton}, {Soyumer},
  {Vaughan}, \& {Whitney}}]{baliunas95}
{Baliunas}, S.~L., {Donahue}, R.~A., {Soon}, W.~H., {et~al.} 1995, \apj, 438,
  269

\bibitem[{{Baliunas} \& {Vaughan}(1985)}]{baliunas85}
{Baliunas}, S.~L. \& {Vaughan}, A.~H. 1985, \araa, 23, 379

\bibitem[{{Barnes}(2003)}]{barnes03}
{Barnes}, S.~A. 2003, \apj, 586, 464

\bibitem[{{Beeck} {et~al.}(2013{\natexlab{a}}){Beeck}, {Cameron}, {Reiners}, \&
  {Sch{\"u}ssler}}]{beeck13a}
{Beeck}, B., {Cameron}, R.~H., {Reiners}, A., \& {Sch{\"u}ssler}, M.
  2013{\natexlab{a}}, \aap, 558, A48

\bibitem[{{Beeck} {et~al.}(2013{\natexlab{b}}){Beeck}, {Cameron}, {Reiners}, \&
  {Sch{\"u}ssler}}]{beeck13}
{Beeck}, B., {Cameron}, R.~H., {Reiners}, A., \& {Sch{\"u}ssler}, M.
  2013{\natexlab{b}}, \aap, 558, A49

\bibitem[{{Beichman} {et~al.}(2014){Beichman}, {Benneke}, {Knutson}, {Smith},
  {Lagage}, {Dressing}, {Latham}, {Lunine}, {Birkmann}, {Ferruit}, {Giardino},
  {Kempton}, {Carey}, {Krick}, {Deroo}, {Mandell}, {Ressler}, {Shporer},
  {Swain}, {Vasisht}, {Ricker}, {Bouwman}, {Crossfield}, {Greene}, {Howell},
  {Christiansen}, {Ciardi}, {Clampin}, {Greenhouse}, {Sozzetti}, {Goudfrooij},
  {Hines}, {Keyes}, {Lee}, {McCullough}, {Robberto}, {Stansberry}, {Valenti},
  {Rieke}, {Rieke}, {Fortney}, {Bean}, {Kreidberg}, {Ehrenreich}, {Deming},
  {Albert}, {Doyon}, \& {Sing}}]{beichman14}
{Beichman}, C., {Benneke}, B., {Knutson}, H., {et~al.} 2014, \pasp, 126, 1134

\bibitem[{{B{\"o}hm-Vitense}(2007)}]{bohm07}
{B{\"o}hm-Vitense}, E. 2007, \apj, 657, 486

\bibitem[{{Bonfils} {et~al.}(2013){Bonfils}, {Delfosse}, {Udry}, {Forveille},
  {Mayor}, {Perrier}, {Bouchy}, {Gillon}, {Lovis}, {Pepe}, {Queloz}, {Santos},
  {S{\'e}gransan}, \& {Bertaux}}]{bonfils13}
{Bonfils}, X., {Delfosse}, X., {Udry}, S., {et~al.} 2013, \aap, 549, A109

\bibitem[{{Bonfils} {et~al.}(2007){Bonfils}, {Mayor}, {Delfosse}, {Forveille},
  {Gillon}, {Perrier}, {Udry}, {Bouchy}, {Lovis}, {Pepe}, {Queloz}, {Santos},
  \& {Bertaux}}]{bonfils07}
{Bonfils}, X., {Mayor}, M., {Delfosse}, X., {et~al.} 2007, \aap, 474, 293

\bibitem[{{Boro Saikia} {et~al.}(2018){Boro Saikia}, {Marvin}, {Jeffers},
  {Cameron}, {Marsden}, {Petit}, \& {Yadav}}]{borosaikia18}
{Boro Saikia}, S., {Marvin}, C.~J., {Jeffers}, S.~V.~and{Reiners}, A., {et~al.}
  2018, \aap, 616, A108

\bibitem[{{Browning}(2008)}]{browning08}
{Browning}, M.~K. 2008, \apj, 676, 1262

\bibitem[{{Buccino} {et~al.}(2011){Buccino}, {D{\'\i}az}, {Luoni}, {Abrevaya},
  \& {Mauas}}]{buccino11}
{Buccino}, A.~P., {D{\'\i}az}, R.~F., {Luoni}, M.~L., {Abrevaya}, X.~C., \&
  {Mauas}, P. J.~D. 2011, \aj, 141, 34

\bibitem[{{Buccino} {et~al.}(2006){Buccino}, {Lemarchand}, \&
  {Mauas}}]{buccino06}
{Buccino}, A.~P., {Lemarchand}, G.~A., \& {Mauas}, P. J.~D. 2006, \icarus, 183,
  491

\bibitem[{{Buccino} {et~al.}(2007){Buccino}, {Lemarchand}, \&
  {Mauas}}]{buccino07}
{Buccino}, A.~P., {Lemarchand}, G.~A., \& {Mauas}, P. J.~D. 2007, \icarus, 192,
  582

\bibitem[{{Carmona} {et~al.}(2023){Carmona}, {Delfosse}, {Belloti},
  Cort\'es-Zuleta, Mignon, Heidari, Artigau, Cook, Moutou, Fouque, Donati,
  Petit, Morin, Boisse, Martioli, \& the SPIRou~consortium}]{carmona22}
{Carmona}, A., {Delfosse}, X., {Belloti}, S., {et~al.} 2023, in preparation

\bibitem[{{Chabrier} \& {Baraffe}(1997)}]{chabrier97}
{Chabrier}, G. \& {Baraffe}, I. 1997, \aap, 327, 1039

\bibitem[{{Chabrier} \& {K{\"u}ker}(2006)}]{chabrier06}
{Chabrier}, G. \& {K{\"u}ker}, M. 2006, \aap, 446, 1027

\bibitem[{{Charbonneau}(2010)}]{charbonneau10}
{Charbonneau}, P. 2010, Living Reviews in Solar Physics, 7, 3

\bibitem[{{Charbonneau} \& {MacGregor}(1997)}]{charbonneau97}
{Charbonneau}, P. \& {MacGregor}, K.~B. 1997, \apj, 486, 502

\bibitem[{{Cincunegui} {et~al.}(2007){Cincunegui}, {D{\'{\i}}az}, \&
  {Mauas}}]{cincunegui07}
{Cincunegui}, C., {D{\'{\i}}az}, R.~F., \& {Mauas}, P.~J.~D. 2007, \aap, 469,
  309

\bibitem[{{Cortés-Zuleta} {et~al.}(2023){Cortés-Zuleta}, {Boisse}, {Klein},
  {Martioli}, {Cristofari}, {Antoniadis-Karnavas}, {Donati}, {Delfosse},
  {Cadieux}, {Heidari}, {Artigau}, {Belloti}, {Bonfils}, {Carmona}, {Cook},
  {Díaz}, {Moutou}, {Vandal}, team, \& team}]{cortes23}
{Cortés-Zuleta}, P., {Boisse}, I., {Klein}, B., {et~al.} 2023

\bibitem[{{Cumming} {et~al.}(1999){Cumming}, {Marcy}, \& {Butler}}]{cumming99}
{Cumming}, A., {Marcy}, G.~W., \& {Butler}, R.~P. 1999, \apj, 526, 890

\bibitem[{{Cutri} {et~al.}(2003){Cutri}, {Skrutskie}, {van Dyk}, {Beichman},
  {Carpenter}, {Chester}, {Cambresy}, {Evans}, {Fowler}, {Gizis}, {Howard},
  {Huchra}, {Jarrett}, {Kopan}, {Kirkpatrick}, {Light}, {Marsh}, {McCallon},
  {Schneider}, {Stiening}, {Sykes}, {Weinberg}, {Wheaton}, {Wheelock}, \&
  {Zacarias}}]{cutri03}
{Cutri}, R.~M., {Skrutskie}, M.~F., {van Dyk}, S., {et~al.} 2003, VizieR Online
  Data Catalog, II/246

\bibitem[{{Delfosse} {et~al.}(2013){Delfosse}, {Bonfils}, {Forveille}, {Udry},
  {Mayor}, {Bouchy}, {Gillon}, {Lovis}, {Neves}, {Pepe}, {Perrier}, {Queloz},
  {Santos}, \& {S{\'e}gransan}}]{delfosse13}
{Delfosse}, X., {Bonfils}, X., {Forveille}, T., {et~al.} 2013, \aap, 553, A8

\bibitem[{{Delfosse} {et~al.}(1998){Delfosse}, {Forveille}, {Perrier}, \&
  {Mayor}}]{delfosse98}
{Delfosse}, X., {Forveille}, T., {Perrier}, C., \& {Mayor}, M. 1998, \aap, 331,
  581

\bibitem[{{Delfosse} {et~al.}(2000){Delfosse}, {Forveille}, {S{\'e}gransan},
  {Beuzit}, {Udry}, {Perrier}, \& {Mayor}}]{delfosse00}
{Delfosse}, X., {Forveille}, T., {S{\'e}gransan}, D., {et~al.} 2000, \aap, 364,
  217

\bibitem[{{Delorme} {et~al.}(2011){Delorme}, {Collier Cameron}, {Hebb},
  {Rostron}, {Lister}, {Norton}, {Pollacco}, \& {West}}]{delorme11}
{Delorme}, P., {Collier Cameron}, A., {Hebb}, L., {et~al.} 2011, \mnras, 413,
  2218

\bibitem[{{Dikpati} {et~al.}(2005){Dikpati}, {Gilman}, \&
  {MacGregor}}]{dikpati05}
{Dikpati}, M., {Gilman}, P.~A., \& {MacGregor}, K.~B. 2005, \apj, 631, 647

\bibitem[{{Doyon} {et~al.}(2014){Doyon}, {Lafreni{\`e}re}, {Albert}, {Artigau},
  {Meyer}, \& {Jayawardhana}}]{doyon14}
{Doyon}, R., {Lafreni{\`e}re}, D., {Albert}, L., {et~al.} 2014, in Search for
  Life Beyond the Solar System. Exoplanets, Biosignatures \& Instruments, ed.
  D.~{Apai} \& P.~{Gabor}, 3.6

\bibitem[{{Dressing} \& {Charbonneau}(2015)}]{dressing15}
{Dressing}, C.~D. \& {Charbonneau}, D. 2015, \apj, 807, 45

\bibitem[{{Ducati}(2002)}]{ducati02}
{Ducati}, J.~R. 2002, VizieR Online Data Catalog

\bibitem[{{Dumusque} {et~al.}(2014){Dumusque}, {Boisse}, \&
  {Santos}}]{dumusque14}
{Dumusque}, X., {Boisse}, I., \& {Santos}, N.~C. 2014, \apj, 796, 132

\bibitem[{{Dumusque} {et~al.}(2011){Dumusque}, {Udry}, {Lovis}, {Santos}, \&
  {Monteiro}}]{dumusque11b}
{Dumusque}, X., {Udry}, S., {Lovis}, C., {Santos}, N.~C., \& {Monteiro},
  M.~J.~P.~F.~G. 2011, \aap, 525, A140

\bibitem[{{Faria} {et~al.}(2022){Faria}, {Su{\'a}rez Mascare{\~n}o},
  {Figueira}, {Silva}, {Damasso}, {Demangeon}, {Pepe}, {Santos}, {Rebolo},
  {Cristiani}, {Adibekyan}, {Alibert}, {Allart}, {Barros}, {Cabral},
  {D'Odorico}, {Di Marcantonio}, {Dumusque}, {Ehrenreich}, {Gonz{\'a}lez
  Hern{\'a}ndez}, {Hara}, {Lillo-Box}, {Lo Curto}, {Lovis}, {Martins},
  {M{\'e}gevand}, {Mehner}, {Micela}, {Molaro}, {Nunes}, {Pall{\'e}},
  {Poretti}, {Sousa}, {Sozzetti}, {Tabernero}, {Udry}, \& {Zapatero
  Osorio}}]{faria22}
{Faria}, J.~P., {Su{\'a}rez Mascare{\~n}o}, A., {Figueira}, P., {et~al.} 2022,
  \aap, 658, A115

\bibitem[{{Finch} {et~al.}(2018){Finch}, {Zacharias}, \& {Jao}}]{finch18}
{Finch}, C., {Zacharias}, N., \& {Jao}, W.-C. 2018, arXiv e-prints,
  arXiv:1802.08272

\bibitem[{{Fontenla} {et~al.}(2016){Fontenla}, {Linsky}, {Witbrod}, {France},
  {Buccino}, {Mauas}, {Vieytes}, \& {Walkowicz}}]{fontenla16}
{Fontenla}, J.~M., {Linsky}, J.~L., {Witbrod}, J., {et~al.} 2016, \apj, 830,
  154

\bibitem[{{Fuhrmeister} {et~al.}(2005){Fuhrmeister}, {Schmitt}, \&
  {Hauschildt}}]{fuhrmeister05}
{Fuhrmeister}, B., {Schmitt}, J.~H.~M.~M., \& {Hauschildt}, P.~H. 2005, \aap,
  439, 1137

\bibitem[{{Gaia Collaboration}(2018)}]{gaia18}
{Gaia Collaboration}. 2018, VizieR Online Data Catalog, I/345

\bibitem[{Gaidos {et~al.}(2014)Gaidos, Mann, L{\'e}pine, Buccino, James,
  Ansdell, Petrucci, Mauas, \& Hilton}]{gaidos2014}
Gaidos, E., Mann, A., L{\'e}pine, S., {et~al.} 2014, Monthly Notices of the
  Royal Astronomical Society, 443, 2561

\bibitem[{{Gomes da Silva} {et~al.}(2022){Gomes da Silva}, {Bensabat},
  {Monteiro}, \& {Santos}}]{gomes22}
{Gomes da Silva}, J., {Bensabat}, A., {Monteiro}, T., \& {Santos}, N.~C. 2022,
  arXiv e-prints, arXiv:2210.06903

\bibitem[{{Gomes da Silva} {et~al.}(2014){Gomes da Silva}, {Santos}, {Boisse},
  {Dumusque}, \& {Lovis}}]{gomes14}
{Gomes da Silva}, J., {Santos}, N.~C., {Boisse}, I., {Dumusque}, X., \&
  {Lovis}, C. 2014, \aap, 566, A66

\bibitem[{{Gomes da Silva} {et~al.}(2011){Gomes da Silva}, {Santos}, {Bonfils},
  {Delfosse}, {Forveille}, \& {Udry}}]{gomes11}
{Gomes da Silva}, J., {Santos}, N.~C., {Bonfils}, X., {et~al.} 2011, \aap, 534,
  A30

\bibitem[{{Gomes da Silva} {et~al.}(2012){Gomes da Silva}, {Santos}, {Bonfils},
  {Delfosse}, {Forveille}, {Udry}, {Dumusque}, \& {Lovis}}]{gomes12}
{Gomes da Silva}, J., {Santos}, N.~C., {Bonfils}, X., {et~al.} 2012, \aap, 541,
  A9

\bibitem[{H{\'e}brard {et~al.}(2016)H{\'e}brard, Donati, Delfosse, Morin,
  Moutou, \& Boisse}]{hebrard2016modelling}
H{\'e}brard, {\'E}., Donati, J.-F., Delfosse, X., {et~al.} 2016, Monthly
  Notices of the Royal Astronomical Society, 461, 1465

\bibitem[{{Henry} {et~al.}(2018){Henry}, {Jao}, {Winters}, {Dieterich},
  {Finch}, {Ianna}, {Riedel}, {Silverstein}, {Subasavage}, \&
  {Vrijmoet}}]{henry2018}
{Henry}, T.~J., {Jao}, W.-C., {Winters}, J.~G., {et~al.} 2018, \aj, 155, 265

\bibitem[{{Houdebine}(2009)}]{houdebine09}
{Houdebine}, E.~R. 2009, \mnras, 397, 2133

\bibitem[{{Hsu} {et~al.}(2020){Hsu}, {Ford}, \& {Terrien}}]{hsu2020}
{Hsu}, D.~C., {Ford}, E.~B., \& {Terrien}, R. 2020, \mnras, 498, 2249

\bibitem[{{Iba{\~n}ez Bustos} {et~al.}(2019{\natexlab{a}}){Iba{\~n}ez Bustos},
  {Buccino}, {Flores}, {Martinez}, {Maizel}, {Messina}, \& {Mauas}}]{ibanez19}
{Iba{\~n}ez Bustos}, R.~V., {Buccino}, A.~P., {Flores}, M., {et~al.}
  2019{\natexlab{a}}, \mnras, 483, 1159

\bibitem[{{Iba{\~n}ez Bustos} {et~al.}(2019{\natexlab{b}}){Iba{\~n}ez Bustos},
  {Buccino}, {Flores}, \& {Mauas}}]{ibanez19b}
{Iba{\~n}ez Bustos}, R.~V., {Buccino}, A.~P., {Flores}, M., \& {Mauas},
  P.~J.~D. 2019{\natexlab{b}}, \aap, 628, L1

\bibitem[{{Iba{\~n}ez Bustos} {et~al.}(2020){Iba{\~n}ez Bustos}, {Buccino},
  {Messina}, {Lanza}, \& {Mauas}}]{ibanez20}
{Iba{\~n}ez Bustos}, R.~V., {Buccino}, A.~P., {Messina}, S., {Lanza}, A.~F., \&
  {Mauas}, P.~J.~D. 2020, \aap, 644, A2

\bibitem[{{Ip} {et~al.}(2004){Ip}, {Kopp}, \& {Hu}}]{ip00}
{Ip}, W.-H., {Kopp}, A., \& {Hu}, J.-H. 2004, \apjl, 602, L53

\bibitem[{{Jardine} {et~al.}(2013){Jardine}, {Vidotto}, {van Ballegooijen},
  {Donati}, {Morin}, {Fares}, \& {Gombosi}}]{jardine13}
{Jardine}, M., {Vidotto}, A.~A., {van Ballegooijen}, A., {et~al.} 2013, \mnras,
  431, 528

\bibitem[{{Kiraga} \& {Stepien}(2007)}]{kiraga07}
{Kiraga}, M. \& {Stepien}, K. 2007, \actaa, 57, 149

\bibitem[{{Koen} {et~al.}(2010){Koen}, {Kilkenny}, {van Wyk}, \&
  {Marang}}]{koen10}
{Koen}, C., {Kilkenny}, D., {van Wyk}, F., \& {Marang}, F. 2010, \mnras, 403,
  1949

\bibitem[{{Kopparapu}(2013)}]{kopparapu13}
{Kopparapu}, R.~K. 2013, \apjl, 767, L8

\bibitem[{{K{\"u}ker} {et~al.}(2019){K{\"u}ker}, {R{\"u}diger}, {Olah}, \&
  {Strassmeier}}]{kuker19}
{K{\"u}ker}, M., {R{\"u}diger}, G., {Olah}, K., \& {Strassmeier}, K.~G. 2019,
  \aap, 622, A40

\bibitem[{{K{\"u}rster} {et~al.}(2003){K{\"u}rster}, {Endl}, {Rouesnel}, {Els},
  {Kaufer}, {Brillant}, {Hatzes}, {Saar}, \& {Cochran}}]{kurster03}
{K{\"u}rster}, M., {Endl}, M., {Rouesnel}, F., {et~al.} 2003, \aap, 403, 1077

\bibitem[{{Leenaarts} {et~al.}(2012){Leenaarts}, {Carlsson}, \& {Rouppe van der
  Voort}}]{leenaarts12}
{Leenaarts}, J., {Carlsson}, M., \& {Rouppe van der Voort}, L. 2012, \apj, 749,
  136

\bibitem[{{Lockwood} {et~al.}(2007){Lockwood}, {Skiff}, {Henry}, {Henry},
  {Radick}, {Baliunas}, {Donahue}, \& {Soon}}]{lockwood07}
{Lockwood}, G.~W., {Skiff}, B.~A., {Henry}, G.~W., {et~al.} 2007, \apjs, 171,
  260

\bibitem[{{Lovis} {et~al.}(2011){Lovis}, {Dumusque}, {Santos}, {Bouchy},
  {Mayor}, {Pepe}, {Queloz}, {S{\'e}gransan}, \& {Udry}}]{lovis11b}
{Lovis}, C., {Dumusque}, X., {Santos}, N.~C., {et~al.} 2011, ArXiv e-prints
  1107.5325 [\eprint[arXiv]{1107.5325}]

\bibitem[{{Lovis} {et~al.}(2017){Lovis}, {Snellen}, {Mouillet}, {Pepe},
  {Wildi}, {Astudillo-Defru}, {Beuzit}, {Bonfils}, {Cheetham}, {Conod},
  {Delfosse}, {Ehrenreich}, {Figueira}, {Forveille}, {Martins}, {Quanz},
  {Santos}, {Schmid}, {S{\'e}gransan}, \& {Udry}}]{lovis17}
{Lovis}, C., {Snellen}, I., {Mouillet}, D., {et~al.} 2017, \aap, 599, A16

\bibitem[{{Makarov} {et~al.}(2010){Makarov}, {Parker}, \& {Ulrich}}]{makarov10}
{Makarov}, V.~V., {Parker}, D., \& {Ulrich}, R.~K. 2010, \apj, 717, 1202

\bibitem[{{Mauas}(2000)}]{mauas00}
{Mauas}, P. J.~D. 2000, \apj, 539, 858

\bibitem[{{Mauas} \& {Falchi}(1994)}]{mauas94}
{Mauas}, P. J.~D. \& {Falchi}, A. 1994, \aap, 281, 129

\bibitem[{{Mayor} {et~al.}(2003){Mayor}, {Pepe}, {Queloz}, {Bouchy},
  {Rupprecht}, {Lo Curto}, {Avila}, {Benz}, {Bertaux}, {Bonfils}, {Dall},
  {Dekker}, {Delabre}, {Eckert}, {Fleury}, {Gilliotte}, {Gojak}, {Guzman},
  {Kohler}, {Lizon}, {Longinotti}, {Lovis}, {Megevand}, {Pasquini}, {Reyes},
  {Sivan}, {Sosnowska}, {Soto}, {Udry}, {van Kesteren}, {Weber}, \&
  {Weilenmann}}]{mayor03}
{Mayor}, M., {Pepe}, F., {Queloz}, D., {et~al.} 2003, The Messenger, 114, 20

\bibitem[{{Meunier}(2021)}]{meunier21b}
{Meunier}, N. 2021, arXiv e-prints, arXiv:2104.06072

\bibitem[{{Meunier} {et~al.}(2022){Meunier}, {Kretzschmar}, {Gravet}, {Mignon},
  \& {Delfosse}}]{meunier22}
{Meunier}, N., {Kretzschmar}, M., {Gravet}, R., {Mignon}, L., \& {Delfosse}, X.
  2022, \aap, 658, A57

\bibitem[{{Meunier} \& {Lagrange}(2019)}]{meunier19e}
{Meunier}, N. \& {Lagrange}, A.~M. 2019, \aap, 625, L6

\bibitem[{{Meunier} \& {Lagrange}(2020{\natexlab{a}})}]{meunier20c}
{Meunier}, N. \& {Lagrange}, A.~M. 2020{\natexlab{a}}, \aap, 638, A54

\bibitem[{{Meunier} \& {Lagrange}(2020{\natexlab{b}})}]{meunier20b}
{Meunier}, N. \& {Lagrange}, A.~M. 2020{\natexlab{b}}, \aap, 642, A157

\bibitem[{{Meunier} {et~al.}(2015){Meunier}, {Lagrange}, {Borgniet}, \&
  {Rieutord}}]{meunier15}
{Meunier}, N., {Lagrange}, A.-M., {Borgniet}, S., \& {Rieutord}, M. 2015, \aap,
  583, A118

\bibitem[{{Meunier} {et~al.}(2019){Meunier}, {Lagrange}, {Boulet}, \&
  {Borgniet}}]{meunier19}
{Meunier}, N., {Lagrange}, A.~M., {Boulet}, T., \& {Borgniet}, S. 2019, \aap,
  627, A56

\bibitem[{{Meunier} {et~al.}(2010){Meunier}, {Lagrange}, \&
  {Desort}}]{meunier10}
{Meunier}, N., {Lagrange}, A.-M., \& {Desort}, M. 2010, \aap, 519, A66

\bibitem[{{Meunier} {et~al.}(2017{\natexlab{a}}){Meunier}, {Lagrange}, {Mbemba
  Kabuiku}, {Alex}, {Mignon}, \& {Borgniet}}]{meunier17}
{Meunier}, N., {Lagrange}, A.-M., {Mbemba Kabuiku}, L., {et~al.}
  2017{\natexlab{a}}, \aap, 597, A52

\bibitem[{{Meunier} {et~al.}(2017{\natexlab{b}}){Meunier}, {Mignon}, \&
  {Lagrange}}]{meunier17b}
{Meunier}, N., {Mignon}, L., \& {Lagrange}, A.-M. 2017{\natexlab{b}}, \aap,
  607, A124

\bibitem[{{Mignon} {et~al.}(2023){Mignon}, {Delfosse}, {Bonfils}, \&
  {Meunier}}]{mignon22a}
{Mignon}, L., {Delfosse}, X., {Bonfils}, X., \& {Meunier}, N. 2023, in
  preparation

\bibitem[{{Morin} {et~al.}(2008){Morin}, {Donati}, {Petit}, {Delfosse},
  {Forveille}, {Albert}, {Auri{\`e}re}, {Cabanac}, {Dintrans}, {Fares},
  {Gastine}, {Jardine}, {Ligni{\`e}res}, {Paletou}, {Ramirez Velez}, \&
  {Th{\'e}ado}}]{morin08}
{Morin}, J., {Donati}, J.~F., {Petit}, P., {et~al.} 2008, \mnras, 390, 567

\bibitem[{{Newton} {et~al.}(2018){Newton}, {Mondrik}, {Irwin}, {Winters}, \&
  {Charbonneau}}]{newton18}
{Newton}, E.~R., {Mondrik}, N., {Irwin}, J., {Winters}, J.~G., \&
  {Charbonneau}, D. 2018, \aj, 156, 217

\bibitem[{{Parker}(1955)}]{parker55}
{Parker}, E.~N. 1955, \apj, 122, 293

\bibitem[{{Perdelwitz} {et~al.}(2021){Perdelwitz}, {Mittag}, {Tal-Or},
  {Schmitt}, {Caballero}, {Jeffers}, {Reiners}, {Schweitzer}, {Trifonov},
  {Ribas}, {Quirrenbach}, {Amado}, {Seifert}, {Cifuentes},
  {Cort{\'e}s-Contreras}, {Montes}, {Revilla}, \& {Skrzypinski}}]{perdelwitz21}
{Perdelwitz}, V., {Mittag}, M., {Tal-Or}, L., {et~al.} 2021, \aap, 652, A116

\bibitem[{{Pizzolato} {et~al.}(2003){Pizzolato}, {Maggio}, {Micela},
  {Sciortino}, \& {Ventura}}]{pizzolato03}
{Pizzolato}, N., {Maggio}, A., {Micela}, G., {Sciortino}, S., \& {Ventura}, P.
  2003, \aap, 397, 147

\bibitem[{{Pojmanski}(1997)}]{pojmanski97}
{Pojmanski}, G. 1997, \actaa, 47, 467

\bibitem[{{Pojmanski}(2002)}]{pojmanski02}
{Pojmanski}, G. 2002, \actaa, 52, 397

\bibitem[{{Radick} {et~al.}(2018){Radick}, {Lockwood}, {Henry}, {Hall}, \&
  {Pevtsov}}]{radick18}
{Radick}, R.~R., {Lockwood}, G.~W., {Henry}, G.~W., {Hall}, J.~C., \&
  {Pevtsov}, A.~A. 2018, \apj, 855, 75

\bibitem[{{Radick} {et~al.}(1998){Radick}, {Lockwood}, {Skiff}, \&
  {Baliunas}}]{radick98}
{Radick}, R.~R., {Lockwood}, G.~W., {Skiff}, B.~A., \& {Baliunas}, S.~L. 1998,
  \apjs, 118, 239

\bibitem[{{Reiners} \& {Basri}(2010)}]{reiners10}
{Reiners}, A. \& {Basri}, G. 2010, \apj, 710, 924

\bibitem[{{Reiners} {et~al.}(2012){Reiners}, {Joshi}, \& {Goldman}}]{reiners12}
{Reiners}, A., {Joshi}, N., \& {Goldman}, B. 2012, \aj, 143, 93

\bibitem[{Riaz {et~al.}(2006)Riaz, Gizis, \& Harvin}]{riaz2006identification}
Riaz, B., Gizis, J.~E., \& Harvin, J. 2006, The Astronomical Journal, 132, 866

\bibitem[{{Riedel} {et~al.}(2014){Riedel}, {Finch}, {Henry}, {Subasavage},
  {Jao}, {Malo}, {Rodriguez}, {White}, {Gies}, {Dieterich}, {Winters},
  {Davison}, {Nelan}, {Blunt}, {Cruz}, {Rice}, \& {Ianna}}]{riedel14}
{Riedel}, A.~R., {Finch}, C.~T., {Henry}, T.~J., {et~al.} 2014, \aj, 147, 85

\bibitem[{{Robertson} {et~al.}(2013){Robertson}, {Endl}, {Cochran}, \&
  {Dodson-Robinson}}]{robertson13}
{Robertson}, P., {Endl}, M., {Cochran}, W.~D., \& {Dodson-Robinson}, S.~E.
  2013, \apj, 764, 3

\bibitem[{{Robertson} \& {Mahadevan}(2014)}]{robertson14}
{Robertson}, P. \& {Mahadevan}, S. 2014, \apjl, 793, L24

\bibitem[{{Robertson} {et~al.}(2014){Robertson}, {Mahadevan}, {Endl}, \&
  {Roy}}]{robertson14b}
{Robertson}, P., {Mahadevan}, S., {Endl}, M., \& {Roy}, A. 2014, Science, 345,
  440

\bibitem[{{Robertson} {et~al.}(2020){Robertson}, {Stefansson}, {Mahadevan},
  {Endl}, {Cochran}, {Beard}, {Bender}, {Diddams}, {Duong}, {Ford}, {Fredrick},
  {Halverson}, {Hearty}, {Holcomb}, {Juan}, {Kanodia}, {Lubin}, {Metcalf},
  {Monson}, {Ninan}, {Palafoutas}, {Ramsey}, {Roy}, {Schwab}, {Terrien}, \&
  {Wright}}]{robertson20}
{Robertson}, P., {Stefansson}, G., {Mahadevan}, S., {et~al.} 2020, \apj, 897,
  125

\bibitem[{{Robinson} \& {Durney}(1982)}]{robinson82}
{Robinson}, R.~D. \& {Durney}, B.~R. 1982, \aap, 108, 322

\bibitem[{{Saar}(2011)}]{saar11}
{Saar}, S.~H. 2011, in Physics of Sun and Star Spots, ed. D.~{Prasad Choudhary}
  \& K.~G. {Strassmeier}, Vol. 273, 61--67

\bibitem[{{Saar} \& {Brandenburg}(1999)}]{saar99}
{Saar}, S.~H. \& {Brandenburg}, A. 1999, \apj, 524, 295

\bibitem[{{Santos} {et~al.}(2014){Santos}, {Mortier}, {Faria}, {Dumusque},
  {Adibekyan}, {Delgado-Mena}, {Figueira}, {Benamati}, {Boisse}, {Cunha},
  {Gomes da Silva}, {Lo Curto}, {Lovis}, {Martins}, {Mayor}, {Melo}, {Oshagh},
  {Pepe}, {Queloz}, {Santerne}, {S{\'e}gransan}, {Sozzetti}, {Sousa}, \&
  {Udry}}]{santos14}
{Santos}, N.~C., {Mortier}, A., {Faria}, J.~P., {et~al.} 2014, \aap, 566, A35

\bibitem[{{Savanov}(2012)}]{savanov12}
{Savanov}, I.~S. 2012, Astronomy Reports, 56, 716

\bibitem[{{Snellen} {et~al.}(2015){Snellen}, {de Kok}, {Birkby}, {Brandl},
  {Brogi}, {Keller}, {Kenworthy}, {Schwarz}, \& {Stuik}}]{snellen15}
{Snellen}, I., {de Kok}, R., {Birkby}, J.~L., {et~al.} 2015, \aap, 576, A59

\bibitem[{{Spiegel} \& {Zahn}(1992)}]{spiegel92}
{Spiegel}, E.~A. \& {Zahn}, J.~P. 1992, \aap, 265, 106

\bibitem[{Stauffer {et~al.}(2010)Stauffer, Tanner, Bryden, Ramirez, Berriman,
  Ciardi, Kane, Mizusawa, Payne, Plavchan, {et~al.}}]{stauffer2010accurate}
Stauffer, J., Tanner, A.~M., Bryden, G., {et~al.} 2010, Publications of the
  Astronomical Society of the Pacific, 122, 885

\bibitem[{{Steenbeck} \& {Krause}(1969)}]{steenbeck69}
{Steenbeck}, M. \& {Krause}, F. 1969, Astronomische Nachrichten, 291, 49

\bibitem[{{Strugarek}(2016)}]{strugarek16}
{Strugarek}, A. 2016, \apj, 833, 140

\bibitem[{{Su{\'a}rez Mascare{\~n}o} {et~al.}(2016){Su{\'a}rez Mascare{\~n}o},
  {Rebolo}, \& {Gonz{\'a}lez Hern{\'a}ndez}}]{suarez16}
{Su{\'a}rez Mascare{\~n}o}, A., {Rebolo}, R., \& {Gonz{\'a}lez Hern{\'a}ndez},
  J.~I. 2016, \aap, 595, A12

\bibitem[{{Su{\'a}rez Mascare{\~n}o} {et~al.}(2018){Su{\'a}rez Mascare{\~n}o},
  {Rebolo}, {Gonz{\'a}lez Hern{\'a}ndez}, {Toledo-Padr{\'o}n}, {Perger},
  {Ribas}, {Affer}, {Micela}, {Damasso}, {Maldonado}, {Gonz{\'a}lez-Alvarez},
  {Leto}, {Pagano}, {Scandariato}, {Sozzetti}, {Lanza}, {Malavolta}, {Claudi},
  {Cosentino}, {Desidera}, {Giacobbe}, {Maggio}, {Rainer}, {Esposito},
  {Benatti}, {Pedani}, {Morales}, {Herrero}, {Lafarga}, {Rosich}, \&
  {Pinamonti}}]{suarez18}
{Su{\'a}rez Mascare{\~n}o}, A., {Rebolo}, R., {Gonz{\'a}lez Hern{\'a}ndez},
  J.~I., {et~al.} 2018, \aap, 612, A89

\bibitem[{{Tremblay} {et~al.}(2013){Tremblay}, {Ludwig}, {Freytag}, {Steffen},
  \& {Caffau}}]{tremblay13}
{Tremblay}, P.-E., {Ludwig}, H.-G., {Freytag}, B., {Steffen}, M., \& {Caffau},
  E. 2013, \aap, 557, A7

\bibitem[{{van Leeuwen}(2007)}]{vanleeuwen07}
{van Leeuwen}, F. 2007, \aap, 474, 653

\bibitem[{{Vida} {et~al.}(2013){Vida}, {Kriskovics}, \& {Ol{\'a}h}}]{vida13}
{Vida}, K., {Kriskovics}, L., \& {Ol{\'a}h}, K. 2013, Astronomische
  Nachrichten, 334, 972

\bibitem[{{Vida} {et~al.}(2014){Vida}, {Ol{\'a}h}, \& {Szab{\'o}}}]{vida14}
{Vida}, K., {Ol{\'a}h}, K., \& {Szab{\'o}}, R. 2014, \mnras, 441, 2744

\bibitem[{{Vidotto} {et~al.}(2015){Vidotto}, {Fares}, {Jardine}, {Moutou}, \&
  {Donati}}]{vidotto15}
{Vidotto}, A.~A., {Fares}, R., {Jardine}, M., {Moutou}, C., \& {Donati}, J.~F.
  2015, \mnras, 449, 4117

\bibitem[{{Vidotto} {et~al.}(2013){Vidotto}, {Jardine}, {Morin}, {Donati},
  {Lang}, \& {Russell}}]{vidotto13}
{Vidotto}, A.~A., {Jardine}, M., {Morin}, J., {et~al.} 2013, \aap, 557, A67

\bibitem[{{von Bloh} {et~al.}(2007){von Bloh}, {Bounama}, {Cuntz}, \&
  {Franck}}]{vonbloh07}
{von Bloh}, W., {Bounama}, C., {Cuntz}, M., \& {Franck}, S. 2007, \aap, 476,
  1365

\bibitem[{{Weinberger} {et~al.}(2016){Weinberger}, {Boss}, {Keiser},
  {Anglada-Escud{\'e}}, {Thompson}, \& {Burley}}]{weinberger16}
{Weinberger}, A.~J., {Boss}, A.~P., {Keiser}, S.~A., {et~al.} 2016, \aj, 152,
  24

\bibitem[{{Wilson}(1978)}]{wilson78}
{Wilson}, O.~C. 1978, \apj, 226, 379

\bibitem[{{Winters} {et~al.}(2021){Winters}, {Charbonneau}, {Henry}, {Irwin},
  {Jao}, {Riedel}, \& {Slatten}}]{winters2021}
{Winters}, J.~G., {Charbonneau}, D., {Henry}, T.~J., {et~al.} 2021, \aj, 161,
  63

\bibitem[{Winters {et~al.}(2014)Winters, Henry, Lurie, Hambly, Jao, Bartlett,
  Boyd, Dieterich, Finch, Hosey, {et~al.}}]{winters2014solar}
Winters, J.~G., Henry, T.~J., Lurie, J.~C., {et~al.} 2014, The Astronomical
  Journal, 149, 5

\bibitem[{{Winters} {et~al.}(2017){Winters}, {Sevrinsky}, {Jao}, {Henry},
  {Riedel}, {Subasavage}, {Lurie}, {Ianna}, \& {Finch}}]{winters17}
{Winters}, J.~G., {Sevrinsky}, R.~A., {Jao}, W.-C., {et~al.} 2017, \aj, 153, 14

\bibitem[{{Wright} {et~al.}(2004){Wright}, {Marcy}, {Butler}, \&
  {Vogt}}]{wright04}
{Wright}, J.~T., {Marcy}, G.~W., {Butler}, R.~P., \& {Vogt}, S.~S. 2004, \apjs,
  152, 261

\bibitem[{{Wright} \& {Drake}(2016)}]{wright16}
{Wright}, N.~J. \& {Drake}, J.~J. 2016, \nat, 535, 526

\bibitem[{{Wright} {et~al.}(2011){Wright}, {Drake}, {Mamajek}, \&
  {Henry}}]{wright11}
{Wright}, N.~J., {Drake}, J.~J., {Mamajek}, E.~E., \& {Henry}, G.~W. 2011,
  \apj, 743, 48

\bibitem[{{Wright} {et~al.}(2018){Wright}, {Newton}, {Williams}, {Drake}, \&
  {Yadav}}]{wright18}
{Wright}, N.~J., {Newton}, E.~R., {Williams}, P. K.~G., {Drake}, J.~J., \&
  {Yadav}, R.~K. 2018, \mnras, 479, 2351

\bibitem[{{Yadav} {et~al.}(2015){Yadav}, {Christensen}, {Morin}, {Gastine},
  {Reiners}, {Poppenhaeger}, \& {Wolk}}]{yadav15}
{Yadav}, R.~K., {Christensen}, U.~R., {Morin}, J., {et~al.} 2015, \apjl, 813,
  L31

\bibitem[{{Yadav} {et~al.}(2016){Yadav}, {Christensen}, {Wolk}, \&
  {Poppenhaeger}}]{yadav16}
{Yadav}, R.~K., {Christensen}, U.~R., {Wolk}, S.~J., \& {Poppenhaeger}, K.
  2016, \apjl, 833, L28

\bibitem[{{Zechmeister} \& {K{\"u}rster}(2009)}]{zechmeister09b}
{Zechmeister}, M. \& {K{\"u}rster}, M. 2009, \aap, 496, 577

\bibitem[{{Zechmeister} {et~al.}(2009){Zechmeister}, {K{\"u}rster}, \&
  {Endl}}]{zechmeister09}
{Zechmeister}, M., {K{\"u}rster}, M., \& {Endl}, M. 2009, \aap, 505, 859

\end{thebibliography}

\clearpage
\onecolumn

\begin{appendix}

\section{Selection process}\label{selection_app}


\subsection{Time coverage}

As we are interested in analysing long-term variability, 
we selected stars with a temporal coverage longer than six times the maximum rotation period, P$_{\rm max}$, and with observations on at least ten independent nights. We confirmed that all published cycle periods were above that minimum span: therefore, we do not expect that many cycles are missed with this selection. 
For each star, P$_{\rm max}$ was derived from the upper envelope in the empirical law established by \cite{astudillo17}, which relates the mean activity level (average $\log R'_{HK}$) to the rotation period of star. GJ~9360 is just below this span criterion, but was kept in our sample because of the large number of measurements: this star is very quiet and beyond the validity level of the empirical law, so that its rotation rate is very uncertain.

 \subsection{Selection based on the quality of the spectra}

 The second  step of selection focuses on the quality of the spectra and is based on the S/N.
 We chose a lower threshold of 10 on the S/N in the median order (order 50, $\sim$ 550 nm), as in a previous work (\cite[][]{astudillo17} and in \cite[][]{mignon22a}).
Some spectra exhibit  a very low S/N in the range used for the Ca II H \& K index: We did not eliminate them at this stage because it would also eliminate many stars with an interesting Na or H$\alpha$ variability, but we performed a number of checks a posteriori. We first verified that  the Ca index computed with a very low S/N had no bias. 
 To be conservative, we also removed  stars from the lists of interesting stars whose S/N values were lower than one and that affected the classifications described in Sect.~\ref{sectRes}.

 For each star, we then rejected measurements with uncertainties larger than the median value plus 3$\sigma$.
 As they are dominated by photon noise, high uncertainty values reflect very poor spectral quality. In addition, some observations correspond to studies of the  Rossiter-McLaughlin effect, with many points per night, usually with a low S/N, which need to be averaged: they were kept even when the individual S/N values were low. 
 We confirmed that these particular points in the time series of five stars that remained in the final sample had very little impact on the whole time series. 
 We finally binned all measurements over each observation night (by using an average value weighted on uncertainties values) and repeated the 3$\sigma$ selection process. 
 
 Finally, all measurements rejected from the time series obtained for a given activity index were also rejected for the other activity indices to keep the same temporal sampling for the three different indices. 
 For each star in the sample, we therefore produced a time series corresponding to each activity indicator, with only one binned point per night; hereafter called night time series. 
 
\subsection{Elimination of outliers}

The third step of selection focuses on outliers that can be due to instrumental effects and flares.
To reject strong outliers from a time series, we computed the histogram of each index to identify them based on the gap between the bulk of the values and the isolated ones that should correspond to outliers. Uncertainties on the measured indices
were taken into account to detect and measure the gap used to reject  highest values and/or lowest values (often corresponding to problems in the spectra).
This method is more robust than applying a simple 3$\sigma$ threshold on the nightly indices, for example, because the rms is not representative of the bulk of the values when there are strong outliers, which prevents us from removing them.
This selection step was repeated a second time to remove remaining outliers and a few lower-intensity flares. 
Flares are stochastic events and can be present at different phases in the activity cycle and with different intensity amplitudes: some are therefore far above the other values 
and were removed during this selection process because we focus on the long-term variability caused by spots and plages, but a few may be 
indistinguishable from the bulk of the values.

\section{Stellar sample}\label{sample}

\begin{longtable}{lllllllll}
\caption{\label{tab_targets} Stellar sample}\\
\hline
Name  &  V-K & T$_{\rm eff}$  & Mass  & Dist. & N$_{\rm night}$ & $\log R'_{HK}$ & rms & Flag  \\
          &       & (K) &     (solar) &  (pc) & & & $R'_{HK}10^5$ & seasons \\
 \hline
CD-246144 &  3.546 & 3922 &  0.669 &  17.80 &   12 &  -4.619 &   0.243  & -   \\
CD-4114656 &  3.405 & 3925 &  0.600 &  23.66 &   18 &  -4.668 &   0.152  & -   \\
GJ~1 &  4.027 & 3589 &  0.394 &   4.35 &   38 &  -5.506 &   0.104  & -   \\
GJ~43 &  6.310 & 3616 &  0.494 &  29.82 &   17 &  -6.383 &   0.038 & -   \\
GJ~54.1 &  5.654 & 3200 &  0.135 &   3.71 &  202 &  -4.588 &   1.354  & X   \\
GJ~87 &  3.965 & 3700 &  0.459 &  10.51 &  136 &  -5.359 &   0.151  & -   \\
GJ~91 &  4.231 & 3757 &  0.534 &  12.79 &   21 &  -4.889 &   0.172 &  -   \\
GJ~93 &  3.780 & 3860 &  0.534 &  25.83 &   12 &  -5.168 &   0.506 &  -   \\
GJ~105B &  5.028 (d) & 3200 &  0.254 &   7.22 &   20 &  -5.514 &   0.072 &  -   \\
GJ~126 &  3.869 & 3830 &  0.505 &  25.12 &   29 &  -4.997 &   0.168 &  -   \\
\hline
\end{longtable}
\tablefoot{Stellar properties of our sample of 177 stars. 
V-K are from the CDS when available, or derived from the relation between G-K versus V-K when V is not available (indicated by "d"). Most distances are from \cite{gaia18}. When not available, distances were derived from the following parallaxes: parallaxes of GJ~257, GJ~2033, GJ~4206, GJ~9163, and HD196982 are from \cite{vanleeuwen07},
the parallax of GJ~406 is  from \cite{weinberger16}, the parallaxes of GJ~3813 and GJ~4038 are from \cite{finch18}, the parallax of L43-72 is from \cite{winters17}, and the parallax of LP~993-116 is from \cite{riaz2006identification}; the distance of  GJ~3332 is from \cite{riedel14}. Masses are from the mass-luminosity relation of \cite{delfosse00}, with a weighted average of the masses derived from $H$, $K,$ and $J$ from the CDS: these band magnitudes were taken from the 2MASS database \cite[][]{cutri03} with the exception of $H$ and $J$ for GJ~205 from \cite{ducati02}, and $H$ of GJ~1 and $K$ of GJ~803 from \cite{koen10}. T$_{\rm eff}$ is from the CDS.
The flag indicates whether the star is in the season sub-sample. Only the beginning of the table is shown for information. The full table is available online at the CDS. 
}

\section{Global variability}\label{globvar}

\begin{longtable}{llllllllll}
\caption{\label{tab_nuits} Stars in the nightly sample}\\
\hline
Name  & N$_{\rm nights}$   & Span &  P$_{\rm rotmax}$ &   SCa & rms  & SNa & rms & SH$\alpha$ & rms  \\
      &     &   (days)    & (days) &   & SCa & &  SNa & &  SH$\alpha$\\
 \hline
CD-246144 &    12 &  1845 &  23.5 &  1.184 &  0.102 &  0.135 &  0.003 &  0.659 &  0.010   \\
CD-4114656 &    18 &  4123 &  25.3 &  0.970 &  0.052 &  0.139 &  0.003 &  0.672 &  0.007   \\
GJ~1 &    38 &  2050 &  91.0 &  0.406 &  0.049 &  0.103 &  0.004 &  0.824 &  0.012   \\
GJ~43 &    17 &  5278 & 346.8 &  0.650 &  0.180 &  0.129 &  0.008 &  0.787 &  0.009   \\
GJ~54.1 &   202 &  5131 &  22.4 &  7.068 &  2.375 &  0.483 &  0.123 &  1.845 &  0.370   \\
GJ~87 &   136 &  4666 &  72.7 &  0.491 &  0.086 &  0.130 &  0.005 &  0.806 &  0.014   \\
GJ~91 &    21 &  1867 &  35.5 &  1.169 &  0.101 &  0.166 &  0.004 &  0.763 &  0.009   \\
GJ~93 &    12 &  2643 &  54.4 &  0.614 &  0.062 &  0.150 &  0.003 &  0.754 &  0.007   \\
GJ~105B &    20 &  2408 &  92.1 &  0.749 &  0.115 &  0.099 &  0.009 &  0.889 &  0.017   \\
GJ~126 &    29 &   485 &  41.8 &  0.761 &  0.090 &  0.147 &  0.005 &  0.756 &  0.008   \\
GJ~149B &    29 &  5046 &  26.3 &  0.544 &  0.083 &  0.113 &  0.003 &  0.540 &  0.006   \\
GJ~157B &    10 &  1449 &  17.8 &  6.832 &  0.911 &  0.369 &  0.024 &  2.086 &  0.084   \\
GJ~163 &   167 &  5067 &  77.5 &  0.720 &  0.096 &  0.115 &  0.006 &  0.836 &  0.015   \\
GJ~173 &    11 &  1428 &  50.5 &  0.805 &  0.058 &  0.134 &  0.002 &  0.772 &  0.009   \\
GJ~176 &    93 &  4777 &  31.6 &  1.522 &  0.200 &  0.188 &  0.012 &  0.818 &  0.039   \\
GJ~179 &    16 &  1529 &  57.9 &  1.230 &  0.122 &  0.127 &  0.008 &  0.910 &  0.022   \\
GJ~180 &    52 &  2907 &  51.2 &  0.788 &  0.085 &  0.125 &  0.006 &  0.813 &  0.017   \\
GJ~182 &    11 &  3258 &   7.8 &  6.903 &  0.254 &  0.278 &  0.011 &  1.769 &  0.069   \\
GJ~191 &   139 &  5220 & 147.6 &  0.281 &  0.033 &  0.060 &  0.005 &  0.847 &  0.020   \\
GJ~205 &    66 &  1400 &  20.9 &  2.058 &  0.169 &  0.174 &  0.008 &  0.751 &  0.025   \\
GJ~208 &    15 &  3976 &  10.5 &  3.748 &  0.273 &  0.188 &  0.014 &  1.030 &  0.065   \\
GJ~213 &   114 &  4676 & 160.2 &  0.507 &  0.101 &  0.080 &  0.007 &  0.903 &  0.008   \\
GJ~221 &    92 &  2970 &  27.4 &  0.869 &  0.073 &  0.126 &  0.002 &  0.617 &  0.005   \\
GJ~229 &   173 &  4319 &  25.1 &  1.500 &  0.099 &  0.162 &  0.004 &  0.714 &  0.011   \\
GJ~273 &   237 &  4489 &  86.7 &  0.771 &  0.083 &  0.097 &  0.005 &  0.916 &  0.020   \\
GJ~299 &    20 &  3334 &  92.0 &  0.867 &  0.292 &  0.111 &  0.019 &  0.911 &  0.017   \\
GJ~300 &    32 &  1963 &  67.2 &  1.348 &  0.141 &  0.115 &  0.013 &  0.951 &  0.024   \\
GJ~317 &    90 &  3050 &  62.9 &  1.016 &  0.147 &  0.128 &  0.010 &  0.868 &  0.024   \\
GJ~334 &    22 &  3776 &  16.2 &  2.208 &  0.109 &  0.155 &  0.003 &  0.733 &  0.013   \\
GJ~341 &    51 &  3331 &  25.6 &  1.380 &  0.133 &  0.158 &  0.005 &  0.740 &  0.017   \\
GJ~357 &    45 &  2521 &  95.2 &  0.500 &  0.082 &  0.089 &  0.004 &  0.843 &  0.009   \\
GJ~358 &    33 &  2656 &  23.2 &  3.317 &  0.471 &  0.259 &  0.033 &  1.512 &  0.162   \\
GJ~361 &    90 &  1917 &  33.4 &  1.302 &  0.114 &  0.165 &  0.004 &  0.798 &  0.015   \\
GJ~367 &    20 &  2236 &  43.4 &  1.011 &  0.062 &  0.149 &  0.005 &  0.786 &  0.007   \\
GJ~369 &    44 &  3276 &  39.7 &  0.803 &  0.089 &  0.145 &  0.005 &  0.761 &  0.009   \\
GJ~382 &    31 &  5269 &  23.0 &  1.982 &  0.185 &  0.195 &  0.010 &  0.825 &  0.035   \\
GJ~388 &    40 &  5025 &  12.1 &  9.128 &  0.904 &  0.554 &  0.083 &  3.018 &  0.245   \\
GJ~390 &    37 &  2625 &  24.6 &  1.747 &  0.118 &  0.180 &  0.006 &  0.781 &  0.016   \\
GJ~393 &   162 &  4117 &  44.3 &  0.991 &  0.090 &  0.147 &  0.006 &  0.804 &  0.018   \\
GJ~406 &    47 &  5598 &  16.3 & 66.725 & 48.278 &  3.280 &  0.449 &  6.716 &  0.951   \\
GJ~422 &    40 &  5165 & 115.2 &  0.476 &  0.089 &  0.107 &  0.005 &  0.849 &  0.014   \\
GJ~433 &    77 &  3011 &  47.4 &  0.832 &  0.073 &  0.139 &  0.005 &  0.760 &  0.010   \\
GJ~436 &   103 &  1526 &  72.1 &  0.662 &  0.053 &  0.114 &  0.004 &  0.819 &  0.009   \\
GJ~438 &    18 &  2978 &  56.7 &  0.607 &  0.057 &  0.116 &  0.005 &  0.777 &  0.009   \\
GJ~443 &    18 &  4421 &  41.2 &  1.329 &  0.105 &  0.182 &  0.010 &  0.820 &  0.018   \\
GJ~447 &   135 &  3911 &  73.1 &  1.304 &  0.236 &  0.102 &  0.014 &  0.939 &  0.030   \\
GJ~465 &    16 &  2240 & 121.5 &  0.415 &  0.058 &  0.083 &  0.005 &  0.830 &  0.010   \\
GJ~479 &    55 &  1060 &  28.5 &  2.173 &  0.205 &  0.209 &  0.014 &  1.077 &  0.050   \\
GJ~480 &    32 &   747 &  55.1 &  1.082 &  0.098 &  0.152 &  0.008 &  0.828 &  0.012   \\
GJ~510 &    12 &  1050 &  20.7 &  2.284 &  0.247 &  0.202 &  0.007 &  0.963 &  0.039   \\
GJ~514 &   142 &  4336 &  33.8 &  1.063 &  0.076 &  0.152 &  0.004 &  0.727 &  0.008   \\
GJ~526 &    30 &  1790 &  48.1 &  0.757 &  0.064 &  0.149 &  0.005 &  0.751 &  0.008   \\
GJ~536 &   173 &  4287 &  32.7 &  1.141 &  0.103 &  0.165 &  0.004 &  0.748 &  0.012   \\
GJ~551 &   140 &  5492 &  27.5 & 13.276 &  3.341 &  0.972 &  0.231 &  2.713 &  0.623   \\
GJ~569 &    23 &  2824 &  13.5 &  4.881 &  0.461 &  0.323 &  0.028 &  1.580 &  0.094   \\
GJ~581 &   219 &  2908 & 114.5 &  0.508 &  0.062 &  0.081 &  0.006 &  0.888 &  0.012   \\
GJ~588 &    59 &  5112 &  51.4 &  0.999 &  0.093 &  0.149 &  0.008 &  0.831 &  0.021   \\
GJ~606 &    18 &  1859 &  24.4 &  1.725 &  0.115 &  0.177 &  0.004 &  0.814 &  0.020   \\
GJ~618A &    17 &  2118 &  62.5 &  0.790 &  0.050 &  0.118 &  0.004 &  0.822 &  0.006   \\
GJ~620 &    19 &   846 &  21.1 &  1.617 &  0.061 &  0.162 &  0.002 &  0.716 &  0.008   \\
GJ~628 &   165 &  4088 &  92.4 &  0.726 &  0.098 &  0.096 &  0.005 &  0.899 &  0.014   \\
GJ~637 &    15 &   879 &  54.7 &  0.682 &  0.093 &  0.111 &  0.005 &  0.780 &  0.003   \\
GJ~645 &    12 &  3095 &  31.2 &  1.350 &  0.107 &  0.166 &  0.004 &  0.773 &  0.008   \\
GJ~654 &   162 &  3090 &  63.2 &  0.589 &  0.059 &  0.147 &  0.004 &  0.767 &  0.011   \\
GJ~660 &    14 &  1230 &  48.4 &  1.262 &  0.282 &  0.119 &  0.015 &  0.932 &  0.054   \\
GJ~667C &   220 &  5474 &  74.2 &  0.539 &  0.073 &  0.096 &  0.006 &  0.824 &  0.014   \\
GJ~674 &   192 &  5470 &  33.2 &  1.678 &  0.246 &  0.158 &  0.014 &  1.043 &  0.075   \\
GJ~676A &   106 &  3284 &  22.4 &  1.518 &  0.099 &  0.130 &  0.006 &  0.679 &  0.012   \\
GJ~680 &    34 &  5029 &  45.2 &  0.911 &  0.060 &  0.144 &  0.005 &  0.775 &  0.008   \\
GJ~682 &    17 &  2882 &  74.1 &  1.117 &  0.110 &  0.106 &  0.012 &  0.921 &  0.019   \\
GJ~686 &    19 &  2298 &  50.5 &  0.688 &  0.068 &  0.121 &  0.004 &  0.779 &  0.012   \\
GJ~693 &   159 &  4137 &  94.5 &  0.631 &  0.128 &  0.101 &  0.008 &  0.869 &  0.012   \\
GJ~696 &    37 &  2804 &  19.3 &  1.631 &  0.109 &  0.130 &  0.003 &  0.669 &  0.010   \\
GJ~699 &   101 &  2231 & 106.7 &  0.700 &  0.114 &  0.085 &  0.011 &  0.885 &  0.022   \\
GJ~701 &   143 &  4444 &  37.1 &  0.982 &  0.088 &  0.147 &  0.004 &  0.746 &  0.008   \\
GJ~707 &    19 &  4498 &  18.0 &  1.200 &  0.074 &  0.122 &  0.002 &  0.620 &  0.011   \\
GJ~724 &    22 &   490 &  26.4 &  1.591 &  0.119 &  0.172 &  0.007 &  0.729 &  0.015   \\
GJ~729 &    18 &  4096 &  15.1 &  7.074 &  0.510 &  0.452 &  0.032 &  2.209 &  0.126   \\
GJ~739 &    19 &  1403 &  49.4 &  0.957 &  0.146 &  0.152 &  0.009 &  0.818 &  0.025   \\
GJ~740 &    53 &  3570 &  23.2 &  1.542 &  0.111 &  0.155 &  0.004 &  0.706 &  0.014   \\
GJ~752A &   135 &  4502 &  46.6 &  1.014 &  0.088 &  0.159 &  0.006 &  0.799 &  0.017   \\
GJ~754 &   157 &  4496 &  76.8 &  1.173 &  0.386 &  0.107 &  0.026 &  0.938 &  0.046   \\
GJ~784 &    38 &  5021 &  21.5 &  1.561 &  0.089 &  0.160 &  0.003 &  0.708 &  0.009   \\
GJ~803 &    29 &  5050 &   7.5 &  9.252 &  0.849 &  0.378 &  0.046 &  2.340 &  0.225   \\
GJ~816 &    12 &   679 &  51.3 &  0.939 &  0.079 &  0.147 &  0.004 &  0.812 &  0.009   \\
GJ~832 &    99 &  5095 &  53.9 &  0.712 &  0.086 &  0.127 &  0.005 &  0.776 &  0.009   \\
GJ~846 &    48 &  2345 &  20.3 &  1.775 &  0.088 &  0.162 &  0.003 &  0.719 &  0.010   \\
GJ~849 &    63 &  5298 &  58.1 &  0.988 &  0.063 &  0.150 &  0.009 &  0.793 &  0.011   \\
GJ~855 &    33 &  2148 &  22.9 &  1.714 &  0.079 &  0.173 &  0.003 &  0.735 &  0.008   \\
GJ~864 &    13 &  2912 &  30.0 &  1.022 &  0.073 &  0.142 &  0.003 &  0.681 &  0.005   \\
GJ~876 &   212 &  5304 &  77.5 &  0.978 &  0.116 &  0.109 &  0.010 &  0.907 &  0.026   \\
GJ~877 &    35 &  3016 &  65.5 &  0.771 &  0.066 &  0.125 &  0.004 &  0.828 &  0.015   \\
GJ~880 &   126 &  4449 &  25.6 &  1.665 &  0.137 &  0.177 &  0.008 &  0.764 &  0.022   \\
GJ~887 &   108 &  5674 &  26.8 &  1.227 &  0.164 &  0.157 &  0.009 &  0.763 &  0.024   \\
GJ~891 &    28 &   526 &  41.2 &  1.040 &  0.112 &  0.141 &  0.006 &  0.806 &  0.015   \\
GJ~908 &    76 &  3059 &  64.0 &  0.531 &  0.068 &  0.118 &  0.004 &  0.782 &  0.010   \\
GJ~1001A &    10 &  2684 &  82.7 &  0.950 &  0.162 &  0.092 &  0.009 &  0.897 &  0.007   \\
GJ~1061 &    86 &  5578 & 119.6 &  1.686 &  0.312 &  0.161 &  0.042 &  0.985 &  0.035   \\
GJ~1075 &    31 &  1328 &  15.4 &  1.928 &  0.110 &  0.156 &  0.004 &  0.720 &  0.014   \\
GJ~1100 &    27 &  1111 &  34.4 &  0.862 &  0.084 &  0.155 &  0.003 &  0.716 &  0.007   \\
GJ~1132 &   103 &   744 &  63.2 &  1.252 &  0.315 &  0.115 &  0.016 &  0.914 &  0.027   \\
GJ~1135 &    17 &   777 &  28.1 &  1.323 &  0.047 &  0.149 &  0.002 &  0.720 &  0.009   \\
GJ~1214 &   121 &  3319 & 104.6 &  1.325 &  0.408 &  0.219 &  0.163 &  0.966 &  0.034   \\
GJ~1236 &    11 &  2902 &  79.7 &  0.680 &  0.187 &  0.105 &  0.033 &  0.877 &  0.011   \\
GJ~2049 &    46 &  3731 &  25.2 &  1.257 &  0.074 &  0.141 &  0.006 &  0.660 &  0.007   \\
GJ~2056 &    10 &  2159 &  28.8 &  0.914 &  0.079 &  0.120 &  0.002 &  0.612 &  0.003   \\
GJ~2066 &   100 &  4108 &  48.8 &  0.885 &  0.063 &  0.144 &  0.005 &  0.786 &  0.010   \\
GJ~2121 &    28 &  1446 &  77.4 &  0.723 &  0.248 &  0.128 &  0.012 &  0.860 &  0.025   \\
GJ~2126 &    94 &  5405 &  15.4 &  1.448 &  0.093 &  0.120 &  0.003 &  0.628 &  0.010   \\
GJ~3018 &    17 &  1018 &  26.9 &  1.345 &  0.067 &  0.163 &  0.003 &  0.749 &  0.009   \\
GJ~3053 &   128 &   783 &  88.7 &  0.962 &  0.271 &  0.144 &  0.076 &  0.930 &  0.029   \\
GJ~3082 &    40 &  1567 &  26.5 &  1.577 &  0.163 &  0.171 &  0.006 &  0.818 &  0.029   \\
GJ~3090 &    21 &  3734 &  16.3 &  3.012 &  0.199 &  0.230 &  0.009 &  0.993 &  0.040   \\
GJ~3135 &   143 &  2837 &  79.8 &  0.687 &  0.269 &  0.089 &  0.013 &  0.864 &  0.024   \\
GJ~3148 &    58 &  1403 &  15.9 &  6.139 &  1.561 &  0.360 &  0.055 &  2.345 &  0.218   \\
GJ~3192 &    15 &  5568 &  52.3 &  0.662 &  0.113 &  0.099 &  0.009 &  0.852 &  0.012   \\
GJ~3218 &    39 &   833 &  22.0 &  2.357 &  0.202 &  0.222 &  0.010 &  0.970 &  0.041   \\
GJ~3221 &    25 &  1028 &  19.1 &  3.663 &  0.478 &  0.231 &  0.012 &  1.534 &  0.099   \\
GJ~3256 &    23 &   699 &  22.3 &  2.167 &  0.205 &  0.193 &  0.010 &  0.836 &  0.027   \\
GJ~3293 &   191 &  2291 &  43.8 &  1.122 &  0.151 &  0.158 &  0.011 &  0.809 &  0.018   \\
GJ~3307 &    18 &  1763 &  27.3 &  1.621 &  0.102 &  0.171 &  0.007 &  0.743 &  0.014   \\
GJ~3323 &   120 &  4332 &  24.8 &  5.057 &  1.898 &  0.274 &  0.081 &  1.502 &  0.335   \\
GJ~3341 &   114 &  1437 &  62.2 &  0.748 &  0.180 &  0.136 &  0.009 &  0.769 &  0.010   \\
GJ~3403 &    11 &  4338 &  24.0 &  1.340 &  0.073 &  0.157 &  0.002 &  0.727 &  0.010   \\
GJ~3404 &    12 &  1492 &  86.6 &  0.629 &  0.083 &  0.098 &  0.006 &  0.860 &  0.014   \\
GJ~3440 &    28 &  3000 &  24.8 &  1.940 &  0.289 &  0.198 &  0.011 &  0.804 &  0.016   \\
GJ~3470 &   108 &  3391 &  29.7 &  1.624 &  0.261 &  0.192 &  0.017 &  0.811 &  0.020   \\
GJ~3502 &    24 &  1548 &  65.4 &  0.696 &  0.117 &  0.143 &  0.012 &  0.799 &  0.012   \\
GJ~3528 &    14 &  1601 &  46.6 &  0.976 &  0.144 &  0.163 &  0.009 &  0.819 &  0.017   \\
GJ~3543 &    83 &  2330 &  22.6 &  2.183 &  0.186 &  0.202 &  0.009 &  0.933 &  0.040   \\
GJ~3563 &    12 &   873 &  75.7 &  0.621 &  0.085 &  0.101 &  0.003 &  0.808 &  0.003   \\
GJ~3591 &    16 &  4798 &  25.8 &  1.106 &  0.148 &  0.150 &  0.004 &  0.672 &  0.012   \\
GJ~3634 &    60 &  1438 &  48.6 &  0.950 &  0.124 &  0.140 &  0.008 &  0.777 &  0.009   \\
GJ~3643 &    21 &  1419 &  68.1 &  0.658 &  0.216 &  0.097 &  0.012 &  0.837 &  0.010   \\
GJ~3708 &    22 &   845 &  73.4 &  0.736 &  0.153 &  0.105 &  0.013 &  0.868 &  0.019   \\
GJ~3728 &    16 &   320 &  16.3 &  3.215 &  0.457 &  0.246 &  0.009 &  1.202 &  0.039   \\
GJ~3737 &    10 &  3153 & 160.1 &  0.614 &  0.293 &  0.080 &  0.009 &  0.907 &  0.013   \\
GJ~3779 &    17 &  1804 &  78.3 &  0.889 &  0.119 &  0.136 &  0.007 &  0.764 &  0.010   \\
GJ~3804 &    15 &  1082 &  97.5 &  0.664 &  0.109 &  0.098 &  0.008 &  0.870 &  0.006   \\
GJ~3813 &    13 &   376 &  35.2 &  1.404 &  0.116 &  0.166 &  0.008 &  0.758 &  0.012   \\
GJ~3822 &    61 &  1770 &  18.8 &  2.156 &  0.157 &  0.175 &  0.005 &  0.774 &  0.020   \\
GJ~3871 &    28 &  3606 &  29.0 &  1.893 &  0.146 &  0.206 &  0.012 &  0.820 &  0.015   \\
GJ~3874 &    20 &  1395 &  38.8 &  0.849 &  0.084 &  0.168 &  0.004 &  0.729 &  0.012   \\
GJ~3915 &    16 &   364 &  41.3 &  0.908 &  0.102 &  0.157 &  0.005 &  0.760 &  0.005   \\
GJ~3962 &    24 &  1708 &  20.4 &  1.827 &  0.109 &  0.155 &  0.005 &  0.714 &  0.013   \\
GJ~3969 &    30 &  1548 &  27.4 &  1.420 &  0.176 &  0.154 &  0.008 &  0.752 &  0.013   \\
GJ~4001 &    17 &  2100 &  25.1 &  1.655 &  0.108 &  0.157 &  0.004 &  0.796 &  0.018   \\
GJ~4079 &    17 &  1492 &  14.7 &  3.198 &  0.286 &  0.229 &  0.010 &  1.011 &  0.044   \\
GJ~4092 &    10 &   280 &  27.5 &  1.074 &  0.057 &  0.139 &  0.001 &  0.651 &  0.003   \\
GJ~4100 &   114 &  2744 &  23.9 &  1.879 &  0.286 &  0.177 &  0.011 &  0.768 &  0.024   \\
GJ~4160 &    17 &   663 &  36.6 &  0.963 &  0.090 &  0.168 &  0.005 &  0.727 &  0.006   \\
GJ~4206 &    28 &  1416 &  25.0 &  1.544 &  0.197 &  0.154 &  0.006 &  0.692 &  0.015   \\
GJ~4254 &    27 &  1561 &  21.0 &  1.561 &  0.075 &  0.148 &  0.003 &  0.715 &  0.006   \\
GJ~4303 &    67 &  2554 &  15.6 &  3.221 &  0.464 &  0.224 &  0.012 &  0.973 &  0.047   \\
GJ~4332 &    20 &  3804 &  20.8 &  1.899 &  0.177 &  0.164 &  0.004 &  0.723 &  0.015   \\
GJ~9018 &    41 &   436 &  25.7 &  1.478 &  0.108 &  0.168 &  0.005 &  0.740 &  0.014   \\
GJ~9050 &    10 &  3654 &  33.4 &  1.267 &  0.106 &  0.181 &  0.007 &  0.783 &  0.015   \\
GJ~9066 &    24 &  1780 &  23.2 &  6.693 &  1.265 &  0.570 &  0.042 &  2.035 &  0.130   \\
GJ~9133 &    13 &  1187 &  53.6 &  0.603 &  0.046 &  0.119 &  0.004 &  0.777 &  0.011   \\
GJ~9137 &    18 &  4760 &  59.7 &  0.658 &  0.111 &  0.143 &  0.005 &  0.779 &  0.009   \\
GJ~9201 &    13 &  2251 &  46.2 &  0.863 &  0.091 &  0.162 &  0.004 &  0.758 &  0.008   \\
GJ~9206 &    11 &  1535 &  37.6 &  1.176 &  0.077 &  0.159 &  0.002 &  0.708 &  0.004   \\
GJ~9210 &    10 &  1764 &  24.6 &  0.739 &  0.018 &  0.119 &  0.001 &  0.580 &  0.004   \\
GJ~9349 &    17 &  1829 &  49.3 &  0.896 &  0.120 &  0.138 &  0.008 &  0.778 &  0.016   \\
GJ~9360 &    27 &  4420 & 548.7 &  0.395 &  0.148 &  0.095 &  0.007 &  0.817 &  0.010   \\
GJ~9425 &    81 &  4204 &  22.6 &  1.137 &  0.112 &  0.107 &  0.003 &  0.612 &  0.006   \\
GJ~9443 &    15 &  4422 &  45.1 &  0.727 &  0.085 &  0.153 &  0.006 &  0.729 &  0.005   \\
GJ~9482 &    39 &  4466 &  18.2 &  1.693 &  0.101 &  0.140 &  0.003 &  0.670 &  0.011   \\
GJ~9568 &    15 &  1136 &  99.1 &  0.531 &  0.197 &  0.108 &  0.006 &  0.832 &  0.009   \\
GJ~9588 &    34 &  2609 &  66.9 &  0.461 &  0.099 &  0.100 &  0.004 &  0.791 &  0.010   \\
GJ~9592 &   138 &  4494 &  28.6 &  1.331 &  0.133 &  0.159 &  0.006 &  0.736 &  0.018   \\
L32-8 &    12 &  3374 &  63.5 &  1.002 &  0.112 &  0.116 &  0.005 &  0.868 &  0.008   \\
LP816-60 &   133 &  4504 &  60.1 &  1.381 &  0.226 &  0.113 &  0.012 &  0.935 &  0.031   \\
LTT1349 &    44 &  1849 &  17.3 &  1.714 &  0.107 &  0.136 &  0.003 &  0.682 &  0.013   \\
HIP77518 &    11 &  1747 &  21.8 &  1.464 &  0.209 &  0.153 &  0.007 &  0.701 &  0.016   \\
HIP95903 &    31 &  1869 &  24.7 &  1.034 &  0.088 &  0.120 &  0.004 &  0.631 &  0.005   \\
BD-13 6424 &    12 &   317 &   8.2 &  8.347 &  0.350 &  0.347 &  0.016 &  2.112 &  0.080   \\
\hline
\end{longtable}
\tablefoot{Properties of the 177 stars in the nightly sample. 
}

\section{Estimate of $\tau_{\rm min}$ from the linear and polynomial criteria}\label{pmin}

We followed an approach very similar to that of \cite{buccino11} and \cite{ibanez19}  to attempt to estimate a minimum timescale that would be compatible with the observed variability. We used a monthly smoothed time series of the spot number from the Solar Influences Data analysis Center (SIDC) since 1749. We scaled this time series to periods over a  range (300 days to 20 years) instead of the 11~yr solar cycle, with a step of 100~days. For each star and each period, we applied the observed sampling at a random phase, added some noise (corresponding to the observed short-timescale dispersion for that star inside each season) and scaled the time series in amplitude so that it corresponded to its observed global dispersion and average level (for a given activity index). We made 1000 realisations for each period, and we carried out linear and quadratic fits for all of them and computed the F$_{\rm red}$ in each case. This was done for the nightly time series and, when possible, on the season time series. We then compared these F$_{\rm red}$ values, obtained at a given period, to the observed values and computed the percentage of simulations for which the synthetic F$_{\rm red}$ was higher than the observed value for that period. 
More specifically, for a star in the nightly sample alone, we computed the number of simulations for which F$_{\rm red\:1/0}$  was higher than the observed F$_{\rm red\:1/0}$  (if the latter was higher than 0.2) and F$_{\rm red\:2/0}$  was higher than the observed F$_{\rm red\:2/0}$  (if the latter was higher than 0.2). When the star was also in the season sample, the test was made on four values of F$_{\rm red}$ when one of the four observed F$_{\rm red}$ was higher than 0.2. We then obtained for each star and activity index the percentage of realisations for which a value of F$_{\rm red}$ as high as the observed value can be reached as a function of the period of the synthetic signal. 

\begin{figure}[h]
\includegraphics{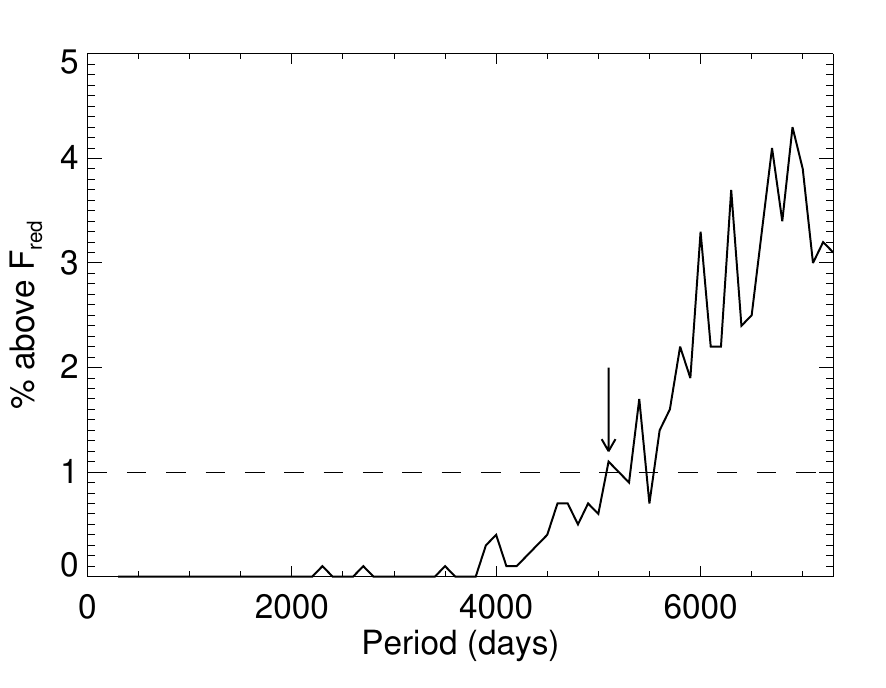}
\caption{
Example of the percentage of realisations with F$_{\rm red}$ values above the observed value at each tested period in the Ca time series of GJ~87. The arrow indicates the derived $\tau_{\rm min}$.
}
\label{ex_pc_pmin}
\end{figure}

An example is shown in Fig.~\ref{ex_pc_pmin}. Below 5100 d, fewer than 1\% of the realisations of the synthetic series can reach or are above the observed F$_{\rm red}$  values. On the other hand, above 5100 d,  more than 1\% of the realisations verify this condition. 
In this case (for any period longer than 5100 d), we considered that even if this tested period is not the true one, a signal at this period would have a likelihood of more than 1\% to give F$_{\rm red}$ values as high as the observed values for that star and index, and that therefore, we would not be able to rule out this period (if this period is not the true period, it would therefore correspond to a fap, and we set this fap level at 1\%).
The shortest period for which the percentage is higher than 1\% is therefore considered to be $\tau_{\rm min}$. This estimation could be made for stars: we interpret this by the likely presence of a long-term variability on a timescale longer than $\tau_{\rm min}$ for those stars, which are therefore of interest for long-term variability studies. Values of $\tau_{\rm min}$ blow the temporal coverage of the observations are indicated in Table~\ref{tab_Pcyc}. Those that are higher than the temporal coverage are indicated in Table~\ref{tab_elim}.

\section{Table of retrieved timescales from our photospheric and chromospheric analysis}

\begin{longtable}{llllll|llllll}
\caption{\label{tab_Pcyc} Long-term variability timescales in our sample}\\
\hline
Name  & V-K & $\log R'_{HK}$ &  $\tau_{\rm cyc}$  & Indicator & Span & Name  & V-K & $\log R'_{HK}$ & $\tau_{\rm cyc}$  & Indicator & Span \\
      &      &              & (days) & & (days)        &  &       &               & (days)         &  & (days) \\
\hline
GJ87 (N) & 3.96 &   -5.36 & 2948$\pm$89 & H$\alpha$ &  4666 &   &   &   & 2716$\pm$51 & Phot &  3189   \\   
  &   &   & $>$4600 & Na &  4666 & GJ680  & 4.30 &   -5.05 & 867$\pm$20 & Phot &  3189   \\   
  &   &   & $>$900 & H$\alpha$ &  4666 &   &   &   & 1918$\pm$48 & Phot &  3189   \\   
  &   &   & 1277$\pm$34 & Phot &  3256 & GJ693 (N) & 4.77 &   -5.53 & 1162$\pm$52 & Na &  4137   \\   
GJ149B (N) & 2.66 &   -4.69 & $>$2200 & Ca &  5046 & GJ701 (N) & 4.05 &   -4.92 & 3205$\pm$269 & Na &  4444   \\   
GJ176  & 4.34 &   -4.81 & $>$2900 & Ca &  4777 &   &   &   & $>$3400 & Ca &  4444   \\   
  &   &   & $>$2200 & Na &  4777 & GJ752A  & 4.44 &   -5.07 & 1255$\pm$36 & Na &  4502   \\   
  &   &   & $>$2500 & H$\alpha$ &  4777 &   &   &   & $>$4100 & Ca &  4502   \\   
GJ180 (N) & 4.30 &   -5.13 & 2748$\pm$20 & Phot &  3299 &   &   &   & $>$1200 & Na &  4502   \\   
GJ191 (N) & 3.80 &   -5.82 & 891$\pm$7 & Na &  5220 &   &   &   & $>$2300 & H$\alpha$ &  4502   \\   
  &   &   & 417$\pm$7 & Phot &  3299 &   &   &   & 1186$\pm$53 & Phot &  2944   \\   
GJ213 (N) & 5.12 &   -5.88 & $>$1300 & Ca &  4676 & GJ754 (N) & 5.38 &   -5.40 & 1739$\pm$64 & Na &  4496   \\   
GJ221 (N) & 3.39 &   -4.72 & 1008$\pm$65 & Phot &  3297 &   &   &   & 1522$\pm$74 & H$\alpha$ &  4496   \\   
GJ229  & 3.96 &   -4.66 & 713$\pm$8 & H$\alpha$ &  4319 & GJ832  & 4.17 &   -5.16 & 2357$\pm$69 & Ca &  5095   \\   
  &   &   & $>$1000 & Ca &  4319 &   &   &   & 2346$\pm$76 & Na &  5095   \\   
  &   &   & 1376$\pm$38 & Phot &  3299 &   &   &   & 1114$\pm$74 & Phot &  3256   \\   
  &   &   & 3234$\pm$92 & Phot &  3299 & GJ849  & 4.77 &   -5.21 & $>$2400 & Na &  5298   \\   
GJ273  & 5.01 &   -5.52 & 1975$\pm$49 & Ca &  4489 & GJ876 (N) & 5.18 &   -5.40 & 2065$\pm$80 & H$\alpha$ &  5304   \\   
  &   &   & $>$1600 & Ca &  4489 &   &   &   & 1647$\pm$92 & Phot &  3286   \\   
  &   &   & 2258$\pm$266 & Phot &  2933 & GJ880  & 4.12 &   -4.68 & $>$2300 & Ca &  4449   \\   
GJ317  & 4.95 &   -4.57 & 2133$\pm$60 & H$\alpha$ &  3050 &   &   &   & $>$3900 & Na &  4449   \\   
  &   &   & $>$1500 & Na &  3050 &   &   &   & $>$2100 & H$\alpha$ &  4449   \\   
GJ341 (N) & 3.88 &   -4.67 & $>$3100 & Ca &  3331 &   &   &   & 387$\pm$6 & Phot &  2345   \\   
  &   &   & $>$3000 & H$\alpha$ &  3331 &   &   &   & 909$\pm$33 & Phot &  2345   \\   
GJ393 (N) & 4.34 &   -5.03 & $>$1700 & Ca &  4117 & GJ887 (N) & 3.88 &   -4.70 & 2411$\pm$53 & Na &  5674   \\   
  &   &   & 726$\pm$3 & Phot &  3152 &   &   &   & $>$1500 & Na &  5674   \\   
GJ433 (N) & 4.19 &   -5.08 & 1094$\pm$31 & H$\alpha$ &  3011 &   &   &   & $>$1300 & H$\alpha$ &  5674   \\   
GJ447  & 5.50 &   -5.36 & 2064$\pm$120 & Ca &  3911 & GJ908 (N) & 3.95 &   -5.28 & 1303$\pm$123 & Phot &  3279   \\   
  &   &   & 806$\pm$16 & Na &  3911 & GJ1075 (N) & 3.49 &   -4.34 & $>$800 & Ca &  1328   \\   
  &   &   & 1556$\pm$71 & Phot &  3168 &   &   &   & 1092$\pm$27 & Phot &  3248   \\   
GJ514  & 3.99 &   -4.86 & 980$\pm$26 & Ca &  4336 & GJ2056 (N) & 3.49 &   -4.75 & 1616$\pm$26 & Phot &  3300   \\   
  &   &   & 1141$\pm$71 & Phot &  2926 & GJ2066 (N) & 4.33 &   -5.10 & 1224$\pm$33 & Phot &  3277   \\   
GJ536 (N) & 4.02 &   -4.83 & 793$\pm$16 & Ca &  4287 &   &   &   & 2759$\pm$620 & Phot &  3277   \\   
  &   &   & 746$\pm$21 & Na &  4287 & GJ2126 (N) & 3.12 &   -4.34 & 1465$\pm$90 & Ca &  5405   \\   
  &   &   & $>$3700 & Ca &  4287 & GJ3135 (N) & 4.61 &   -5.42 & $>$1800 & Na &  2837   \\   
GJ551  & 6.75 &   -4.72 & $>$5100 & Na &  5492 & GJ3148 (N) & 4.90 &   -4.36 & 1813$\pm$26 & Phot &  3296   \\   
  &   &   & 2561$\pm$67 & Phot &  3179 & GJ3341 (N) & 4.35 &   -5.26 & $>$1100 & Na &  1437   \\   
GJ581  & 4.72 &   -5.66 & 1419$\pm$72 & Ca &  2908 & GJ3822  & 3.93 &   -4.47 & $>$1600 & Na &  1770   \\   
  &   &   & 2270$\pm$59 & Phot &  3179 & GJ9425 (N) & 3.42 &   -4.59 & 3031$\pm$173 & Ca &  4204   \\   
GJ588  & 4.55 &   -5.13 & 2977$\pm$174 & Na &  5112 &   &   &   & $>$2500 & Na &  4204   \\   
GJ628  & 5.00 &   -5.61 & 1460$\pm$44 & Na &  4088 &   &   &   & 1022$\pm$60 & Phot &  3213   \\   
  &   &   & 1726$\pm$72 & H$\alpha$ &  4088 &   &   &   & 2645$\pm$164 & Phot &  3213   \\   
  &   &   & $>$1200 & Ca &  4088 & GJ9482 (N) & 3.59 &   -4.45 & $>$800 & H$\alpha$ &  4466   \\   
  &   &   & $>$1000 & H$\alpha$ &  4088 & GJ9592 (N) & 4.01 &   -4.75 & 3527$\pm$170 & Ca &  4494   \\   
  &   &   & 395$\pm$4 & Phot &  3160 &   &   &   & 3734$\pm$186 & Na &  4494   \\   
  &   &   & 1272$\pm$22 & Phot &  3160 &   &   &   & 1647$\pm$41 & H$\alpha$ &  4494   \\   
GJ654 (N) & 4.10 &   -5.27 & $>$2100 & Ca &  3090 &   &   &   & $>$3500 & Ca &  4494   \\   
  &   &   & $>$2900 & H$\alpha$ &  3090 &   &   &   & $>$2800 & Na &  4494   \\   
GJ676A  & 3.76 &   -4.59 & 740$\pm$15 & Na &  3284 &   &   &   & $>$3200 & H$\alpha$ &  4494   \\   
  &   &   & 2982$\pm$269 & H$\alpha$ &  3284 & LP816-60  & 5.26 &   -5.23 & 606$\pm$14 & Na &  4504   \\   
\hline
\end{longtable}
\tablefoot{Typical timescales $\tau_{\rm cyc}$ from our analysis of chromospheric time series and photometric time series. The lower estimate of timescales derived from linear and quadratic analysis (Appendix~\ref{pmin}) is  indicated. The V-K values are from Table~\ref{tab_targets} for stars in our sample, and are either from the CDS or based on a G-K law as for some of the stars in our sample. 
The $\log R'_{HK}$ is  from our analysis.
(N) after the name highlights stars without a previously published cycle period.
}

\begin{longtable}{llllll|llllll}
\caption{\label{tab_elim} Long-term variability of interest for exoplanet searches}\\
\hline
Name  & V-K & $\log R'_{HK}$ &  $\tau$  & Indicator & Span & Name  & V-K & $\log R'_{HK}$ & $\tau$  & Indicator & Span \\
    &      &              & (days) & & (days)        &  &       &               & (days)         &  & (days) \\
\hline
GJ87 (N) & 3.96 &   -5.36 & 8190$\pm$1823 & Ca &  4666 &   &   &   & 4947$\pm$31 & H$\alpha$ &  4502   \\   
  &   &   & $>$5100 & Ca &  4666 &   &   &   & 5200$\pm$303 & Phot &  2944   \\   
  &   &   & 5015$\pm$186 & Phot &  3256 & GJ832  & 4.17 &   -5.16 & 5331$\pm$306 & Phot &  3256   \\   
GJ208 (N) & 3.63 &   -4.09 & 7008$\pm$219 & Phot &  3186 & GJ849  & 4.77 &   -5.21 & 3815$\pm$364 & Phot &  3236   \\   
GJ221 (N) & 3.39 &   -4.72 & 5783$\pm$1555 & Ca &  2970 & GJ876 (N) & 5.18 &   -5.40 & 5966$\pm$775 & Na &  5304   \\   
  &   &   & $>$3800 & Ca &  2970 & GJ880  & 4.12 &   -4.68 & 4685$\pm$354 & Ca &  4449   \\   
  &   &   & 4125$\pm$783 & Phot &  3297 &   &   &   & 5330$\pm$418 & Na &  4449   \\   
GJ229  & 3.96 &   -4.66 & 5640$\pm$506 & Ca &  4319 &   &   &   & 5203$\pm$584 & H$\alpha$ &  4449   \\   
GJ341 (N) & 3.88 &   -4.67 & $>$3500 & Na &  3331 & GJ908 (N) & 3.95 &   -5.28 & $>$3700 & Ca &  3059   \\   
GJ406  & 7.42 &   -4.38 & 3373$\pm$12 & Phot &  2926 &   &   &   & $>$3400 & Na &  3059   \\   
GJ436 (N) & 4.54 &   -5.35 & $>$1700 & Na &  1526 & GJ1075 (N) & 3.49 &   -4.34 & 4037$\pm$319 & Phot &  3248   \\   
GJ514  & 3.99 &   -4.86 & 3801$\pm$333 & Phot &  2926 & GJ2126 (N) & 3.12 &   -4.34 & 3236$\pm$331 & Phot &  3166   \\   
GJ628  & 5.00 &   -5.61 & 4155$\pm$258 & Ca &  4088 & GJ3634 (N) & 4.46 &   -5.09 & 3505$\pm$60 & Phot &  3299   \\   
GJ696 (N) & 3.61 &   -4.49 & 7360$\pm$171 & Phot &  3163 & GJ4001 (N) & 3.86 &   -4.66 & 4427$\pm$1449 & Phot &  3154   \\   
GJ752A  & 4.44 &   -5.07 & 5822$\pm$562 & Ca &  4502 & GJ9482 (N) & 3.59 &   -4.45 & 4215$\pm$154 & Phot &  3187   \\   
\hline
\end{longtable}
\tablefoot{
Stars eliminated due to periods longer than the coverage of the observations (step 4 for the chromospheric analysis). Typical timescales $\tau$ and lower estimate of timescales derived from linear and quadratic analysis (Appendix~\ref{pmin}), even though they are not robust because they are longer than the temporal coverage, are indicated as possible timescales that may have to be considered when searching for exoplanets. The V-K values are from Table~\ref{tab_targets} for stars in our sample, and are either from the CDS or based on a G-K law as for some of the stars in our sample. The $\log R'_{HK}$ is  from our analysis.
(N) after the name highlights stars without a previously published cycle period.
}

\section{Figures of time series of a selected sample of stars}\label{plot_serie}

\begin{figure}
\includegraphics{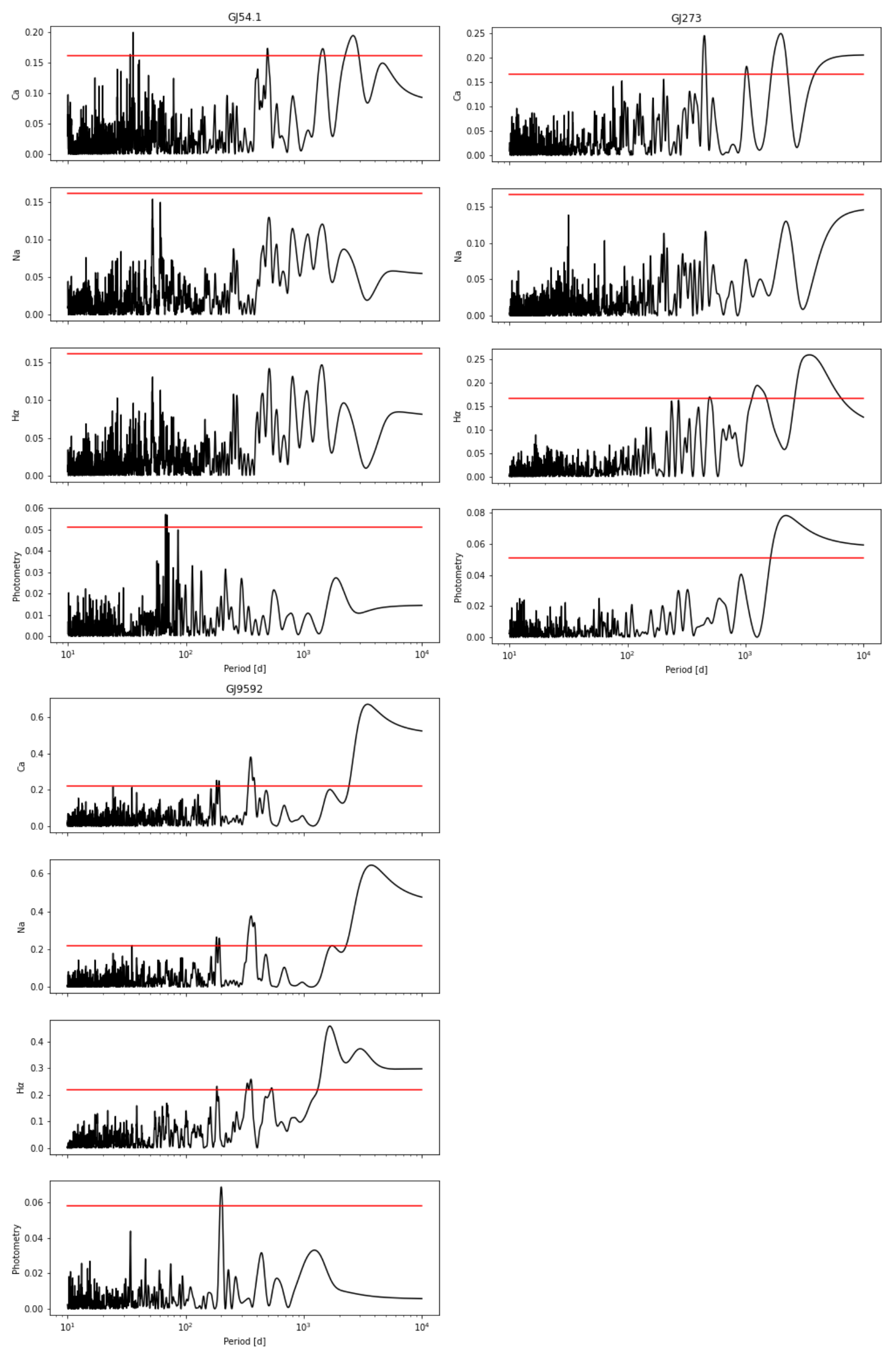}
\caption{Chromospheric and photometric periodograms for GJ~54.1, GJ~273, and GJ~9592. The red horizontal lines correspond to a 1\% fap level. 
}
\label{fig_periodo}
\end{figure}

\begin{figure}[h!]
    \centering
    \subfigure{%
     \label{plot_serie}
       \includegraphics[width=0.85\textwidth]{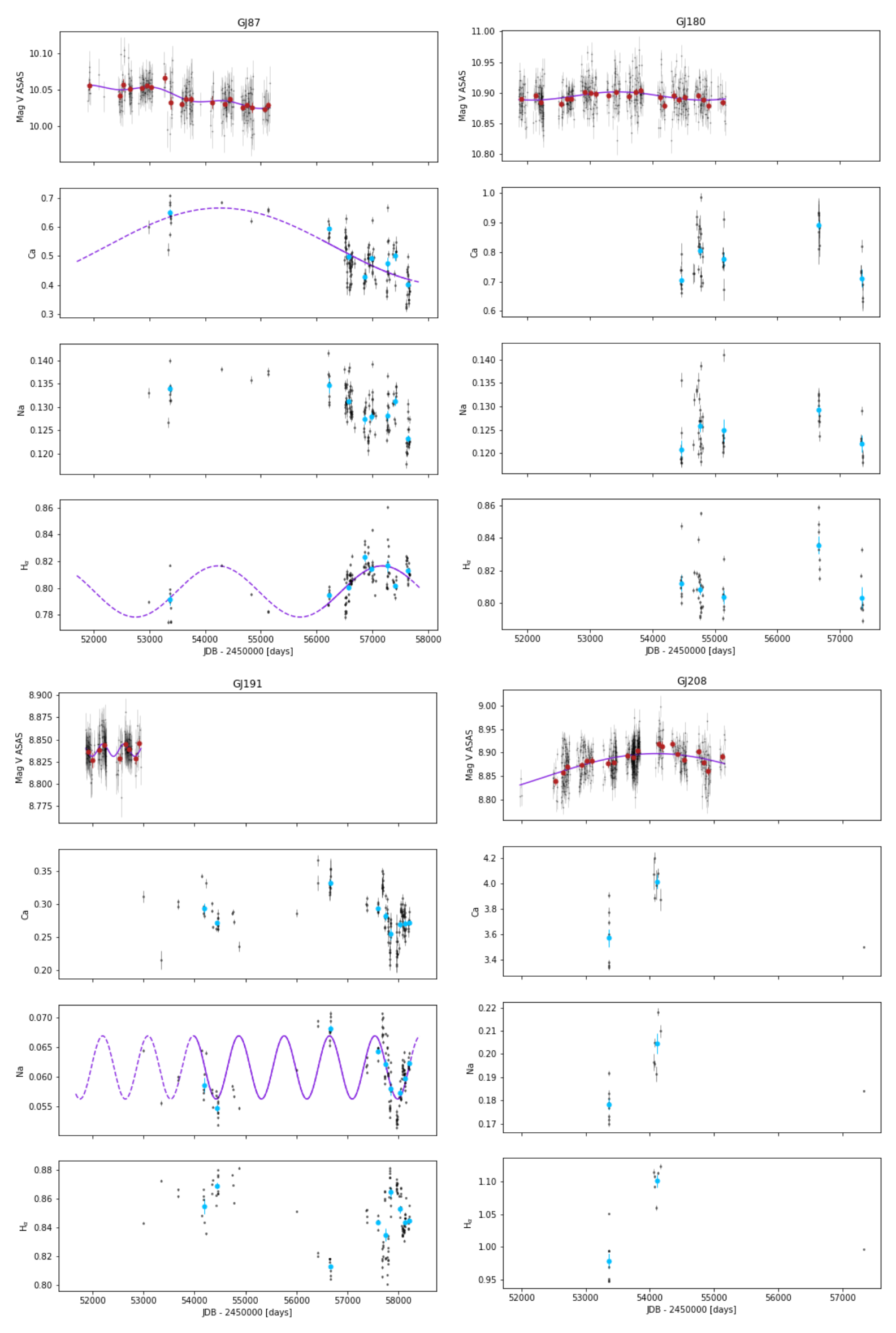}}%
    \caption{ Photometric and chromospheric (Ca, Na, and H$\alpha$) time series for the 43 stars with a fitted sinusoidal model (see Table~\ref{tab_recap}). Average photometry in 100-day bins is shown in red. Average chromospheric indices over seasons (see Sect.~\ref{SectSeason}) are shown in blue. The represented models (in purple) were obtained on the whole temporal span: the time range where the model is best constrained is shown as the solid line, and the time range in which it is poorly constrained is shown as the dashed line. Some of the photometric time series were fitted by two sinusoidal components (see Table~\ref{tab_recap} and Sect.~\ref{SectPhot}). }   
  \end{figure}
  
  \addtocounter{figure}{-1}
  \begin{figure}[h!]
     \centering
    \subfigure{%
       \includegraphics[width=0.85\textwidth]{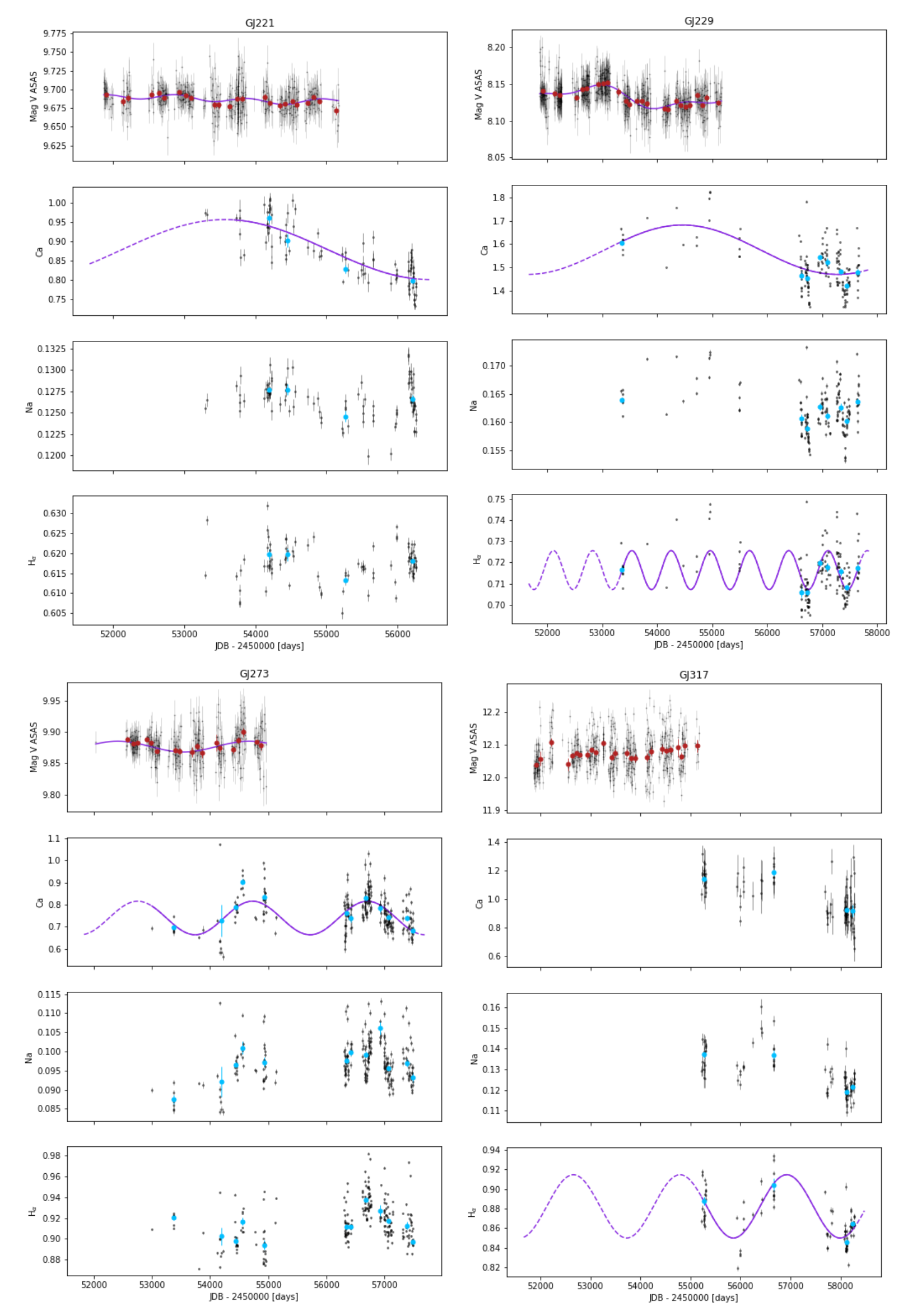}}%
    \caption{Photometric and chromospheric time series  (continued).}%
      \end{figure}
      \addtocounter{figure}{-1}
 
  \begin{figure}[h!]
       \centering
    \subfigure{%
       \includegraphics[width=0.85\textwidth]{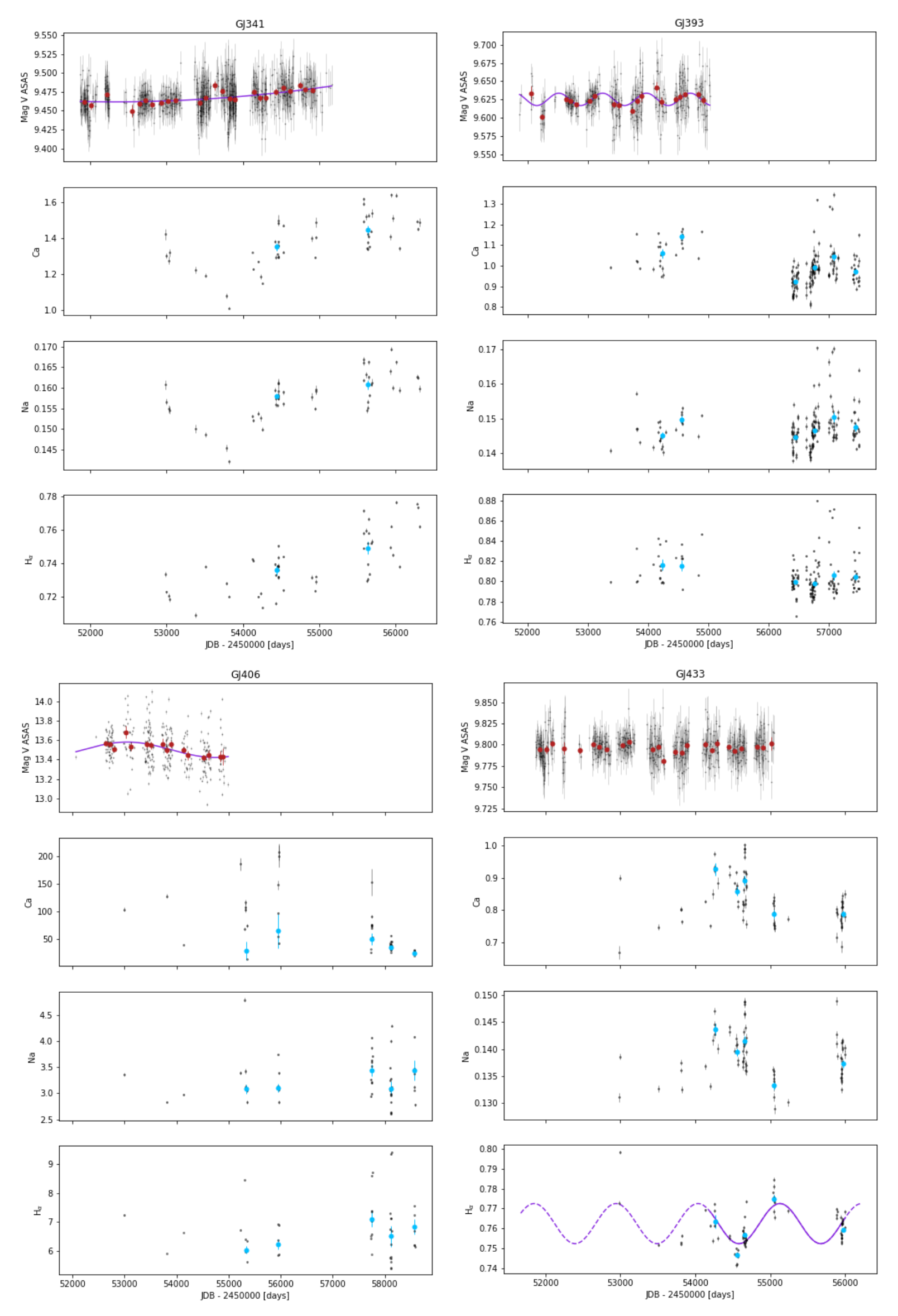}}%
    \caption{Photometric and chromospheric time series (continued).}%
      \end{figure}
 
      \addtocounter{figure}{-1}
  \begin{figure}[h!]
     \centering
    \subfigure{%
       \includegraphics[width=0.85\textwidth]{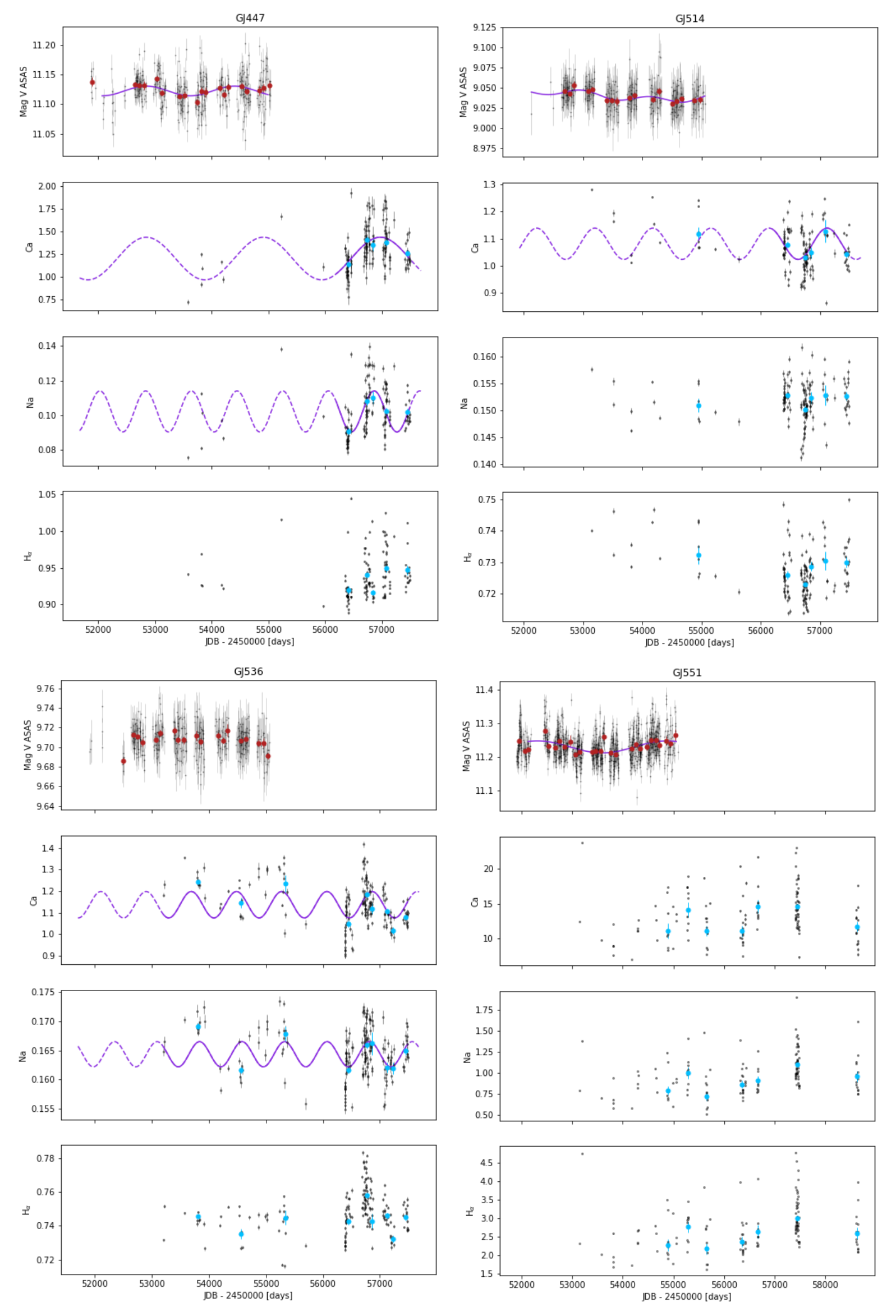}}%
    \caption{Fits of chromospheric time series (continued).}%
  \end{figure}
  
       \addtocounter{figure}{-1}
  \begin{figure}[h!]
     \centering
    \subfigure{%
       \includegraphics[width=0.85\textwidth]{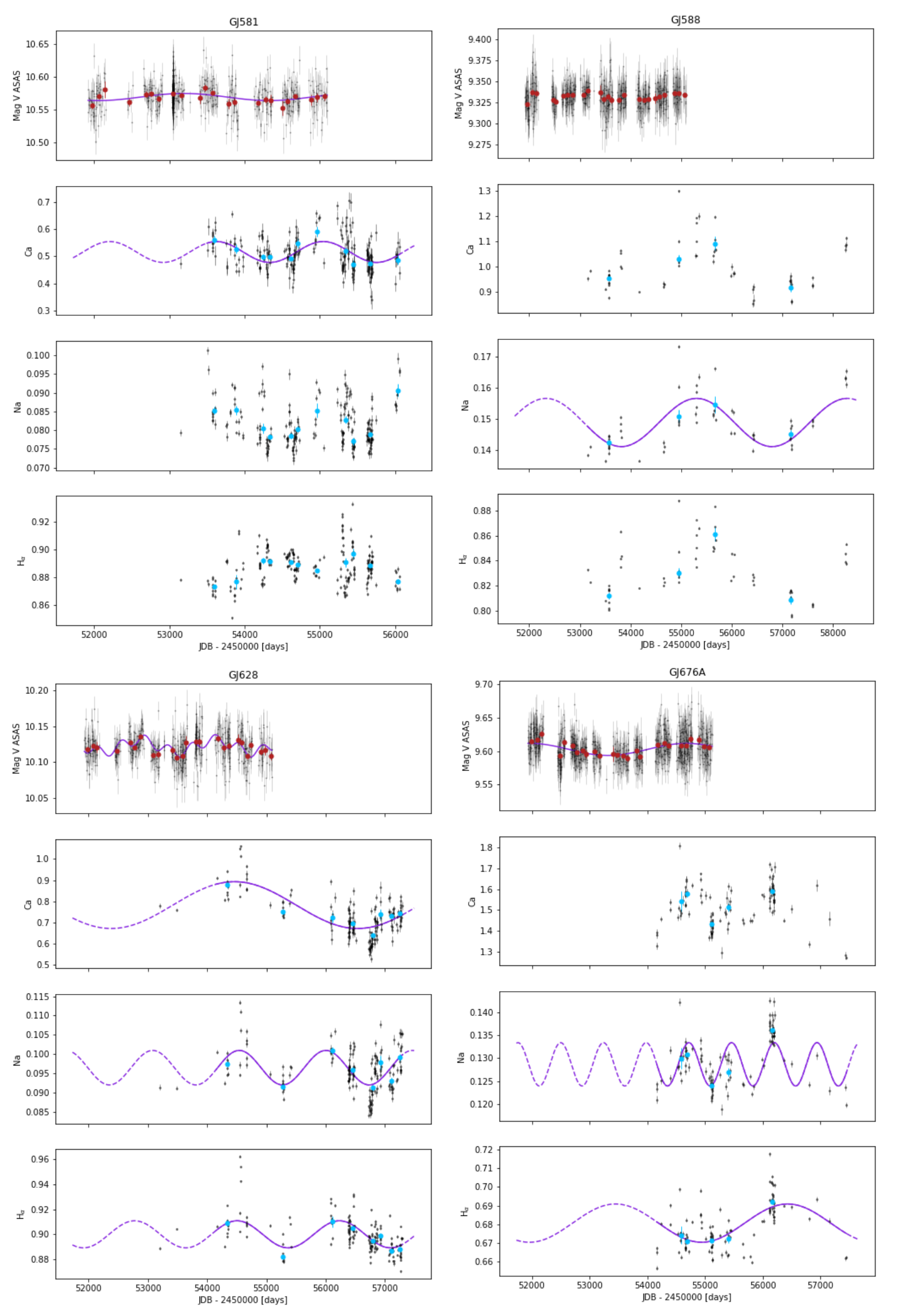}}%
    \caption{Fits of chromospheric time series (continued).}%
  \end{figure}
  
       \addtocounter{figure}{-1}
  \begin{figure}[h!]
     \centering
    \subfigure{%
       \includegraphics[width=0.85\textwidth]{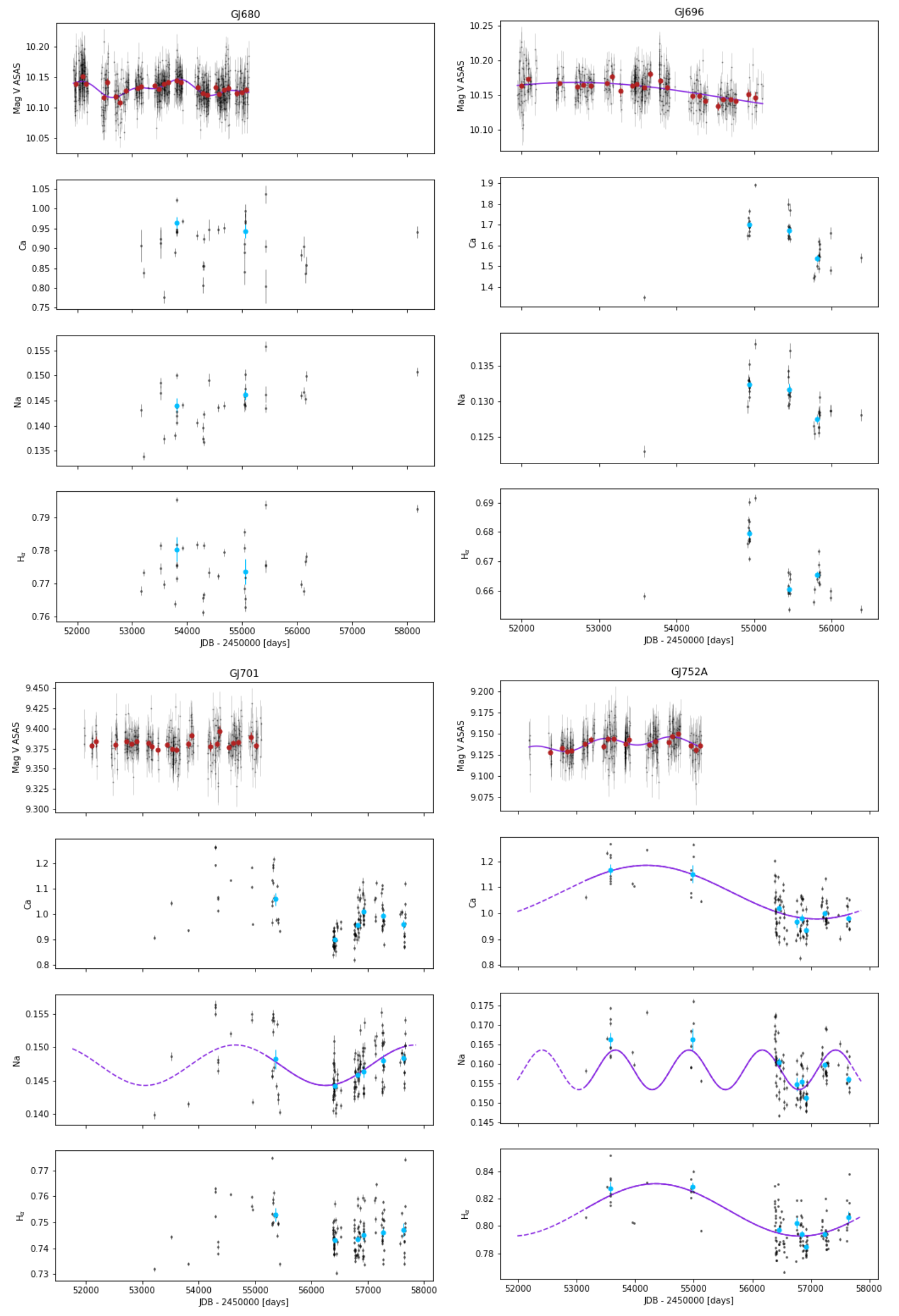}}%
    \caption{Fits of chromospheric time series (continued).}%
  \end{figure}
  
       \addtocounter{figure}{-1}
  \begin{figure}[h!]
     \centering
    \subfigure{%
       \includegraphics[width=0.85\textwidth]{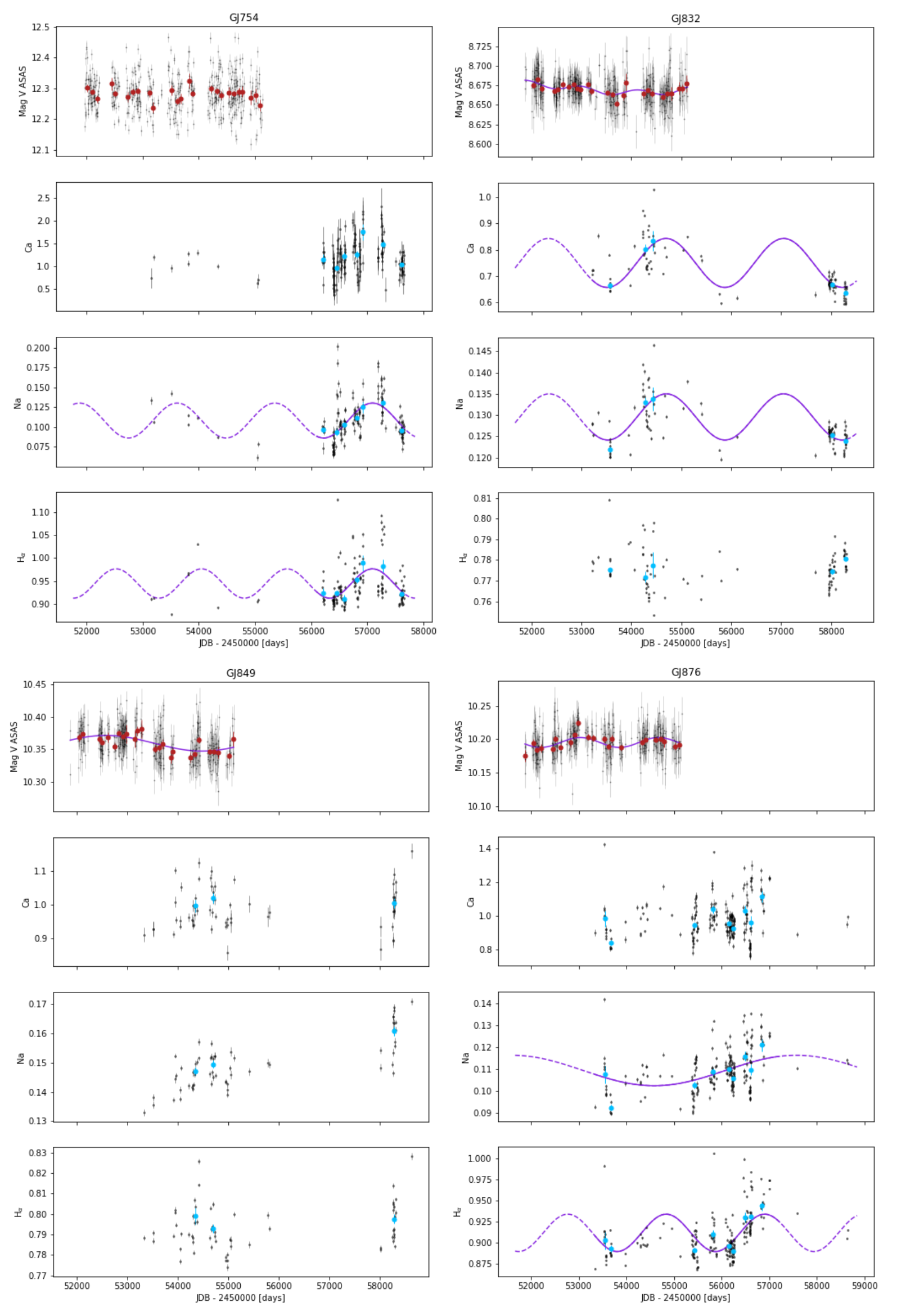}}%
    \caption{Fits of chromospheric time series (continued).}%
  \end{figure}
  
       \addtocounter{figure}{-1}
  \begin{figure}[h!]
     \centering
    \subfigure{%
       \includegraphics[width=0.85\textwidth]{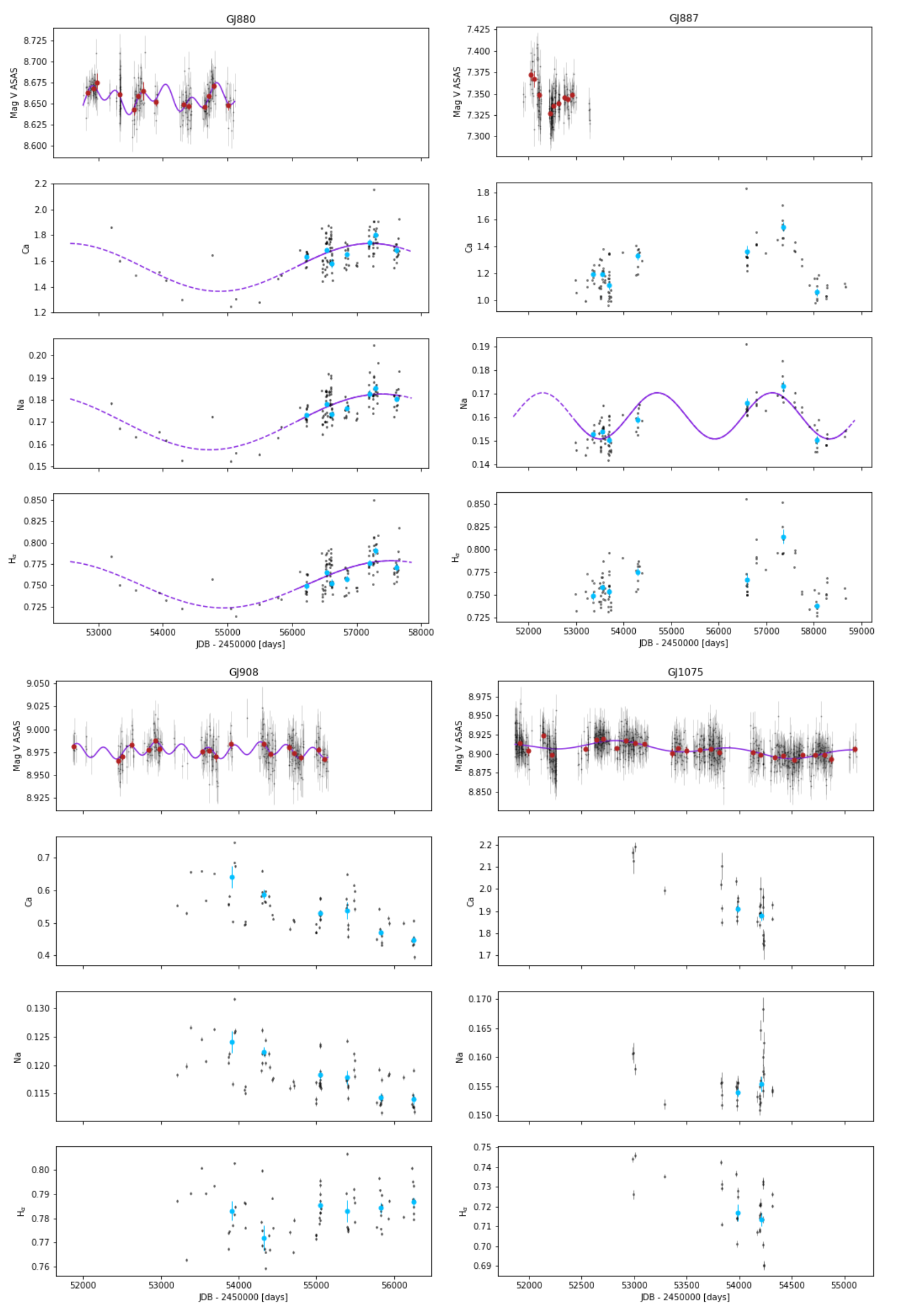}}%
    \caption{Fits of chromospheric time series (continued).}%
  \end{figure}
  
       \addtocounter{figure}{-1}
  \begin{figure}[h!]
     \centering
    \subfigure{%
       \includegraphics[width=0.85\textwidth]{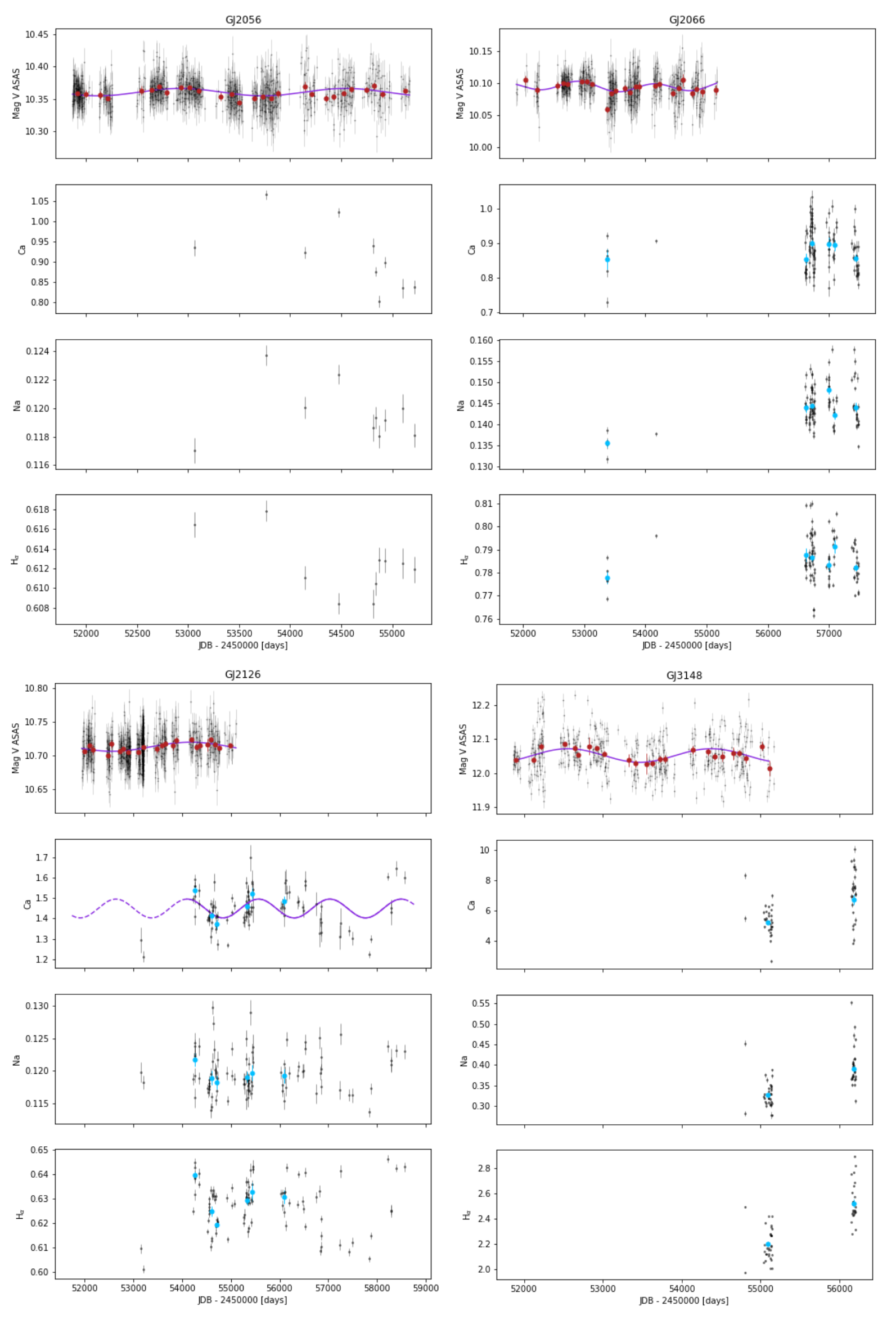}}%
    \caption{Fits of chromospheric time series (continued).}%
  \end{figure}
  
       \addtocounter{figure}{-1}
  \begin{figure}[h!]
     \centering
    \subfigure{%
       \includegraphics[width=0.85\textwidth]{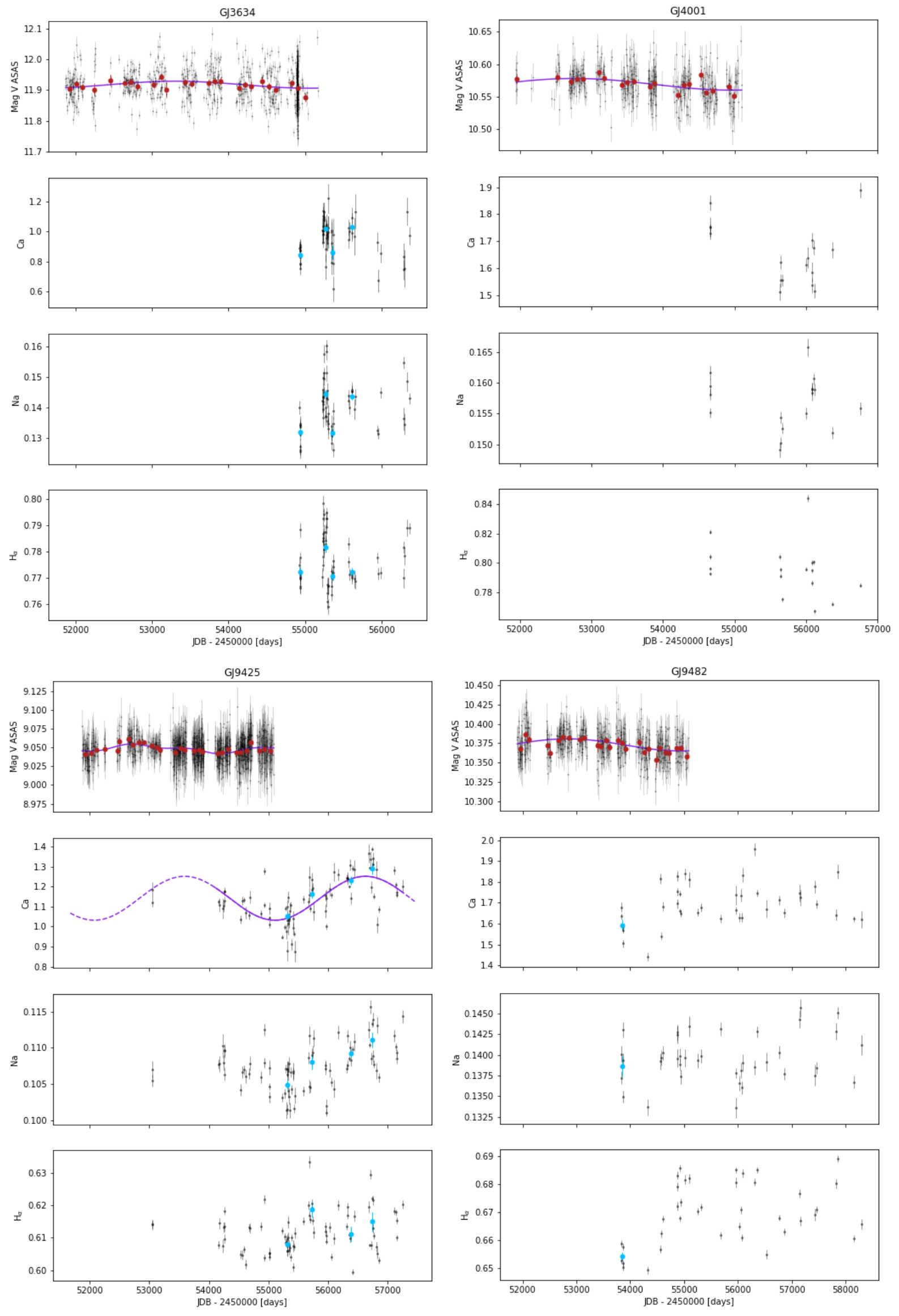}}%
    \caption{Fits of chromospheric time series (continued).}%
  \end{figure}
  
       \addtocounter{figure}{-1}
  \begin{figure}[h!]
     \centering
    \subfigure{%
       \includegraphics[width=0.85\textwidth]{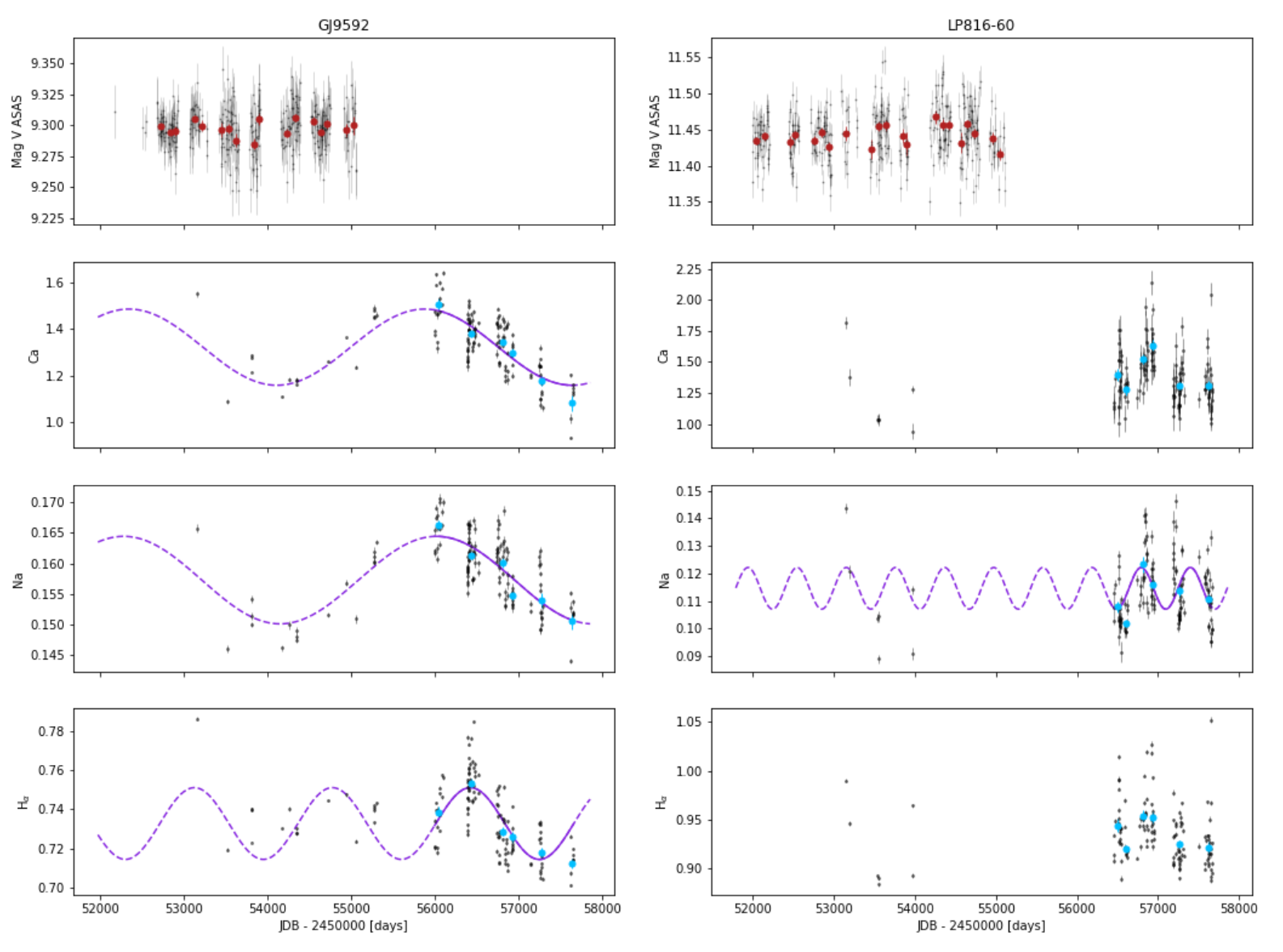}}%
    \caption{Fits of chromospheric time series (continued).}%
  \end{figure}

\end{appendix}

\end{document}